\documentclass[prd,aps,preprint, tightenlines, showpacs,nofootinbib,superscriptaddress]{revtex4-1}

\usepackage{epsfig}
\usepackage{amsmath}

\begin{document}
\title{Transverse Momentum Resummation for Dijet Correlation in Hadronic Collisions}

\author{Peng Sun}
\affiliation{Nuclear Science Division, Lawrence Berkeley National
Laboratory, Berkeley, CA 94720, USA}
\author{C.-P. Yuan}
\affiliation{Department of Physics and Astronomy, Michigan State University,
East Lansing, MI 48824, USA}
\author{Feng Yuan}
\affiliation{Nuclear Science Division, Lawrence Berkeley National
Laboratory, Berkeley, CA 94720, USA}

\begin{abstract}
We study the transverse momentum resummation for dijet correlation in hadron
collisions based on the Collins-Soper-Sterman formalism. The complete one-loop
calculations are carried out in the collinear factorization framework for the
differential cross sections at low imbalance transverse momentum between the
two jets. Important cross checks are performed to demonstrate that the soft
divergences cancelled out between different diagrams, and in particular, those
associated with final state jets. The leading and sub-leading logarithms are identified.
All order resummation is derived following the transverse momentum dependent
factorization at this order. Its phenomenological applications are also presented.
\end{abstract}
\maketitle

\section{Introduction}

Dijet production in hadronic collisions is one of the golden channels to
study perturbative QCD and hadron physics in high energy experiments.
In dijet events, the two jets are
produced mainly in the back-to-back configuration
in the transverse plane,
\begin{equation}
A+B\to Jet_1+Jet_2+X \ , \label{dijet}
\end{equation}
where $A$ and $B$ represent the two incoming hadrons
with momenta $P$ and $\overline{P}$, respectively,
the azimuthal angle between the two jets
is defined as $\phi=\phi_1-\phi_2$ with $\phi_{1,2}$
being the azimuthal angles of the two jets.
There have been comprehensive analyses of
the azimuthal angular correlation (or decorrelation)
in dijet events produced at
hadron colliders~\cite{Abazov:2004hm,Khachatryan:2011zj,daCosta:2011ni}.
In the leading order naive parton picture, the Born diagram
contributes to a Delta function at $\phi=\pi$.
One gluon radiation will lead to
a singular distribution around $\phi=\pi$,
which will persist at even higher orders~\cite{Nagy:2001fj}.
This divergence arises when the
total transverse momentum of dijet (imbalance) is much smaller
than the individual jet momentum,
$q_\perp=|\vec{P}_{1\perp}+\vec{P}_{2\perp}|\ll |P_{1\perp}|\sim |P_{2\perp}|\sim P_T$,
where large logarithms appear in every order
of perturbative calculations. These large logs
are normally  referred as the Sudakov logarithms,
$\alpha_s^i\ln^{2i-1}(P_T^2/q_\perp^2)$.
Therefore, a QCD resummation has to be
included in order to have a reliable theoretical
prediction. The goal of this paper is to derive the resummation formulas
for dijet production in the kinematics of these large logarithms. A brief
summary of our results has been published in Ref.~\cite{Sun:2014gfa}.

In the kinematics region $q_\perp\ll P_T$, the appropriate
resummation method to apply is the so-called transverse momentum dependent (TMD)
resummation or the Collins-Soper-Sterman (CSS) resummation~\cite{Collins:1984kg}.
The CSS resummation was derived for the massive neutral particle
production in hadronic collisions, such as the electroweak
boson (or Higgs boson) production. However, due to the presence of the colored
final state, the resummation of dijet production will be much more complicated.
There have been theoretical progressed in the leading double logarithmic approximation (LLA),
where it was found that each incoming parton contributes half of its color charge to
Sudakov resummation factor~\cite{Banfi:2003jj,Banfi:2008qs,Hautmann:2008vd,Mueller:2013wwa}. In this paper,
we will go beyond the LLA to perform the resummation calculation at the next-to-leading logarithm (NLL)
level. The threshold resummation for dijet production in hadronic collisions
has been investigated in a series of publications by Sterman and his collaborators~\cite{Kidonakis:1997gm,Botts:1989kf}.
The methodology of our calculations for the TMD resummation
is very similar to those studies.

We will start our derivation by evaluating the differential cross sections
as a function of
the imbalance transverse momentum $q_\perp$ at the complete one-loop order. The leading order
contribution is a Delta function $\delta^{(2)}(q_\perp)$. One-loop
corrections come from four contributions: (a) virtual contributions;
(b) soft gluon radiation (real); (c) jet contributions; (d) collinear gluon
radiation associated with the incoming parton distributions. The virtual
graphs have been studied in the literature~\cite{Ellis:1985er}. The jet contributions are
easy to derive following the examples of inclusive jet production. In order
to calculate the analytical results, we will adopt the narrow jet
approximation (NJA)~\cite{Mukherjee:2012uz,Jager:2004jh}, where explicit dependence on jet sizes $R_1$
and $R_2$ can be evaluated. The soft gluon (real) contribution
is most difficult to calculate. This is because, not all soft gluon
radiation contributes to the finite $q_\perp$. The radiation
inside the jet will be part of the jet contribution at one-loop order and
has to be excluded from soft gluon radiation contribution.
In our calculations, we will apply a small offshellness for the jets
in the final state to calculate the soft gluon radiation. We find that
this method yield the exact same results as that using kinematic cut-off
to regulate soft gluon radiation in the NJA. Collinear gluon radiation associated with
incoming parton distributions also contribute to the finite imbalance
transverse momentum. This part can be formulated according to
the well-known DGLAP splitting. In the end, for finite and soft imbalance
transverse momentum $q_\perp$, we add the soft gluon (real) and collinear
gluon (real) contributions together, which leads to the so-called asymptotic
behavior for the differential cross sections at low imbalance transverse
momentum $q_\perp$.
An important cross check of our derivation is
to compare its numerical results
with the dijet production codes available in public.
We will carry out these
comparisons in this paper.
Another cross check is to demonstrate the soft divergences cancellation
between real and virtual graphs. We will show these cancellations in details
for the hard partonic channels in our calculations. After the cancellation,
we are left with only collinear divergences associated incoming parton
distributions. This indicates that we do have a consistent results
for dijet production at one-loop calculations.

The large logarithms mentioned above will be evident from the
complete one-loop results. Resummation of these large logarithms
is the main goal of current paper. To perform the resummation, we first
show that the differential cross sections at one-loop order
can be factorized into the transverse momentum distributions, soft factor, and hard factors, respectively.
The transverse momentum distributions follow
the definitions for Drell-Yan or Higgs boson production at low transverse
momentum. The soft factor will have to take into account the additional
effect of gluon radiation associated with the two
final state jets. The idea to
construct the soft factor follows the examples of threshold
resummation studied by Sterman et al. \cite{Kidonakis:1997gm}.

Our resummation formula can be summarized as
\begin{eqnarray}
\frac{d^4\sigma}
{dy_1 dy_2 d P_T^2
d^2q_{\perp}}=\sum_{ab}\sigma_0\left[\int\frac{d^2\vec{b}_\perp}{(2\pi)^2}
e^{-i\vec{q}_\perp\cdot
\vec{b}_\perp}W_{ab\to cd}(x_1,x_2,b_\perp)+Y_{ab\to cd}\right] \ ,
\end{eqnarray}
where the first term $W$ contains all order resummation
and the second term $Y$ comes from the fixed order corrections.
$\sigma_0$ represents the overall normalization of the
differential cross section, $y_1$ and $y_2$ are rapidities of the two jets, $P_T$ the leading
jet transverse momentum, and $q_\perp$ the imbalance transverse momentum between
the two jets as defined above. All order resummation for $W$ from each
partonic channel $ab\to cd$ can be written as
\begin{eqnarray}
W_{ab\to cd}\left(x_1,x_2,b\right)&=&x_1\,f_a(x_1,\mu=b_0/b_\perp)
x_2\, f_b(x_2,\mu=b_0/b_\perp) e^{-S_{\rm Sud}(Q^2,b_\perp)} \nonumber\\
&\times& \textmd{Tr}\left[\mathbf{H}_{ab\to cd}
\mathrm{exp}\left[-\int_{b_0/b_\perp}^{Q}\frac{d
\mu}{\mu}\mathbf{\gamma}_{}^{s\dag}\right]\mathbf{S}_{ab\to cd}
\mathrm{exp}\left[-\int_{b_0/b_\perp}^{Q}\frac{d
\mu}{\mu}\mathbf{\gamma}_{}^{s}\right]\right]\ ,\label{resum}
\end{eqnarray}
where $Q^2=\hat s=x_1x_2S$, representing the hard momentum scale. $b_0=2e^{-\gamma_E}$,
with $\gamma_E$ being the Euler constant.
$f_{a,b}(x,\mu)$ are parton distributions for the incoming
partons $a$ and $b$, and $x_{1,2}=P_T\left(e^{\pm y_1}+e^{\pm y_2}\right)/\sqrt{S}$
are momentum fractions of the incoming hadrons carried by the partons.
In the above equation, the hard and soft factors $\mathbf{H}$ and $\mathbf{S}$
are expressed as matrices in the color space of partonic channel $ab\to cd$, and $\gamma^s$
are the associated anomalous dimensions for the soft factor (defined below). The Sudakov
form factor ${\cal S}_{Sud}$ resums the leading double logarithms and
the universal sub-leading logarithms,
\begin{eqnarray}
S_{\rm Sud}(Q^2,b_\perp)=\int^{Q^2}_{b_0^2/b_\perp^2}\frac{d\mu^2}{\mu^2}
\left[\ln\left(\frac{Q^2}{\mu^2}\right)A+B+D_1\ln\frac{Q^2}{P_T^2R_1^2}+
D_2\ln\frac{Q^2}{P_T^2R_2^2}\right]\ , \label{su}
\end{eqnarray}
where $R_{1,2}$ represent the cone sizes for the two jets, respectively.
Here the parameters $A$, $B$, $D_1$, $D_2$ can be expanded
perturbatively in $\alpha_s$. At one-loop order,
$A=C_A \frac{\alpha_s}{\pi}$,
$B=-2C_A\beta_0\frac{\alpha_s}{\pi}$ for gluon-gluon initial state,
$A=C_F \frac{\alpha_s}{\pi}$,
$B=\frac{-3C_F}{2}\frac{\alpha_s}{\pi}$ for quark-quark initial state,
and $A=\frac{(C_F+C_A) }{2}\frac{\alpha_s}{\pi}$,
$B=(\frac{-3C_F}{4}-C_A\beta_0)\frac{\alpha_s}{\pi}$ for gluon-quark initial state.
Here, $\beta_0=(11-2N_f/3)/12$,
with $N_f$ being the number of effective light quarks.
At the next-to-leading logarithmic level, the jet cone size enters as well~\cite{Banfi:2003jj}.
That is the reason we have two additional factors in Eq.~(\ref{su}):
$D_{1,2}=C_A\frac{\alpha_s}{2\pi}$ for gluon jet and $D_{1,2}=C_F\frac{\alpha_s}{2\pi}$ for quark jet.
The cone size $R$ is introduced to regulate the collinear
gluon radiation associated with the final state jets. Only the soft
gluon radiation outside the jet cone will contribute to the imbalance $q_\perp$ between the two jets.

There are two important issues we will not addressed in much details. First,
our resummation formalism is based on a TMD factorization argument~\cite{Qiu:2007ey}. However,
there is a potential contribution at order of $\alpha_s^3$ which violates the
general TMD factorization~\cite{Collins:2007nk,Vogelsang:2007jk,Mulders:2011zt,Catani:2011st,Mitov:2012gt}. 
It was found, in particular, certain
diagrams in dijet production in hadronic collisions can not be factorized
into the simple universal TMDs.
In terms of resummation coefficients, this will affect the
coefficient $A^{(3)}$ in Eq.~(\ref{su}).
In our numeric calculations, we will include both $A^{(1)}$ and $A^{(2)}$ coefficients
in the resummation formula (see detailed discussions in Sec.~IIC).
Though we do not expect that the factorization violating effect
will affect much the results to be presented below,
it will be useful to estimate the size of
factorization breaking effect in some future work.

Second, we derive our results, in both collinear and TMD approaches,
by adopting the narrow jet approximation. With that, we are able to
demonstrate the explicit cancellation of soft divergences from various
parts of analytical calculation,
and to express the resummation Sudakov form factor in an analytic form
in terms of the jet sizes. Although the derivation itself is not limited to NJA, we
find that the differential cross section expressions are much simplified under
the NJA. How to extend our results to the case without NJA is very interesting
subject. It is worthwhile to come back to this question in the future. However,
in current paper, we will carry out our calculations with the NJA.

The rest of this paper is organized as follows. We will dedicate
Secs. II-V to the calculations of the differential cross sections
at low $q_\perp$ in the collinear factorization approach. In Sec. II,
we start with a brief introduction of the leading order results
in which the overall normalization factor for each
production channel is presented.
In Sec. III, we discuss the generic features to evaluate the one-loop
corrections to $W(b_\perp)$. This includes virtual graph contributions,
jet contributions in the real gluon radiation, collinear gluon radiation
associated with the incoming partons, and the soft gluon radiation
which contributes to finite $q_\perp$. Since the soft gluon radiation
is the most important contributions for dijet production, we dedicate
Sec. IV to show the details of our calculations. In Sec. V, we compare
the asymptotic results for dijet production to the fixed order
calculation published in the literature. The asymptotic behavior for
dijet production is calculated by summing the soft gluon radiation
and collinear gluon radiation derived in Secs. III and IV.
In Sec. VI, we derive the TMD factorization and resummation.
In particular, we will show that the collinear calculations from Secs.~III-IV
at the one-loop order can be factorized into the TMDs,
hard and soft factors. The latter factors are expressed in a matrix form,
where they are written in terms of the color space for every partonic
channel. The resummation is achieved by solving the relevant
renormalization equations. In Sec. VII, we perform the phenomenological
studies based on our resummation formalism.
We conclude our paper in Sec. VIII.

\section{Dijet Production at the Leading Order}

\begin{figure}[tbp]
\centering
\includegraphics[width=6cm]{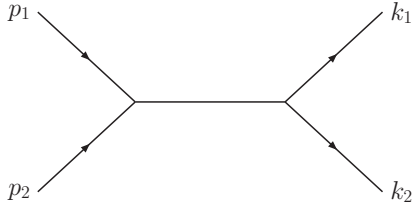}
\caption{Schematic diagram for dijet production at the leading order, with
incoming parton momenta: $p_1$ and $p_2$, and outgoing momenta:
$k_1$ and $k_2$, respectively.}
\label{dijet0}
\end{figure}

Dijet production at the leading order can be calculated
from partonic $2\to 2$ processes,
\begin{equation}
a(p_1)+b(p_2)\to c(k_1)+d(k_2) \  ,
\end{equation}
where $p_{1,2}$ and $k_{1,2}$ are momenta for
incoming and outgoing two partons. Schematically,
we draw the diagrams in Fig.~\ref{dijet0}.

The partonic channels include the following subprocesses:
\begin{eqnarray}
&&q_iq_j\to q_iq_j \ ,\\
&&q_i\bar q_i\to q_j\bar q_j\ , \\
&&q_iq_i\to q_i q_i\ , \\
&&qg\to qg \ ,\\
&&gg\to q\bar q\ ,\\
&&q\bar q\to gg\ , \\
&&gg\to gg\ .
\end{eqnarray}
Their contributions can be summarized as
\begin{eqnarray}
\frac{d^4\sigma}
{dy_1 dy_2 d P_T^2
d^2q_{\perp}}=\sum_{ab}\sigma_0x_1\,f_a(x_1,\mu)
x_2\, f_b(x_2,\mu) {h}_{ab\to cd}^{(0)}\delta^{(2)}(q_\perp) \ ,
\end{eqnarray}
where the overall normalization of the differential cross section
is $\sigma_0=\frac{\alpha_s^2\pi}{s^2}$.
The partonic cross sections $h^{(0)}$ for
all the production channels are listed below.
\begin{eqnarray}
h_{q_iq_j\to q_iq_j}^{(0)}&=&\frac{4}{9}\frac{s^2+u^2}{t^2}\ ,\\
h_{q_i\bar q_i\to q_j\bar q_j}^{(0)}&=&\frac{4}{9}\frac{t^2+u^2}{s^2}\ ,\\
h_{q_iq_i\to q_iq_i}^{(0)}&=&\frac{4}{9}\left(\frac{s^2+u^2}{t^2}+\frac{s^2+t^2}{u^2}\right)-\frac{8}{27}\frac{s^2}{tu} \ ,\\
h_{gg\to q\bar q}^{(0)}&=&\frac{1}{6}\frac{u^2+t^2}{tu}-\frac{3}{8}\frac{u^2+t^2}{s^2}\ ,\\
h_{qg\to qg}^{(0)}&=&\frac{4}{9}\frac{u^2+s^2}{-us}+\frac{u^2+s^2}{t^2}\ ,\\
h_{gg\to gg}^{(0)}&=&\frac{9}{2}\left(3-\frac{ut}{s^2}-\frac{us}{t^2}-\frac{st}{u^2}\right)\ ,\\
\end{eqnarray}
where the kinematic variables
$s = (p_1+p_2)^2$, $t = (p_1-k_1)^2$ and $u = (p_1-k_2)^2$.
As mentioned above, at the leading order, they contribute to a Delta
function of $q_\perp$, which corresponds to the back-to-back configuration
of the two jets in the transverse plane. If we translate this
into the $b_\perp$-space, we will obtain $W(b_\perp)$ at the leading
order take the following form,
\begin{equation}
W_{ab\to cd}^{(0)}(b_\perp)=x_1f_a(x_1,\mu)x_2f_b(x_2,\mu)h_{ab\to cd}^{(0)} \ .
\end{equation}
The goal of the following three sections is to derive the one-loop corrections
to $W_{ab\to cd}^{(1)}(b_\perp)$.

\section{Generic Discussions on One-loop Calculations}

The leading order contributions lead to a Delta function of the imbalance
transverse momentum $q_\perp$. In the following, we will first carry out
one-loop calculations of the differential cross sections at low transverse
momentum. The computation is performed in the $b_\perp$-space, the Fourier
transformed conjugate parameter to the transverse momentum $q_\perp$.
That is to say, we will calculate $W(b_\perp)$ at the one-loop order.

\begin{figure}
\centering
\includegraphics[width=12cm]{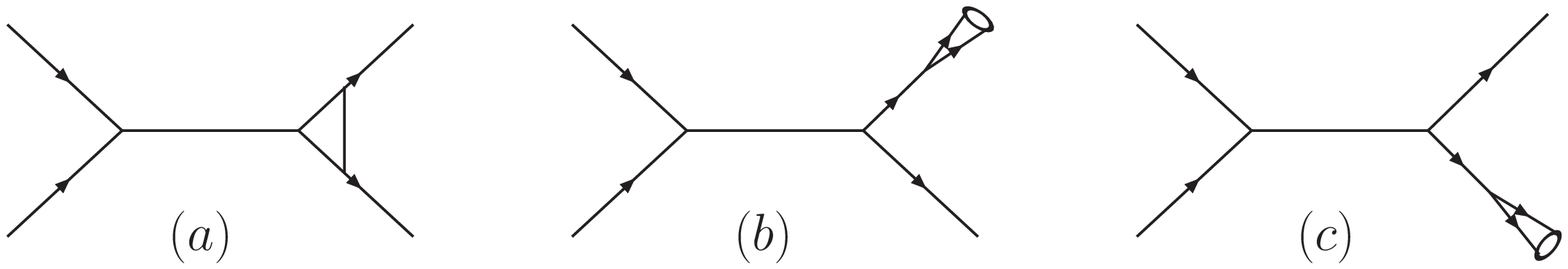}
\caption{Schematic diagrams for virtual graph contribution (a) and
final state jet contributions (b) and (c) at one-loop order.
Both of them are proportional to a Delta function of the imbalance
transverse momentum: $\delta^{(2)}(q_\perp)$.}
\label{dijetv}
\end{figure}

There are four types of radiative contributions at the one-loop order:
\begin{enumerate}
\item Virtual contributions, as shown in Fig.~\ref{dijetv}(a). These contributions
have been calculated in the literature, and they are
proportional to the leading order contributions, leading to a Delta function
of $q_\perp$.
\item Real gluon radiation: Jet contributions. In the real gluon radiation,
one particular contribution is also proportional to the leading order contribution, that is
the jet contribution. In this case, the gluon radiation is within the jet, where its
momentum is collinear to the final state parton. For example, as shown
Fig.~\ref{dijetv}(b), the radiated gluon is collinear to one of the final state parton, and they form a
new jet at one-loop order. Because of momentum conservation, this again
leads to a Delta function of $q_\perp$.
\item Real gluon radiation: collinear gluon associated with incoming partons.
These collinear gluons contribute to a finite $q_\perp$, and is proportional to
$1/q_\perp^2$ multiplied by the splitting kernel of parton distribution functions. Since they
originate from the incident partons, the collinear gluon contributions follow
the generic structure and are easy to calculate.
\item Real gluon radiation: soft gluon contribution. Soft gluon contributions are
more difficult to evaluate. They also contribute to a finite $q_\perp$. To evaluate
this part of contribution, we apply the leading power expansion in the limit
of $q_\perp\ll Q$. However, because the final state jets also carry color, the soft
gluon radiation has to take into account the interactions among initial and final state
partons. In addition, we have to exclude the soft gluon radiation within the final
state (cone) jets whose contributions have already been included
in the final state jet contributions. Detailed calculations will be presented
in the following section.
\end{enumerate}
In the rest of this section, we will go through the first three kinds of contributions, whereas
the soft gluon radiation contribution will be calculated in Sec. IV.

We will carry out our calculations in the collinear factorization formalism, and apply
the dimensional regulation for IR and UV divergences with dimension $D=4-2\epsilon$.
Various divergences appear in individual contributions: $1/\epsilon^2$ represents the soft
divergence, whereas $1/\epsilon$ for either soft or collinear divergence. Since we are dealing with
jet production in the final states, we will also encounter the divergences associated with
the jet size $R$. The explicit calculations of gluon radiation will depend on how we define
the jet, i.e., the jet algorithm will play a role in formulating the one-loop corrections.

Two important cross checks will be performed in the derivations. First, the soft divergences
of $1/\epsilon^2$ will be cancelled out completely among different terms. We notice
that (1,2,4) terms listed above will have $1/\epsilon^2$ contributions. A crucial test of
our calculations is that these $1/\epsilon^2$ cancel out each other. This cancellation is
not trivial, since they come from different diagrams, and some with different color factors.
However, we will show that the total contribution is free of soft divergence of $1/\epsilon^2$.
Second, the divergences associated with the final state jets are also cancelled out among various
terms.  The collinear divergences
associated with the jets are regulated by the jet sizes, and the individual contributions
contain terms of $1/\epsilon \ln (1/R)$ are cancelled out in the final results. We will
show that the cancellation indeed happens at this order.

\begin{figure}[tbp]
\centering
\includegraphics[width=9cm]{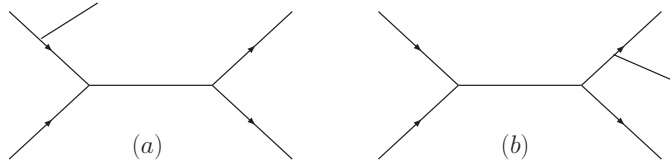}
\caption{Schematic diagrams for real gluon radiation contribution to finite imbalance transverse
momentum: (a) collinear gluon radiation associated with the incoming
partons; (b) soft gluon radiation outside the jet cone of final state jets.}
\label{dijet-r}
\end{figure}

\subsection{Virtual Graphs}
Virtual diagrams for dijet production in hadronic collisions have been calculated in the
classic paper of Ref.~\cite{Ellis:1985er}. In our calculations, we will take their results.

\subsection{Jet Contributions}
Jet contributions contain the collinear gluon radiation and gluon to quark-antiquark
splitting in the final state. The schematic diagrams have been shown in Figs.~\ref{dijetv}(b) and (c).
Because we have two jets in the final state, one-loop jet corrections can come
from either of the jets, as shown in Fig.~\ref{dijetv}(b) and (c), respectively.
The jet contribution needs to be included to capture the collinear gluon radiation (or gluon splitting to
quark-antiquark pair) within the jet cone. These gluon radiations will not change
the kinematics of the parenting parton, and therefore contribute to a Delta function
of $q_\perp$, which is similar to the virtual graph contributions. The requirement
is that the two partons in the splitting process form a jet according to
the jet algorithm adopted in experimental measurements. In order to derive an analytic
expression with the jet cone size dependence to demonstrate the
cancellations in the final results and to derive the resummation formula, we
apply the narrow jet approximation (NJA) in our calculations.
In particular, we follow the technique and scheme of the subtraction developed in
Refs.~\cite{Mukherjee:2012uz,Jager:2004jh}. The basic idea is to note that in the
collinear gluon radiation limit for each final state jet,
\begin{equation}
|{\cal M}(2\to 3)|^2\approx |{\cal M}_0(2\to 2)|^2\times {\cal P}_{1\to 2} \ ,
\end{equation}
where ${\cal M}(2\to 3)$ represents the scattering amplitude of $2\to 3$
subprocess, ${\cal M}_0(2\to 2)$ for the leading order $2\to 2$
subprocess with one of the final state parton branching into two parton
final state represented by the splitting of ${\cal P}_{1\to 2}$.

Therefore, the jet contributions can be
summarized as,
\begin{eqnarray}
\int\frac{d^3k_3}{2E_3(2\pi)^3}\frac{E_J}{E_2}{\cal P}_{1\to 2} \ ,
\end{eqnarray}
where the factor $E_J/E_2$ accounts for the phase space difference
from 3 parton final state to 2 parton final state. The phase space
integral of the above equation is limited that the two partons are
within the jet cone. Here the difference from jet algorithms plays
a role. Using the NJA, the calculations follow what have been
done in Refs.~\cite{Mukherjee:2012uz,Jager:2004jh}, and in particular, we find that the
gluon jet in the final state contributes
\begin{eqnarray}
{\cal J}^g&=&\frac{\alpha_s}{2\pi}\frac{1}{\Gamma(1-\epsilon)}\frac{1}{-2\epsilon}\left(\frac{P_T^2R^2}{4\pi}\right)^{-\epsilon}
\int_0^1 d\xi \left(\xi(1-\xi)\right)^{-2\epsilon}\left[f_{gg}(\xi)+f_{qg}(\xi)\right] \ ,
\end{eqnarray}
for $k_t$-type jet algorithm, where $R$ defines the
jet cone size: $R=\sqrt{(\Delta y)^2+(\Delta\phi)^2}$. Here,
$\Delta y$ and $\Delta\phi$ are the rapidity difference and
azimuthal angle difference between the two partons which define
the jet.  In the above equation, $f_{gg}(\xi)$ and $f_{qg}(\xi)$ are splitting
kernels for gluon to gluon and gluon to quark-antiquark pair~\cite{Mukherjee:2012uz,Jager:2004jh},
respectively.
By applying the $\overline{\rm MS}$ subtraction, we obtain,
\begin{eqnarray}
{\cal J}^g=\frac{\alpha_sC_A}{2\pi}\left[\frac{1}{\epsilon^2}+\frac{1}{\epsilon}
\left(2\beta_0-\ln\frac{P_T^2R^2}{\mu^2}\right)+I^g\right] \ ,
\end{eqnarray}
where we have taken into account the contributions from both
$g\to gg$ and $g\to q\bar q$ splittings. We would like to emphasize
that the singular terms (double and single poles) are
independent of jet algorithm. The jet algorithm dependent contributions arise
from the finite term $I^g$. The above jet function is universal, and
can be used in many other cases as well. Similarly, we find that the
quark jet can be written as
\begin{eqnarray}
{\cal J}^q=\frac{\alpha_sC_F}{2\pi}\left[\frac{1}{\epsilon^2}+\frac{1}{\epsilon}
\left(\frac{3}{2}-\ln\frac{P_T^2R^2}{\mu^2}\right)+I^q\right] \ .
\end{eqnarray}
In the $k_t$-type jet algorithms, the above mentioned $I^q$ and $I^g$ terms
are already available:
\begin{eqnarray}
I^g&=&\frac{1}{2}\left(\ln\frac{P_T^2R^2}{\mu^2}\right)^2-2\beta_0\ln\frac{P_T^2R^2}{\mu^2}
+\frac{67}{9}-\frac{3}{4}\pi^2-\frac{23}{54}N_f\ , \\
I^q&=&\frac{1}{2}\left(\ln\frac{P_T^2R^2}{\mu^2}\right)^2-\frac{3}{2}\ln\frac{P_T^2R^2}{\mu^2}
+\frac{13}{2}-\frac{3}{4}\pi^2\ .
\end{eqnarray}
We note that the double logarithmic terms
are independent of jet algorithm, and it is important
for them to cancel in the final results, as shown in Sec. IV.

\subsection{Collinear Gluon Radiation}

The contributions from the collinear gluon associated with the incoming
partons can be easily evaluated, and they are found to be
proportional to the splitting kernel at the one-loop order:
\begin{equation}
\frac{\alpha_s}{2\pi^2}\frac{1}{q_\perp^2}\int\frac{dx_1'}{x_1'}\frac{dx_2'}{x_2'}
f_{a'}(x_1') f_{b'}(x_2')\left[\delta(\xi_2-1)\xi_1{\cal P}_{a/a'}^{(<)}(\xi_1)+(\xi_1\leftrightarrow \xi_2)\right] \ ,
\end{equation}
where $\xi_i=x_i/x_i'$ and ${\cal P}^{(<)}_{a/a'}$ represents the
splitting kernel part without the Delta function term whose effect is
included in the above mentioned virtual contributions.

\begin{figure}[tbp]
\centering
\includegraphics[width=9cm]{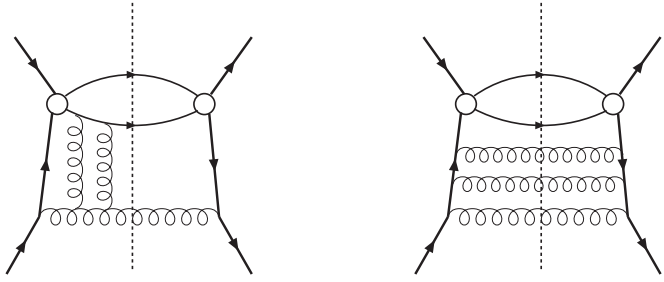}
\caption{Example of Feynman diagrams that contribute to
the TMD factorization breaking effects in two particle production via
hadronic processes at ${\cal O}(\alpha_s^3)$ order~\cite{Collins:2007nk} (left),
where two gluon exchanges between the incoming/outgoing particles
induce non-universality of the TMDs in this process as compared to
other processes such as DIS and Drell-Yan processes. These diagrams
would contribute to $A^{(3)}$ in the resummation formula.
The dominant contribution at this order is illustrated in the right panel,
which are factorizable and come from $A^{(1)}$ in the resummation formula.}
\label{factorizationbreak}
\end{figure}

To evaluate the contribution from soft gluon radiation requires more care. This is because
we have to exclude the collinear contributions associated
with the final state jet which will be factorized into a jet function
and will not contribute to finite $q_\perp$.
Before we discuss its details in the following section, we note that
the TMD factorization breaking effects can appear at higher orders
 in two particle production at hadron colliders~\cite{Collins:2007nk,Vogelsang:2007jk,Mulders:2011zt,Catani:2011st,Mitov:2012gt}.
These effects come
from the diagrams illustrated in Fig.~\ref{factorizationbreak}~\cite{Collins:2007nk},
which belong to a nontrivial contribution at order $\alpha_s^3$ to the
parton distributions. They cannot be factorized into the conventional
transverse momentum dependent parton distributions, though they
could contribute to the $A^{(3)}$ coefficient in the resummation formula.
Since it is beyond the perturbative order (up to $\alpha_s^2$)
discussed in this paper, we shall not discuss it further in this work.

\section{Soft Gluon Radiation at One-loop Order}

\begin{figure}[tbp]
\centering
\includegraphics[width=12cm]{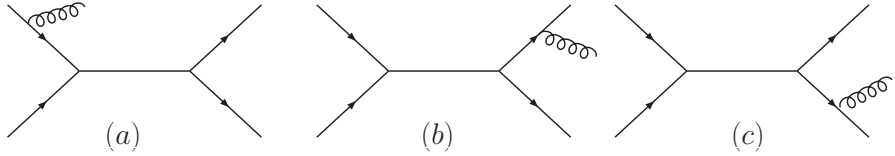}
\caption{Soft gluon radiation contribution to finite imbalance transverse
momentum $q_\perp$: (a) initial state radiation, and (b) and (c) final state radiations.
Because we have chosen the gluon polarization vector along $p_2$, there
is no contribution connecting to $p_2$ line.}
\label{dijet-s}
\end{figure}

For soft gluon radiations, we can apply
the leading power expansion and derive the dominant contribution by
the Eikonal approximation. This analysis has been applied in
Ref.~\cite{Mueller:2013wwa} to obtain the leading double logarithmic contributions
to dijet production. In the current paper, we will extend that analysis
to include the subleading logarithmic contributions as well.
In particular, we will derive the contributions which depend on
the jet cone sizes. The relevant Feynman rules have been listed in Ref.~\cite{Mueller:2013wwa}.
For completeness, we copy these results here.
For outgoing quark, antiquark and gluon lines, we have
\begin{equation}
\frac{2k_i^\mu}{2k_i\cdot k_g+i\epsilon} g\ ,~~-\frac{2k_i^\mu}{2k_i\cdot k_g+i\epsilon}g \ ,~~\frac{2k_i^\mu}{2k_i\cdot k_g+i\epsilon}g \ ,
\end{equation}
respectively,  where $k_i$ represents the momentum of the outgoing particles.
For incoming quark, antiquark and gluon lines, we have,
\begin{equation}
-\frac{2p_1^\mu}{2p_1\cdot k_g-i\epsilon} g\ ,~~\frac{2p_1^\mu}{2p_1\cdot k_g-i\epsilon}g \
,~~\frac{2p_1^\mu}{2p_1\cdot k_g-i\epsilon} g \ ,
\end{equation}
respectively, where $p_1$ represents the momentum for the incoming particle.

Following Ref.~\cite{Mueller:2013wwa}, we choose the physical polarization
for the soft gluon along the incoming particle $p_2$. Therefore, the soft
gluon radiation from the incoming particle $p_2$ vanishes with this
polarization choice. From that,
we can derive the soft gluon radiation contribution easily, with the polarization
tensor for the radiated gluon,
\begin{equation}
\Gamma^{\mu\nu}(k_g)=\left(-g^{\mu\nu}+\frac{k_g^\mu p_2^\nu+k_g^\nu p_2^\mu}{k_g\cdot p_2}\right)  \ .
\end{equation}
For example, from the amplitude squared of the soft gluon radiation
terms in the above, we have
\begin{eqnarray}
\frac{2p_1^\mu}{2p_1\cdot k_g}\frac{2p_1^\nu}{2p_1\cdot k_g}\Gamma_{\mu\nu}
&=&S_g(p_1,p_2)\ ,\\
\frac{2k_1^\mu}{2k_1\cdot k_g}\frac{2k_1^\nu}{2k_1\cdot k_g}\Gamma_{\mu\nu}
&=&S_g(k_1,p_2)\ ,\\
\frac{2k_2^\mu}{2k_2\cdot k_g}\frac{2k_2^\nu}{2k_2\cdot k_g}
\Gamma_{\mu\nu}
&=&S_g(k_2,p_2)\ ,
\end{eqnarray}
where $S_g(p,q)$ is a short-handed notation for
\begin{equation}
S_g(p,q)=\frac{2p\cdot q}{p\cdot k_gq\cdot k_g} \ .
\end{equation}
Similarly, we derive the interferences between them,
\begin{eqnarray}
2\frac{2k_1^\mu}{2k_1\cdot k_g}\frac{2p_1^\nu}{2p_1\cdot k_g}\Gamma_{\mu\nu}
&=&S_g(k_1, p_2)+S_g(p_1, p_2)-S_g(k_1, p_1)\ ,\\
2\frac{2k_2^\mu}{2k_2\cdot k_g}\frac{2p_1^\nu}{2p_1\cdot k_g}\Gamma_{\mu\nu}
&=&S_g(k_2, p_2)+S_g(p_1, p_2)-S_g(k_2, p_1)\ ,\\
2\frac{2k_1^\mu}{2k_1\cdot k_g}\frac{2k_2^\nu}{2k_2\cdot k_g}\Gamma_{\mu\nu}
&=&S_g(k_1, p_2)+S_g(k_2, p_2)-S_g(k_1, k_2)\ .
\end{eqnarray}
In order to evaluate the contributions from soft gluon
radiation, we integrate out the phase space of the
gluon whose transverse momentum leads
to the imbalance between the two jets, e.g.,
\begin{eqnarray}
g^2\int \frac{d^{D-1}k_g}{(2\pi)^{D-1}2E_{k_g}}\delta^{(2)} (q_\perp-k_{g\perp})S_g(p_1,p_2) \ , \label{sg120}
\end{eqnarray}
where we have chosen dimensional regulation for the
phase space integral.
The derivation of the above term is straightforward, by noticing that
the lower limit in the longitudinal momentum fraction integral,
\begin{eqnarray}
\int_{x_{min}}^1\frac{dx}{x}\frac{1}{k_{g\perp}^2} \ ,
\end{eqnarray}
where we have defined $x$ as momentum fraction of $p_1$ carried by the
soft gluon. Because of momentum conservation, we have lower limit
for the $x$-integral: $x_{min}=\frac{k_{g\perp}^2}{Q^2}$. Therefore, the
above integral leads to the following leading contribution,
\begin{equation}
\frac{1}{q_\perp^2}\ln\frac{Q^2}{q_\perp^2} \ .
\end{equation}
Substituting the above equation into Eq.~(\ref{sg120}), we have
\begin{eqnarray}
g^2\int \frac{d^{D-1}k_g}{(2\pi)^{D-1}2E_{k_g}}\delta^{(2)} (q_\perp-k_{g\perp})S_g(p_1,p_2)
=\frac{\alpha_s}{2\pi^2}\frac{1}{q_\perp^2}\left(2\ln\frac{Q^2}{q_\perp^2}\right)\ .
\end{eqnarray}
Because there is no $\epsilon=(4-D)/2$ term in the $dx$ integral, we have
taken $D=4$ in the above equation.
The other terms are more difficult to calculate, because the above
phase space integral will contain jet contributions which have
already been taken into account by the above mentioned jet functions.
To avoid double-counting,
we have to subtract this part of contribution. That means
the phase space integral has to exclude the jet (cone) region.
It is interesting to note that the amount of this exclusion does not
depend on jet algorithm. This is because, here we are considering the
soft gluon radiation, whereas the jet algorithm mainly focuses
on the collinear gluons associated with the jet.

\subsection{Out of the Jet-cone Radiation}

As a general discussion, let us take the example of one term,
\begin{equation}
\int \frac{d^{D-1}k_g}{2E_{k_g}}\delta^{(2)} (q_\perp-k_{g\perp})S_g(k_1,p_1) =\int d^{D-2}k_{g\perp}\delta^{(2)}(q_\perp-k_{g\perp})
\int \frac{d\xi_1}{\xi_1}\frac{2}{(k_{g\perp}-\xi_1 k_{1\perp})^2} \ ,\label{es1}
\end{equation}
where $k_{1\perp}$ represents the transverse momentum for the
final state jet, and $\xi_1=k_g\cdot p_1/k_1\cdot p_1$. Clearly, there is
collinear divergence associated with the jet. That means, if the gluon
radiation is within the jet cone, it will generate a collinear
divergence. In order to regulate this collinear jet divergence, we can
limit the phase space integral to require that the gluon radiation being
outside of the jet cone. With this restriction, there will be no divergence
associated with the jet. Instead, the jet (cone) size $R$ will be introduced to
regulate the collinear divergence from the jet.

There are different ways to regulate the above integral. The main
task is to identify the need of introducing jet (cone) size. In the above example,
the integral diverges when $k_g$ is parallel to $k_1$, where
the invariant mass of $k_1+k_g$ becomes small. The out of cone radiation
requires that the invariant mass has a minimum, say, $\Lambda$, i.e.,
\begin{equation}
(k_1+k_g)^2>\Lambda^2 \ .
\end{equation}
Clearly, $\Lambda$ depends on the jet size. In other words, if $(k_1+k_g)^2$
is smaller than $\Lambda^2$, we have to exclude its contribution, because
it belongs to the jet contribution calculated in previous section.

Following the similar analysis as done for the jet contribution,
we can find out the size of $\Lambda$. For example, if we
substitute the kinematics of $k_1$ and $k_g$ into the above equation,
we will obtain
\begin{equation}
(k_1+k_g)^2\approx k_{1\perp}k_{g\perp}\left(e^{y_1-y_g}+e^{y_g-y_1}\right)-2k_{1\perp}k_{g\perp}\cos(\phi_1-\phi_g)\approx k_{1\perp}k_{g\perp}R_{1g}^2  \ ,
\end{equation}
where $y_1$ and $y_g$ are rapidities for $k_1$ and $k_g$, $\phi_1$ and $\phi_g$
are the azimuthal angles, respectively, and $R_{1g}$ represents the cone size
between $k_1$ and $k_g$. In other words, if $R_{1g}$ is smaller than $R$,
the gluon radiation will be considered inside the jet cone.
Therefore, in the phase space integral of Eq.~(\ref{es1}), we have to impose
the following kinematic restriction: $\Theta(2k_1\cdot k_g-\Lambda^2)$
with $\Lambda^2=k_{1\perp}k_{g\perp}R^2$.
Equivalently, we find it is much easier to adapt
a slight off-shell-ness for the jet momentum $k_1$ to regulate the
divergence: $k_1^2=m_1^2=k_{1\perp}^2R^2$. By doing that we do not need
to impose any kinematic constraints, and the phase space integral
is much easier to carry out. The choice of $m_1^2$ is to make sure that
$(k_1+k_g)^2$ is always larger than $\Lambda^2$. This can be verified
as follows,
\begin{eqnarray}
(k_1+k_g)^2&=&\sqrt{k_{1\perp}^2+m_1^2}k_{g\perp}\left(e^{y_1-y_g}+e^{y_g-y_1}\right)-2k_{1\perp}k_{g\perp}\cos(\phi_1-\phi_g)\nonumber\\
&\approx & \sqrt{k_{1\perp}^2+m_1^2}k_{g\perp}(\Delta y)^2+k_{1\perp}k_{g\perp}(\Delta\phi)^2+2k_{g\perp}\left(\sqrt{k_{1\perp}^2+m_1^2}-k_{1\perp}\right)\ .
\end{eqnarray}
By choosing $m_1^2=k_{1\perp}^2R^2$, it is guaranteed that $(k_1+k_g)^2$ is
larger than $\Lambda^2$ for any values of $\Delta y$ and $\Delta\phi$.

In the narrow jet approximation, i.e., in the $R\to 0$ limit,
the phase space cut-off
technique results into the same leading contributions in terms of $\ln(1/R)$.
After adding an off-shell-ness to the jet momentum, the
integral in Eq. (\ref{es1}) can be written as
\begin{equation}
\int d^{D-2}k_{g\perp}\delta^{(2)}(q_\perp-k_{g\perp})
\int \frac{d\xi_1}{\xi_1}\frac{2}{\xi_1^2k_{1\perp}^2(1+R^2)+k_{g\perp}^2-2\xi_1k_{1\perp}\cdot k_{g\perp}} \ .
\end{equation}
To proceed, we average the azimuthal angle of the jet but fix the
azimuthal angle of $k_{g\perp}$. This corresponds to keeping the
imbalance transverse momentum direction $\vec{q}_\perp$.
With that, we obtain
\begin{eqnarray}
&&\int\frac{d\xi_1}{\xi_1}\int_0^\pi\frac{d\phi}{\pi}\frac{2}{\xi_1^2k_{1\perp}^2(1+R^2)+k_{g\perp}^2-2\xi_1k_{1\perp}k_{g\perp}\cos(\phi)} \ ,\nonumber\\
&=&\int\frac{d\xi_1}{\xi_1}\frac{1}{\sqrt{\left(\xi_1^2k_{1\perp}^2(1+R^2)+k_{g\perp}^2\right)^2-4\xi_1^2k_{1\perp}^2k_{g\perp}^2} }\ ,
\end{eqnarray}
where the lower limit of $\xi_1$ is $\left(\frac{k_{g\perp}^2}{-t}\right)$.
As an illustration, we have taken $D=4$ in the average of $\phi$ angle
in the above equation. There is a $\epsilon$-term correction if we keep
$D$-dimension, which can be calculated accordingly. In the final results shown
below, we have kept those terms for completeness.
After taking the limit of
$q_\perp\ll Q$ and $R\to 0$, we obtain the leading power contribution as
\begin{equation}
\frac{1}{q_\perp^2}\frac{1}{2}\left[\ln\frac{Q^2}{q_\perp^2}+\ln\left(\frac{t}{u}\right)+\ln\frac{1}{R^2}\right] \ .
\end{equation}
Therefore, the final result for the integration of the
$S_g(k_1,p_1)$ term can be written as
\begin{eqnarray}
g^2\int \frac{d^3k_g}{(2\pi)^32E_{k_g}}\delta^{(2)} (q_\perp-k_{g\perp})S_g(k_1,p_1)
=\frac{\alpha_s}{2\pi^2}\frac{1}{q_\perp^2}\left[\ln\frac{Q^2}{q_\perp^2}+\ln\frac{1}{R_1^2}+\ln\left(\frac{t}{u}\right)\right]\ .
\end{eqnarray}
Evaluation of the other terms are similar, we summarize their final results as
\begin{eqnarray}
S_g(p_1,p_2)&\Rightarrow &\frac{\alpha_s}{2\pi^2}\frac{1}{q_\perp^2}\left(2\ln\frac{Q^2}{q_\perp^2}\right)\ ,\\
S_g(k_1,p_1)&\Rightarrow &\frac{\alpha_s}{2\pi^2}\frac{1}{q_\perp^2}\left[\ln\frac{Q^2}{q_\perp^2}+\ln\frac{1}{R_1^2}+\ln\left(\frac{t}{u}\right)
+\epsilon\left(\frac{1}{2}\ln^2\frac{1}{R_1^2}+\frac{\pi^2}{6}\right)\right]\ ,\\
S_g(k_2,p_1)&\Rightarrow &\frac{\alpha_s}{2\pi^2}\frac{1}{q_\perp^2}\left[\ln\frac{Q^2}{q_\perp^2}+\ln\frac{1}{R_2^2}+\ln\left(\frac{u}{t}\right)
+\epsilon\left(\frac{1}{2}\ln^2\frac{1}{R_2^2}+\frac{\pi^2}{6}\right)\right]\ ,\\
S_g(k_1,p_2)&\Rightarrow &\frac{\alpha_s}{2\pi^2}\frac{1}{q_\perp^2}\left[\ln\frac{Q^2}{q_\perp^2}+\ln\frac{1}{R_1^2}+\ln\left(\frac{u}{t}\right)
+\epsilon\left(\frac{1}{2}\ln^2\frac{1}{R_1^2}+\frac{\pi^2}{6}\right)\right]\ ,\\
S_g(k_2,p_2)&\Rightarrow &\frac{\alpha_s}{2\pi^2}\frac{1}{q_\perp^2}\left[\ln\frac{Q^2}{q_\perp^2}+\ln\frac{1}{R_2^2}+\ln\left(\frac{t}{u}\right)
+\epsilon\left(\frac{1}{2}\ln^2\frac{1}{R_2^2}+\frac{\pi^2}{6}\right)\right]\ ,\\
S_g(k_1,k_2)&\Rightarrow &\frac{\alpha_s}{2\pi^2}\frac{1}{q_\perp^2}\left[\ln\frac{1}{R_1^2}+\ln\frac{1}{R_2^2}+2\ln\left(\frac{s^2}{tu}\right)
+\epsilon\left(\frac{1}{2}\ln^2\frac{1}{R_1^2}+\frac{1}{2}\ln^2\frac{1}{R_1^2}+\frac{\pi^2}{3}\right.\right.\nonumber\\
&&\left.\left.-4\ln\frac{s}{-t}\ln\frac{s}{-u}\right)\right]\ ,
\end{eqnarray}
where we have kept the $\epsilon$ terms which will make
finite contributions to the complete one-loop calculation of $W(b)$,
when Fourier transforming from $q_\perp$  to $b_\perp$-space.
Furthermore, the leading double logarithm contributions arise from
the $\frac{1}{q_\perp^2}\ln\frac{Q^2}{q_\perp^2}$ terms in the above equations.

The above results are the basic elements to be used in our calculation
for deriving the low $q_\perp$ behavior of a scattering process,
induced by soft gluon radiation. In the following,
we will apply these results to all the partonic processes which contribute to
the inclusive dijet production in hadronic collisions.

\subsection{$q_iq_j\to q_iq_j$}

Quark-quark channel (with different quark flavors $i$ and $j$)
is the simplest case to calculate. Its leading
Born amplitude can be written as
\begin{equation}
M_0=\bar u (k_1)T^a \gamma^\mu u(p_1) \bar u(k_2) T^a \gamma^\nu u(p_2) \left({G}_{\mu\nu}(k_1-p_1)\right)\ ,
\end{equation}
where $G_{\mu\nu}$ represents the gluon propagator with momentum $k_1-p_1$.
For simplicity, we separate the color factor in the above amplitude,
\begin{equation}
M_0=A_0(p_1,p_2,k_1,k_2)\bar u_k T^a  u_i \bar u_l T^a  u_j \ ,
\end{equation}
where $i,j$ and $k,l$ are color indices for the incoming
and outgoing quarks, respectively.
Soft gluon radiation can be summarized as
\begin{eqnarray}
-\frac{2p_1^\mu}{2p_1\cdot k_g}A_0\bar uT^aT^cu\bar uT^au\\
+\frac{2k_1^\mu}{2k_1\cdot k_g}A_0\bar uT^cT^au\bar uT^au\\
+\frac{2k_2^\mu}{2k_2\cdot k_g}A_0\bar uT^au\bar uT^cT^au \ ,
\end{eqnarray}
where $k_g$ is the radiated gluon momentum, $\mu$ and $c$
for its polarization vector and color index, respectively.
To calculate the soft gluon contribution via this partonic channel, we
need to perform the phase space integration over its
amplitude squared, as discussed in the previous subsection.

Let us first work out the color factors for various terms
in its amplitude squared:
\begin{eqnarray}
&&k_1^\mu k_1^\nu\Rightarrow C_F |M_0|^2\ ,\nonumber\\
&&k_2^\mu k_2^\nu\Rightarrow C_F |M_0|^2\ ,\nonumber\\
&&p_1^\mu p_1^\nu\Rightarrow C_F |M_0|^2\ ,\nonumber\\
&&k_1^\mu k_2^\nu\Rightarrow -\frac{1}{4}C_F |M_0|^2\ ,\nonumber\\
&&k_1^\mu p_1^\nu\Rightarrow -\frac{1}{2N_c}|M_0|^2\ ,\nonumber\\
&&k_2^\mu p_1^\nu\Rightarrow \frac{1}{4}\left(2C_A-C_F\right) |M_0|^2\ .
\end{eqnarray}
Including the proper color factors, the amplitude squared contributes
\begin{eqnarray}
&&C_F\left[S_g(p_1,p_2)+S_g(k_1,p_2)+S_g(k_2,p_2)\right]\nonumber\\
&&+\frac{1}{2N_C}\left[S_g(k_1,p_2)+S_g(p_1,p_2)-S_g(k_1,p_1)\right]\nonumber\\
&&-\frac{1}{4}\left(2C_A-C_F\right)\left[S_g(k_2,p_2)+S_g(p_1,p_2)-S_g(k_2,p_1)\right]\nonumber\\
&&-\frac{1}{4}C_F\left[S_g(k_2,p_2)+S_g(k_1,p_2)-S_g(k_1,k_2)\right]\ .
\end{eqnarray}
After integrating the (restricted) phase space of the radiated gluon, we obtain the
following contribution of the soft gluon radiation to the $q_iq_j\to q_iq_j$
channel:
\begin{eqnarray}
\frac{\alpha_s}{2\pi^2}\frac{1}{q_\perp^2}\left\{h_{q_iq_j\to q_iq_j}^{(0)}\left[
2C_F\ln\frac{Q^2}{q_\perp^2}+C_F\left(\ln\frac{1}{R_1^2}+\ln\frac{1}{R_2^2}\right)\right]
+\Gamma_{sn}^{(qq')}\right\} \ , \label{qqsoft}
\end{eqnarray}
where
\begin{equation}
\Gamma_{sn}^{(qq')}=h_{q_iq_j\to q_iq_j}^{(0)}\left[2\left(C_A-C_F\right)\ln\frac{s}{-t}-2\ln\frac{s}{-u} \right]\ .
\end{equation}
An important cross check of the above result is to show that the
infrared divergences of the soft gluon radiation are cancelled by those
from the virtual diagrams and jet contributions. The only left
divergences are associated with the collinear divergences
for the incoming two quark distributions, which can be absorbed into
the definition of renormalized parton distribution functions.

To check the cancellation, we have to Fourier transform the above expression
into the impact parameter $b_\perp$-space,
\begin{eqnarray}
W^{(s+c)}(b_\perp)&&=\frac{\alpha_s}{2\pi}\int\frac{dx_1'}{x_1'}\frac{dx_2'}{x_2'}f_q(x_1')f_q(x_2')\left\{h_{q_iq_j\to q_iq_j}^{(0)}\right.
 \nonumber\\
&&\times
\left[
\left(-\frac{1}{\epsilon}+\ln\frac{b_0^2}{b_\perp^2\mu^2}\right)C_F\left(\frac{1+\xi^2}{(1-\xi)_+}\delta(1-\xi')+(\xi\leftrightarrow \xi')\right)
+\delta(1-\xi)\delta(1-\xi')\right.\nonumber\\
&&\times \left(\left(-\frac{1}{\epsilon}+\ln\frac{b_0^2}{b_\perp^2\mu^2}\right)
\left(C_F\ln\frac{1}{R_1^2R_2^2}+\epsilon C_F
\left(\frac{1}{2}\ln^2\frac{1}{R_1^2}+\frac{1}{2}\ln^2\frac{1}{R_2^2}+\frac{\pi^2}{3}\right)\right)\right.\nonumber\\
&&\left.\left.+C_F\left(\frac{2}{\epsilon^2}-\frac{2}{\epsilon}\ln\frac{Q^2}{\mu^2}+\ln^2\left(\frac{Q^2}{\mu^2}\right)
-\ln^2\left(\frac{Q^bb_\perp^2}{b_0^2}\right)-\frac{\pi^2}{6}+\ln\frac{s}{-t}\ln\frac{s}{-u}\right)\right)\right]\nonumber\\
&&\left.+\delta(1-\xi)\delta(1-\xi')\left(-\frac{1}{\epsilon}+\ln\frac{b_0^2}{b_\perp^2\mu^2}\right)\Gamma_{sn}^{(qq')}
\right\} \ ,
\end{eqnarray}
where we have also included the collinear gluon radiation contributions associated with
the incoming two quarks.

The virtual graphs have been calculated in the literature, and can be
summarized as follows,
\begin{eqnarray}
\frac{\alpha_s}{2\pi}\left\{C_F\left[-\frac{4}{\epsilon^2}+\frac{1}{\epsilon}\left(4\ln\frac{Q^2}{\mu^2}
-2\ln\frac{Q^2}{P_T^2}+2\ln\frac{u}{t}-6\right)\right]+\frac{1}{\epsilon}\frac{1}{2N_C}4\ln\frac{s^2}{tu}\right\}+\cdots\ ,
\end{eqnarray}
where we only kept the singular terms to check the cancellations between
real and virtual diagrams.
In addition, we have two jets contributions
\begin{eqnarray}
Jet_1+Jet_2=\frac{\alpha_s}{2\pi}C_F\left[\frac{2}{\epsilon^2}+\frac{1}{\epsilon}\left(3-2\ln\frac{P_T^2}{\mu^2}+\ln\frac{1}{R_1^2R_2^2}\right)
+I^{q_1}+I^{q_2}\right] \ ,\label{jet2q}
\end{eqnarray}
where $I^{q_i}$ are finite terms associated with jet functions. Clearly, the $1/\epsilon^2$
and $1/\epsilon$ terms all cancel out after summing up all the above three contributions,
except those associated with the splitting of quark distribution:
\begin{eqnarray}
&&-\frac{1}{\epsilon}C_F\left[\left(\frac{1+\xi^2}{(1-\xi)_+}\delta(1-\xi')+(\xi\leftrightarrow \xi')\right)+3\delta(1-\xi)\delta(1-\xi')\right]\nonumber\\
&&=-\frac{1}{\epsilon}\left[{\cal P}_{qq}(\xi)\delta(1-\xi')+{\cal P}_{qq}(\xi')\delta(1-\xi)\right]\ ,
\end{eqnarray}
where ${\cal P}_{qq}$ is the quark splitting kernel. The complete
expression for the finite terms will be much involved.
To facilitate the discussion on factorization, to be presented in  Sec.~VI,
we show below the most
important terms in the finite contributions, in particular, those with
large logarithms of $\ln({Q^2b_\perp^2}/{b_0^2)}$
and $\ln({\mu^2b_\perp^2}/{b_0^2)}$. This will clear show how
the TMD factorization works.
\begin{eqnarray}
W^{(1)}(b_\perp)|_{logs.}&&=\frac{\alpha_s}{2\pi}\left\{h_{q_iq_j\to q_iq_j}^{(0)}\left[
-\ln\left(\frac{\mu^2b_\perp^2}{b_0^2}\right)\left({\cal P}_{qq}(\xi)\delta(1-\xi')
+{\cal P}_{qq}(\xi')\delta(1-\xi)\right)-\delta(1-\xi)\ \right.\right.\nonumber\\
&&\left.\times\delta(1-\xi')\left(C_F\ln^2\left(\frac{Q^2b_\perp^2}{b_0^2}\right)+\ln\left(\frac{Q^2b_\perp^2}{b_0^2}\right)
\left(-3C_F+C_F\ln\frac{1}{R_1^2}+C_F\ln\frac{1}{R_2^2}\right)\right)\right]\nonumber\\
&&\left.-\delta(1-\xi)\delta(1-\xi')\ln\left(\frac{Q^2b_\perp^2}{b_0^2}\right)\Gamma_{sn}^{(qq')}\right\} \ ,
\end{eqnarray}
where an overall integrand factor of $\int\frac{dx_1'}{x_1'}\frac{dx_2'}{x_2'}f_q(x_1')f_q(x_2')$
was omitted for simplicity.
We would like to emphasize a number of important observations from the above calculations.
First, the factorization scale $\mu$ dependence only exists in terms associated with
the parton splitting kernel. This scale dependence shall be cancelled by the
relevant scale evolution for the integrated parton distributions.
Second, in the final results, $\ln^2(1/R^2)$ terms are cancelled out
between the jet contribution and the soft gluon contribution.
Third, the large logarithms appear in the one-loop calculations
contain three terms: (a) the double logarithms in terms of $\ln^2(Q^2b_\perp^2/b_0^2)$
proportional to incoming partons color factors (here, it is $C_F+C_F$);
(b) single logarithms in terms of $\ln(\mu^2b_\perp^2/b_0^2)$ associated
with parton distributions; (c) the left single logarithms of $\ln(Q^2b_\perp^2/b_0^2)$
contains similar terms as Drell-Yan process (the $-3C_F$ term) and those associated with
dijet production in this particular channel (jet size dependent contributions and
additional contributions which is process-dependent). All these features point to a possible
factorization in terms of TMDs, for which we will discuss in Sec.~VI.

\subsection{$qg\to qg$}

In this process, we have two different color structure at the Born
level,
\begin{equation}
A_1\bar u T^aT^bu+A_2\bar uT^bT^au \ ,
\end{equation}
where $a$ and $b$ represent the color indexes for the incoming
and outgoing gluons, the amplitudes $A_1$ and $A_2$ depend
on momenta of two incoming particles: $p_1$ and $p_2$, and
two outgoing particles: $k_1$ and $k_2$ for the quark and gluons,
respectively.
The leading order amplitude squared reads as,
\begin{equation}
|A_0|^2=C_F\left(A_1+A_2\right)^2-C_AA_1A_2^* \ ,
\end{equation}
where the two terms are separately gauge invariant.
Soft gluon radiation follows previous example, and
can be decomposed into the following three terms,
\begin{eqnarray}
&&\frac{2k_1^\mu}{2k_1\cdot k_g}\left[A_1\bar uT^cT^aT^bu+A_2\bar uT^cT^bT^a u\right] \nonumber\\
&+&\frac{-2p_1^\mu}{2p_1\cdot k_g}\left[A_1\bar uT^aT^bT^cu+A_2\bar uT^bT^aT^c u\right] \nonumber\\
&+&\frac{2k_2^\mu}{2k_2\cdot k_g}\left(-if_{cbd}\right)\left[A_1\bar uT^aT^du+A_2\bar uT^dT^a u\right]  \ ,
\end{eqnarray}
from the initial state and final state radiations,
where $c$ represents the color index for the radiated gluon.

The amplitude squared of the soft gluon radiation can be summarized into
the following form,
\begin{eqnarray}
&&|A_0|^2\left[C_FS_g(p_1,p_2)+C_FS_g(k_1,p_2)+C_AS_g(k_2,p_2)\right]\nonumber\\
&&+\left[\frac{1}{4}\left(A_1+A_2\right)^2-\frac{N_c^2}{4}A_1^2\right]\left(S_g(k_1,p_2)+S_g(k_2,p_2)-S_g(k_1,k_2)\right)\nonumber\\
&&-\left[\frac{1}{4N_c^2}\left(A_1+A_2\right)^2+\frac{1}{4}2A_1A_2^*\right]\left(S_g(k_1,p_2)+S_g(p_1,p_2)-S_g(k_1,p_1)\right)\ ,\nonumber\\
&&- \left[-\frac{1}{4}\left(A_1+A_2\right)^2+\frac{N_c^2}{4}A_2^2\right]\left(S_g(k_2,p_2)+S_g(p_1,p_2)-S_g(k_2,p_1)\right)\ .
\end{eqnarray}
Adding them together and applying the phase space integral, we obtain
the leading contribution induced by soft gluon radiation
in the $qg\to qg$ channel:
\begin{eqnarray}
&&\frac{\alpha_s}{2\pi^2}\frac{1}{q_\perp^2}\left\{h_{qg\to qg}^{(0)}
\left[(C_A+C_F)\ln\frac{Q^2}{q_\perp^2}+C_F\ln\frac{1}{R_1^2}+C_A\ln\frac{1}{R_2^2}\right]+\Gamma_{sn}^{(qg)}\right\}\ ,
\end{eqnarray}
where $\Gamma_{sn}^{(qg)}$ represents additional contribution in the sub-leading logarithm,
\begin{eqnarray}
\Gamma_{sn}^{(qg)}&=&\ln\frac{s}{-u}\left[-\frac{2(N_c^2+1)}{N_c^3}\frac{s^2+u^2}{-su}+\frac{u(s^2+u^2)}{-t^2s}N_c
-\frac{s^2+u^2}{t^2}\left(C_A-C_F\right)\right]\nonumber\\
&&+\ln\frac{s}{-t}\left[\frac{s^2+u^2}{-su}\frac{N_c^2+2}{N_c^3}+\frac{s^2+u^2}{t^2}\left(C_A-C_F\right)\right] \ .
\end{eqnarray}
The Fourier transformation of the above results into the impact parameter
$b_\perp$-space leads to the following contributions,
\begin{eqnarray}
W^{(s+c)}(b_\perp)&=&\frac{\alpha_s}{2\pi}\left\{h_{qg\to qg}^{(0)}\left[\left(-\frac{1}{\epsilon}+\ln\frac{b_0^2}{b_\perp^2\mu^2}\right)
\left(C_F{\cal P}_{qq}(\xi)\delta(1-\xi')+C_A{\cal P}_{gg}(\xi')\delta(1-\xi')\right)\right.\right.\nonumber\\
&&+\delta(1-\xi)\delta(1-\xi')\left(\frac{C_A+C_F}{2}\left(\frac{2}{\epsilon^2}-\frac{2}{\epsilon}\ln\frac{Q^2}{\mu^2}
+\ln^2\left(\frac{Q^2}{\mu^2}\right)-\ln^2\left(\frac{Q^bb_\perp^2}{b_0^2}\right)-\frac{\pi^2}{6}\right)\right. \nonumber\\
&&\left.\left.+\left(-\frac{1}{\epsilon}+\ln\frac{b_0^2}{b_\perp^2\mu^2}\right)\left(-2\beta_0C_A-\frac{3}{2}C_F
+{C_F}\ln\frac{1}{R_1^2}+C_A\ln\frac{1}{R_2^2}\right)\right)\right]\nonumber\\
&&\left.+\left(-\frac{1}{\epsilon}+\ln\frac{b_0^2}{b_\perp^2\mu^2}\right)\delta(1-\xi)\delta(1-\xi')\Gamma_{sn}^{(qg)} \right\}\ .
\end{eqnarray}
Virtual graphs contribute to the following terms in $W(b_\perp)$,
\begin{eqnarray}
W^{(v)}(b_\perp)&=&\frac{\alpha_s}{2\pi}\left\{h_{qg\to qg}^{(0)}\left[(-2C_F-2C_A)\left(-\frac{1}{\epsilon^2}-\frac{1}{\epsilon}\ln\frac{Q^2}{\mu^2}\right)
+\frac{1}{\epsilon}\left(-3C_F-2\beta_0\right)\right]\right.\nonumber\\
&&+\frac{1}{\epsilon}\frac{1}{N_c}\left[\ln\frac{-t}{s}\left((N_c^2-1)\frac{s^2+u^2}{t^2}-\left(\frac{1}{2N_c^2}
+\frac{N_c^2}{2}\right)\frac{s^2+u^2}{su}\right)\right.\nonumber\\
&&\left.\left.+\ln\frac{-u}{s}\left(N_c^2\left(\frac{2s^2}{t^2}-\frac{s}{u}\right)+\frac{s^2+u^2}{su}\right)\right]\right\}\ .
\end{eqnarray}
Furthermore, the jet contribution, including both quark and gluon jets in the final state,
yields
\begin{eqnarray}
W^{(j)}(b_\perp)&=&\frac{\alpha_s}{2\pi}\left[(C_A+C_F)\left(\frac{1}{\epsilon^2}-\frac{1}{\epsilon}\ln\frac{P_T^2}{\mu^2}\right)
+\frac{1}{\epsilon}\left(\frac{3}{2}C_F+2\beta_0C_A+C_F\ln\frac{1}{R_1^2}+C_A\ln\frac{1}{R_2^2}\right)
\right]  \ .
\end{eqnarray}
Clearly, all the divergences are cancelled out between the above terms, except the
collinear divergences associated with the incoming quark and gluon
distribution functions.

Again, the finite contributions take the following form, if we only keep the logarithmic terms,
\begin{eqnarray}
W^{(1)}(b_\perp)|_{logs.}&&=\frac{\alpha_s}{2\pi}h_{qg\to qg}^{(0)}\left\{
-\ln\left(\frac{\mu^2b_\perp^2}{b_0^2}\right)\left[{\cal P}_{qq}(\xi)\delta(1-\xi')
+{\cal P}_{gg}(\xi')\delta(1-\xi)\right] -\delta(1-\xi)\delta(1-\xi')\right.\nonumber\\
&&\times\left[\frac{C_F+C_A}{2}\ln^2\left(\frac{Q^2b_\perp^2}{b_0^2}\right)+\ln\left(\frac{Q^2b_\perp^2}{b_0^2}\right)
\left(-\frac{3}{2}C_F-2\beta_0+C_F\ln\frac{1}{R_1^2}+C_A\ln\frac{1}{R_2^2}\right)\right.\nonumber\\
&&\left.\left.+\ln\left(\frac{Q^2b_\perp^2}{b_0^2}\right)\frac{\Gamma_{sn}^{(qg)}}{h_{qg\to qg}^{(0)}}\right]\right\} \ .
\end{eqnarray}

\subsection{$gg\to q\bar q$}

Similarly, the Born amplitude for the $gg\to q\bar q$ channel is
\begin{equation}
A_1\bar u T^aT^bv+A_2\bar uT^bT^av \ ,
\end{equation}
with two momenta for incoming gluons: $p_1$ and $p_2$, and two momenta
for outgoing quark and antiquark: $k_1$ and $k_2$. The leading
order amplitude squared can be written as
\begin{equation}
|A_0|^2=C_F\left(A_1^2+A_2^2\right)-C_AA_1A_2^* \ ,
\end{equation}
with crossing symmetry to the above $qg\to qg$ channel.
Soft gluon radiation can be derived as
\begin{eqnarray}
&&\frac{2k_1^\mu}{2k_1\cdot k_g}\left[A_1\bar uT^cT^aT^bv+A_2\bar uT^cT^bT^a v\right] \nonumber\\
&+&\frac{-2k_2^\mu}{2k_2\cdot k_g}\left[A_1\bar uT^aT^bT^cv+A_2\bar uT^bT^aT^c v\right] \nonumber\\
&+&\frac{2p_1^\mu}{2p_1\cdot k_g}\left(-if_{cad}\right)\left[A_1\bar uT^dT^bv+A_2\bar uT^bT^d v\right]  \ ,
\end{eqnarray}
where $c$ represents the color index for the radiated gluon.
Its amplitude squared, including proper color factors, yields
\begin{eqnarray}
&&|A_0|^2\left[C_AS_g(p_1,p_2)+C_FS_g(k_1,p_2)+C_FS_g(k_2,p_2)\right]\nonumber\\
&&-\left[\frac{1}{4N_c^2}\left(A_1+A_2\right)^2+\frac{1}{4}2A_1A_2^*\right]\left(S_g(k_1,p_2)+S_g(k_2,p_2)-S_g(k_1,k_2)\right)\nonumber\\
&&+ \left[-\frac{N_c^2}{4}A_1^2+\frac{1}{4}(A_1+A_2)^2\right]\left(S_g(k_1,p_2)+S_g(p_1,p_2)-S_g(k_1,p_1)\right)\ ,\nonumber\\
&&- \left[\frac{N_c^2}{4}A_2^2-\frac{1}{4}(A_1+A_2)^2\right]\left(S_g(k_2,p_2)+S_g(p_1,p_2)-S_g(k_2,p_1)\right)\ ,
\end{eqnarray}
After applying the integral over the phase space of
the radiated gluon, we obtain
\begin{eqnarray}
&&\frac{\alpha_s}{2\pi^2}\frac{1}{q_\perp^2}\left\{h_{gg\to q\bar q}^{(0)}\left[2C_A\ln\frac{Q^2}{q_\perp^2}+C_F\ln\frac{1}{R_1^2}+C_F\ln\frac{1}{R_2^2}\right]
+\Gamma_{sn}^{(q\bar q)}\right\}\ ,
\end{eqnarray}
where
\begin{eqnarray}
\Gamma_{sn}^{(q\bar q)}&=&\frac{1}{4}\ln\frac{s}{-u}\left[\frac{t^2+u^2}{s^2}\frac{u^2-t^2}{tu}\frac{N_c^2}{4}+\frac{1}{2N_c}\frac{1}{2N_c}\frac{t^2+u^2}{tu}
+\frac{1}{2}\frac{t^2+u^2}{s^2}\right]\nonumber\\
&&+\frac{1}{4}\ln\frac{s}{-t}\left[\frac{t^2+u^2}{s^2}\frac{t^2-u^2}{tu}\frac{N_c^2}{4}+\frac{1}{2N_c}\frac{1}{2N_c}\frac{t^2+u^2}{tu}
+\frac{1}{2}\frac{t^2+u^2}{s^2}\right] \ .
\end{eqnarray}
The above result shows the leading contribution at low imbalance transverse momentum $q_\perp$.
After Fourier transformation, we have the following contribution to $W(b_\perp)$:
\begin{eqnarray}
W^{(s+c)}(b_\perp)&=&\frac{\alpha_s}{2\pi}\left\{C_Ah_{gg\to q\bar q}^{(0)}\left[\left(-\frac{1}{\epsilon}+\ln\frac{b_0^2}{b_\perp^2\mu^2}\right)
\left({\cal P}_{gg}(\xi)\delta(1-\xi')+(\xi\leftrightarrow \xi')\right)\right.\right.\nonumber\\
&&+\delta(1-\xi)\delta(1-\xi')\left(\frac{2}{\epsilon^2}-\frac{2}{\epsilon}\ln\frac{Q^2}{\mu^2}
+\ln^2\left(\frac{Q^2}{\mu^2}\right)-\ln^2\left(\frac{Q^bb_\perp^2}{b_0^2}\right)-\frac{\pi^2}{6}\right. \nonumber\\
&&\left.\left.+\left(-\frac{1}{\epsilon}+\ln\frac{b_0^2}{b_\perp^2\mu^2}\right)\left(-4\beta_0+\frac{C_F}{C_A}\ln\frac{1}{R_1^2R_2^2}\right)\right)\right]\nonumber\\
&&\left.+\left(-\frac{1}{\epsilon}+\ln\frac{b_0^2}{b_\perp^2\mu^2}\right)\delta(1-\xi)\delta(1-\xi')\Gamma_{sn}^{(q\bar q)} \right\}\ .
\end{eqnarray}
The virtual graph contribution can be summarized as
\begin{eqnarray}
W^{(v)}(b_\perp)&=&\frac{\alpha_s}{2\pi}\left\{h_{gg\to q\bar q}^{(0)}\left[(-2C_F-2C_A)\left(-\frac{1}{\epsilon^2}-\frac{1}{\epsilon}\ln\frac{Q^2}{\mu^2}\right)
+\frac{1}{\epsilon}\left(-3C_F-2\beta_0\right)\right]\right.\nonumber\\
&&+\frac{1}{\epsilon}\left[N_c\left(\ln\frac{-t}{s}\left(\frac{u}{t}-\frac{2u^2}{s^2}\right)+\ln\frac{-u}{s}\left(\frac{t}{u}-\frac{2t^2}{s^2}\right)\right)
+\frac{1}{N_c}\frac{t^2+u^2}{tu}\ln\frac{s^2}{tu}\right]\ .
\end{eqnarray}
Again, we find out the soft divergences cancel between the above terms and the jet contribution
which is the same as that of Eq.~(\ref{jet2q}). Here, the remaining
collinear divergences are associated with the incoming gluon distributions.

Keeping only the logarithmic terms, we obtain its finite contribution to
$W^{(1)}(b_\perp)$ as follows.
\begin{eqnarray}
W^{(1)}(b_\perp)|_{logs.}&&=\frac{\alpha_s}{2\pi}h_{gg\to q\bar q}^{(0)}\left\{
-\ln\left(\frac{\mu^2b_\perp^2}{b_0^2}\right)\left[{\cal P}_{gg}(\xi)\delta(1-\xi')
+{\cal P}_{gg}(\xi')\delta(1-\xi)\right] -\delta(1-\xi)\delta(1-\xi')\right.\nonumber\\
&&\times\left[C_A\ln^2\left(\frac{Q^2b_\perp^2}{b_0^2}\right)+\ln\left(\frac{Q^2b_\perp^2}{b_0^2}\right)
\left(-4\beta_0+C_F\ln\frac{1}{R_1^2}+C_F\ln\frac{1}{R_2^2}\right)\right.\nonumber\\
&&\left.\left.+\ln\left(\frac{Q^2b_\perp^2}{b_0^2}\right)\frac{\Gamma_{sn}^{(q\bar q)}}{h_{gg\to q\bar q}^{(0)}}\right]\right\} \ .
\end{eqnarray}

\subsection{$gg\to gg$}

For $gg\to gg$ channel, we can simply write down the following decomposition
for the Born amplitude,
\begin{equation}
A_1f_{abe}f_{cde}+A_2f_{ace}f_{bde}+A_3f_{ade}f_{bce} \ ,
\end{equation}
where $a,b,c,d$ are color indices for the gluons, with momenta
for incoming gluons: $p_1$ and $p_2$, and for outgoing gluons: $k_1$ and $k_2$.
The leading order amplitude squared can be written as
\begin{equation}
|A_0|^2=\left(A_1^2+A_2^2+A_3^2+A_1A_2^*-A_1A_3^*+A_2A_3^*\right) \ .
\end{equation}
One soft gluon radiation takes the form,
\begin{eqnarray}
&&\frac{2k_1^\mu}{2k_1\cdot k_g}f_{gcf}\left[A_1f_{abe}f_{fde}+A_2f_{afe}f_{bde}+A_3f_{ade}f_{bfe}\right] \nonumber\\
&+&\frac{2k_2^\mu}{2k_2\cdot k_g}f_{gdf}\left[A_1f_{abe}f_{cfe}+A_2f_{ace}f_{bfe}+A_3f_{afe}f_{bce}\right] \nonumber\\
&+&\frac{2p_1^\mu}{2p_1\cdot k_g}f_{gaf}\left[A_1f_{fbe}f_{cde}+A_2f_{fce}f_{bde}+A_3f_{fde}f_{bce}\right] \ .
\end{eqnarray}
The amplitude squared of the above radiation can be written as,
\begin{eqnarray}
&&|A_0|^2C_A\left[S_g(p_1,p_2)+S_g(k_1,p_2)+S_g(k_2,p_2)\right]\nonumber\\
&&+\left(S_g(k_1,p_2)+S_g(k_2,p_2)-S_g(k_1,k_2)\right)\left[-\frac{N_c}{2}A_1^2-\frac{N_c}{4}\left(A_2^2+A_3^2+2A_1A_2^*-2A_1A_3^*\right)\right] \nonumber\\
&&+ \left(S_g(k_1,p_2)+S_g(p_1,p_2)-S_g(k_1,p_1)\right)\left[-\frac{N_c}{2}A_2^2-\frac{N_c}{4}\left(A_1^2+A_3^2+2A_1A_2^*+2A_2A_3^*\right)\right] \ ,\nonumber\\
&&+\left(S_g(k_2,p_2)+S_g(p_1,p_2)-S_g(k_2,p_1)\right)\left[-\frac{N_c}{2}A_3^2-\frac{N_c}{4}\left(A_1^2+A_2^2+2A_2A_3^*-2A_1A_3^*\right)\right] \ .
\end{eqnarray}
After integrating the phase space, we obtain the leading contribution
for soft gluon radiation in this channel as
\begin{eqnarray}
&&\frac{\alpha_s}{2\pi^2}\frac{1}{q_\perp^2}
\left\{h_{gg\to gg}^{(0)}C_A\left[2\ln\frac{Q^2}{q_\perp^2}+\ln\frac{1}{R_1^2R_2^2}\right]+\Gamma_{sn}^{(gg)}\right\} \ ,
\end{eqnarray}
where
\begin{equation}
\Gamma_{sn}^{(gg)}=C_Ah_{gg\to gg}^{(0)}\left[\frac{t^2}{s^2-tu}\ln\frac{s}{-t}
+\frac{u^2}{s^2-tu}\ln\frac{s}{-u}\right] \ .
\end{equation}
Again, to check the cancellation between different contributions, we
perform the Fourier
transformation to the impact parameter-$b_\perp$ space, and find
\begin{eqnarray}
W^{(s+c)}(b_\perp)&=&\frac{\alpha_s}{2\pi}C_Ah_{gg\to gg}^{(0)}\left\{\left(-\frac{1}{\epsilon}+\ln\frac{b_0^2}{b_\perp^2\mu^2}\right)
\left[{\cal P}_{gg}(\xi)\delta(1-\xi')+(\xi\leftrightarrow \xi')\right]\right.\nonumber\\
&&+\delta(1-\xi)\delta(1-\xi')\left[\frac{2}{\epsilon^2}-\frac{2}{\epsilon}\ln\frac{Q^2}{\mu^2}
+\ln^2\left(\frac{Q^2}{\mu^2}\right)-\ln^2\left(\frac{Q^bb_\perp^2}{b_0^2}\right)-\frac{\pi^2}{6}\right. \nonumber\\
&&\left(-\frac{1}{\epsilon}+\ln\frac{b_0^2}{b_\perp^2\mu^2}\right)\left(-4\beta_0+\ln\frac{1}{R_1^2R_2^2}+\frac{t^2}{s^2-tu}\ln\frac{s}{-t}+\frac{u^2}{s^2-tu}\ln\frac{s}{-u}\right)\nonumber\\
&&\left.\left.+\frac{1}{2}\ln^2\frac{1}{R_1^2}+\frac{1}{2}\ln^2\frac{1}{R_2^2}+\frac{\pi^2}{3}\right]\right\} \ ,
\end{eqnarray}
where we have also included the collinear gluon contributions, and
${\cal P}_{gg}(\xi)$ is the gluon-gluon splitting kernel.
The virtual graphs have been calculated in the literature~\cite{Ellis:1985er}, and they contribute,
\begin{eqnarray}
W^{(v)}(b_\perp)&=&\frac{\alpha_s}{2\pi}C_A\left[-\frac{4}{\epsilon^2}+\left(\frac{1}{\epsilon}-\ln\frac{Q^2}{\mu^2}\right)
\left(4\ln\frac{Q^2}{\mu^2}-8\beta_0-\frac{u^2+s^2}{s^2-tu}\ln\frac{s}{-t}-\frac{t^2+s^2}{s^2-tu}\ln\frac{s}{-u}\right)\right]\nonumber\\
&&+ {\rm finite~terms} \ .
\end{eqnarray}
The two outgoing gluon jets contribute
\begin{eqnarray}
W^{(j)}(b_\perp)&=&\frac{\alpha_s}{2\pi}C_A\left[\frac{2}{\epsilon^2}+\frac{2}{\epsilon}\left(2\beta_0-\ln\frac{P_T^2}{\mu^2}\right)
+\frac{1}{\epsilon}\ln\frac{1}{R_1^2R_2^2}+I^{g_1}+I^{g_2}\right]  \ .
\end{eqnarray}
Comparing the above results, we find that the divergences do cancel each other except the
collinear divergences associated with the gluon splitting functions.

If we only keep the logarithmic terms, we will find the finite contributions as
\begin{eqnarray}
W^{(1)}(b_\perp)|_{logs.}&&=\frac{\alpha_s}{2\pi}h_{gg\to gg}^{(0)}\left\{
-\ln\left(\frac{\mu^2b_\perp^2}{b_0^2}\right)\left[{\cal P}_{gg}(\xi)\delta(1-\xi')
+{\cal P}_{gg}(\xi')\delta(1-\xi)\right] -\delta(1-\xi)\delta(1-\xi')\right.\nonumber\\
&&\times\left[C_A\ln^2\left(\frac{Q^2b_\perp^2}{b_0^2}\right)+\ln\left(\frac{Q^2b_\perp^2}{b_0^2}\right)
\left(-4\beta_0+C_A\ln\frac{1}{R_1^2}+C_A\ln\frac{1}{R_2^2}\right)\right.\nonumber\\
&&\left.\left.+\ln\left(\frac{Q^2b_\perp^2}{b_0^2}\right)\frac{\Gamma_{sn}^{(gg)}}{h_{gg\to gg}^{(0)}}\right]\right\} \ .
\end{eqnarray}

\section{Asymptotic Behavior and Compare to Fixed Order Computations}
An important cross check we will carry out in this section is to compare the
asymptotic result of the dijet differential cross section in low $q_\perp$ region
to the fixed order computation derived in the literature. Through that, we could
be sure that we have captured the most important contributions in the low $q_\perp$
limit.
As discussed in Sec. III, the dominant contributions in the
low $q_\perp$ region arise from collinear
gluon radiation associated with the incoming partons and soft gluon radiation
discussed in Sec. IV. Hence, the asymptotic result is obtained by adding these
two contributions together.

From the derivations of Secs. III and IV, we find that at low transverse momentum
the differential cross section has the following generic form:
\begin{eqnarray}
\frac{d^4\sigma}
{dy_1 dy_2 d P_T^2
d^2q_{\perp}}&=&\frac{\alpha_s}{2\pi^2}\frac{1}{q_\perp^2}
\sum_{ab,a'b'}\sigma_0\int\frac{dx_1'}{x_1'}\frac{dx_2'}{x_2'}x_1'\,f_a(x_1',\mu)
x_2'\, f_b(x_2',\mu) \nonumber\\
&&\times \left\{{h}_{a'b'\to cd}^{(0)}\left[\xi_1{\cal P}_{a'/a}(\xi_1)\delta(1-\xi_2)+\xi_2{\cal P}_{b'/b}(\xi_2)\delta(1-\xi_1)\right.\right.\nonumber\\
&&\left.+\delta(1-\xi_1)\delta(1-\xi_2)\delta_{aa'}\delta_{bb'}\left((C_a+C_b)\ln\frac{Q^2}{q_\perp^2}+C_c\ln\frac{1}{R_1^2}+C_d\ln\frac{1}{R_2^2}\right)\right]\nonumber\\
&&\left.+\delta(1-\xi_1)\delta(1-\xi_2)\delta_{aa'}\delta_{bb'}\Gamma_{sn}^{ab\to cd}\right\} \ , \label{asympto}
\end{eqnarray}
where $C_a$, $C_b$, $C_c$, and $C_d$ are the associated color factors
for the incoming and outgoing partons: $C=C_F$ for quark and $C=C_A$ for gluon.
The calculations in the last section have presented the result of
$\Gamma_{sn}$ for some partonic channels. All other channels can be
found accordingly.

\begin{figure}[tbp]
\centering
\includegraphics[width=6cm]{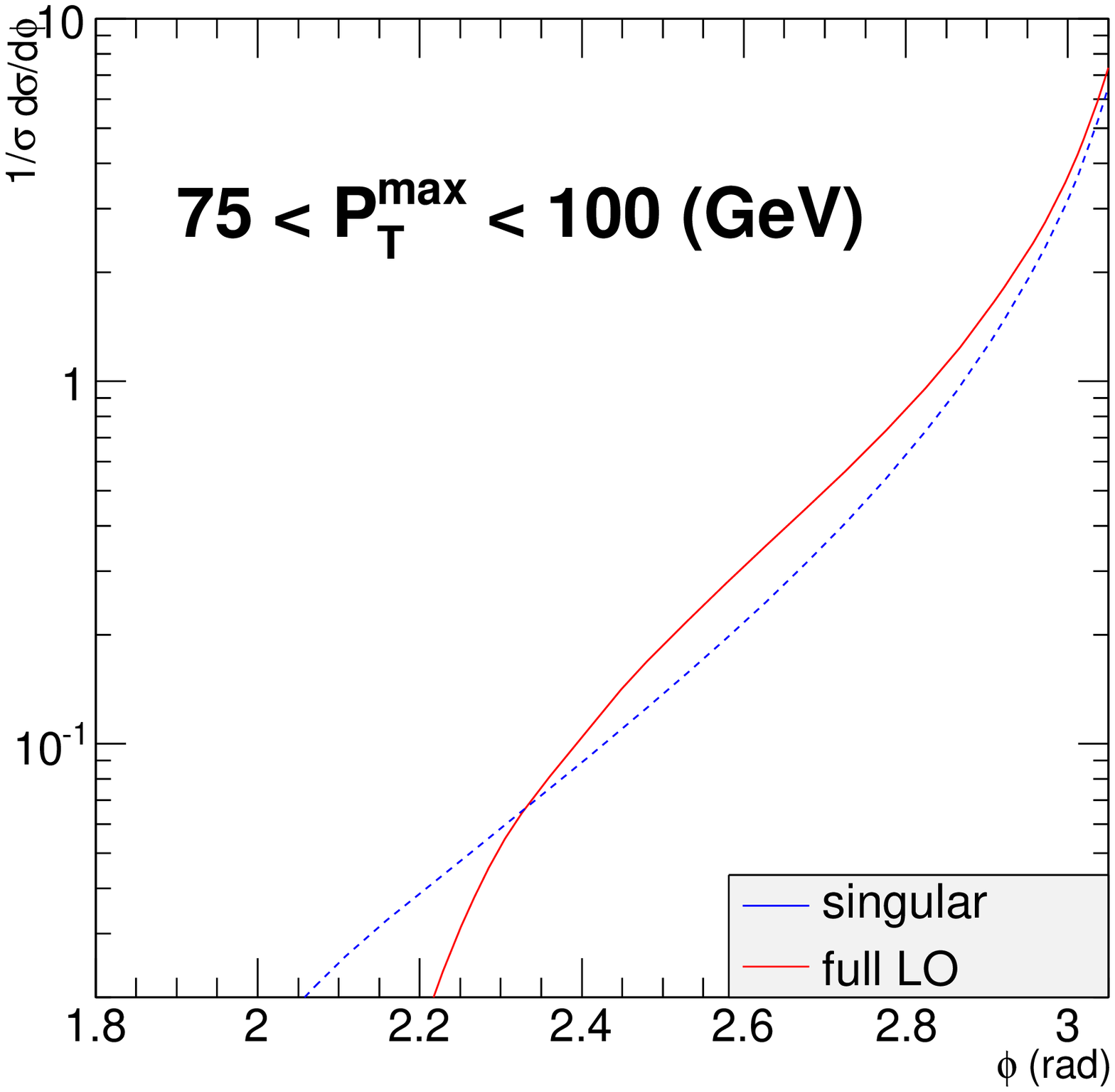}
\includegraphics[width=6cm]{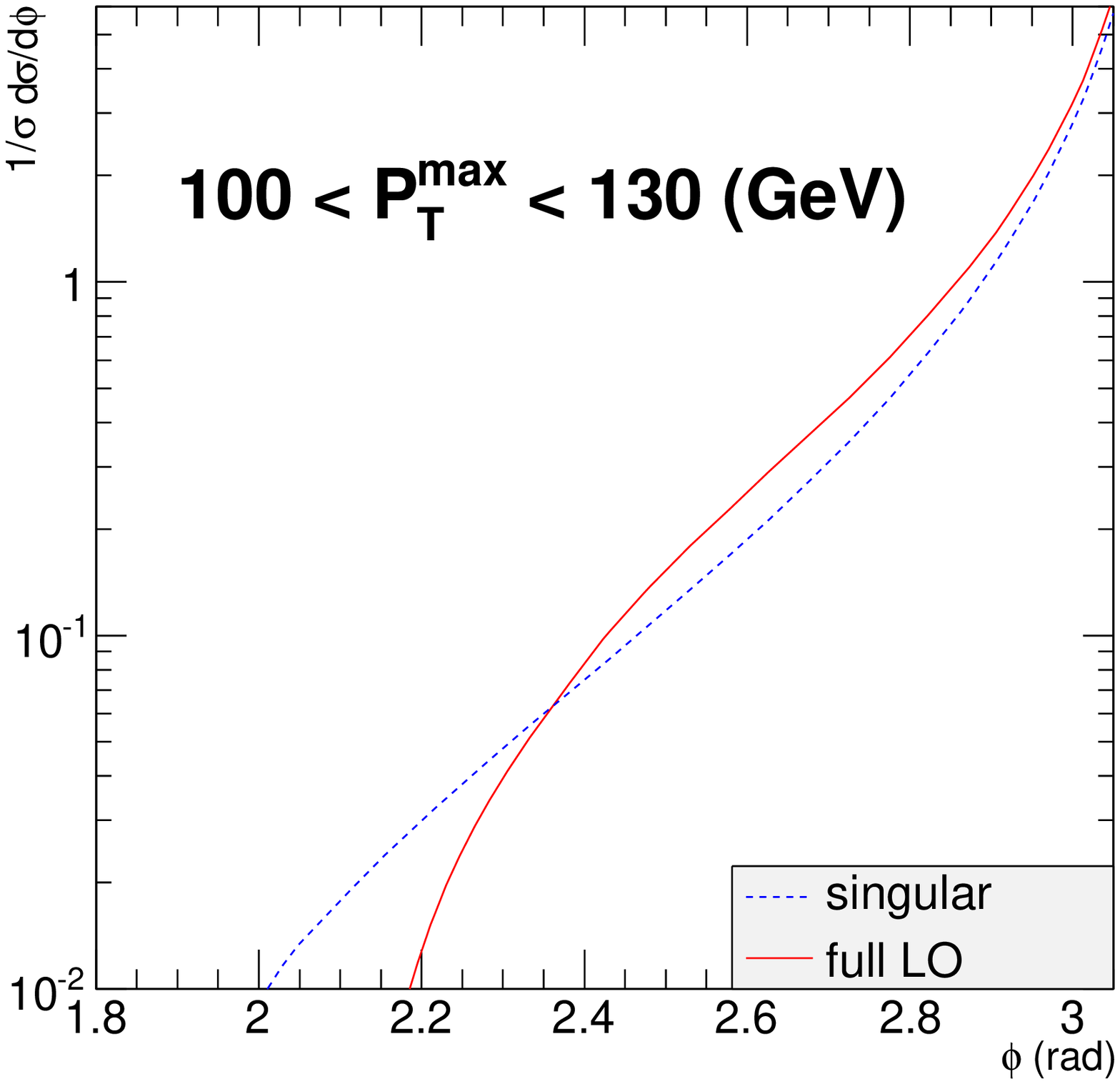}
\includegraphics[width=6cm]{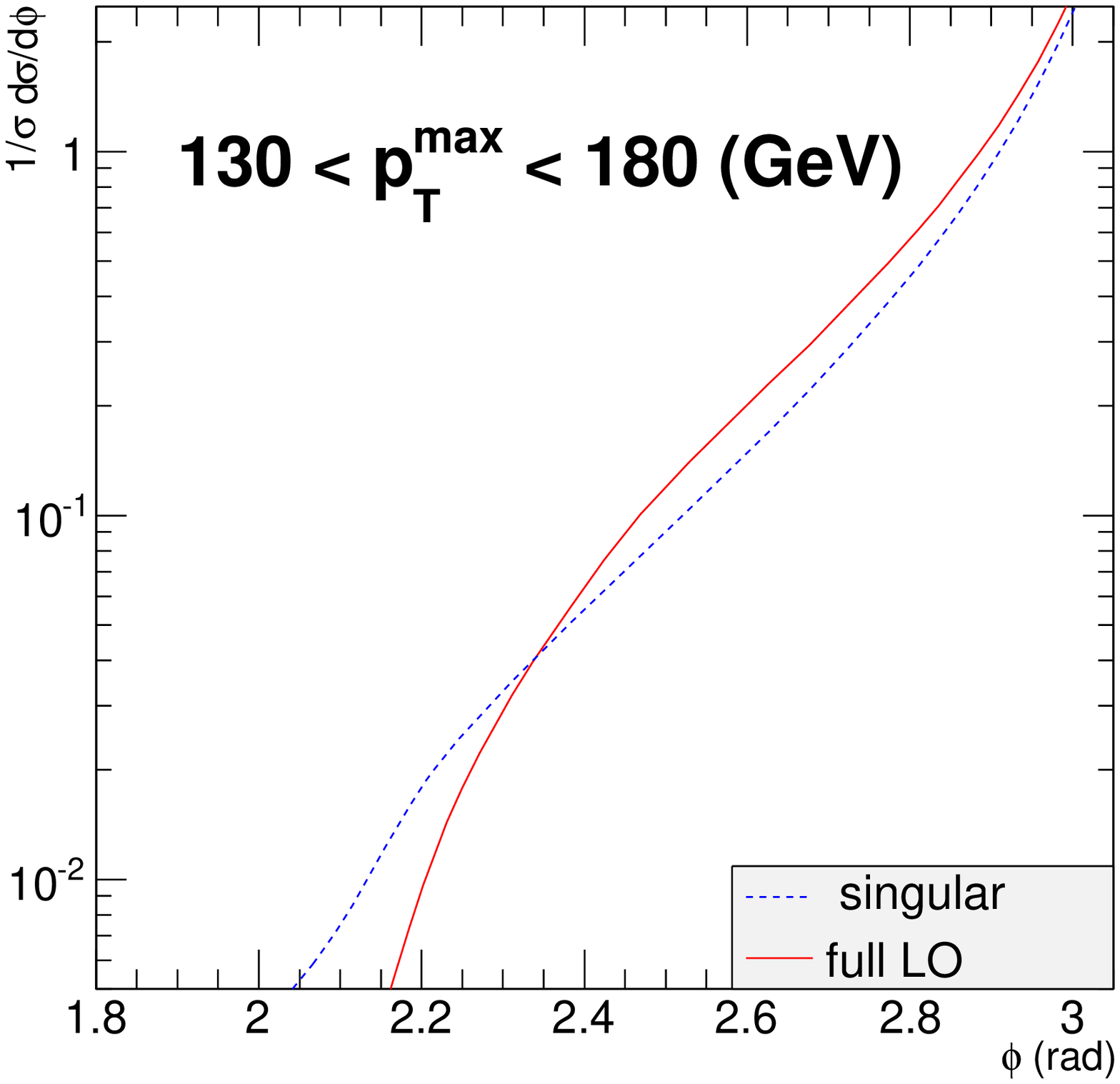}
\includegraphics[width=6cm]{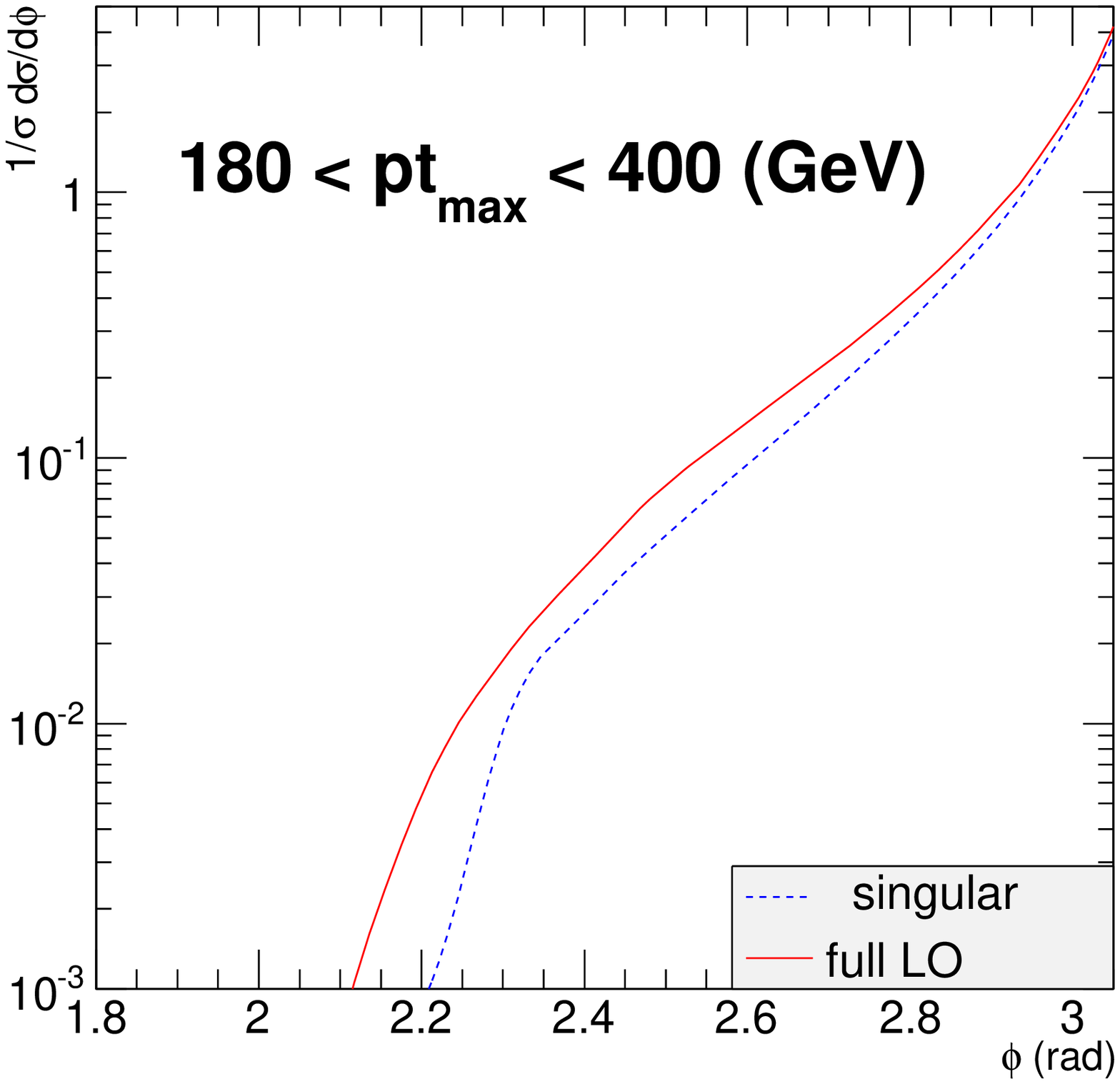}
\caption{The comparisons between the asymptotic derivations
of Eq.~(\ref{asympto}) and the full LO calculations for the
kinematics specified by the D0 collaboration at the Tevatron. The LO
curves come from Ref.~\cite{Abazov:2004hm}.}
\label{asymptotic}
\end{figure}

In Fig.~\ref{asymptotic}, we compare the above differential cross section,
 for dijet production at the Tevatron for the kinematics specified by the
D0 Collaboration~\cite{Abazov:2004hm}, to the fixed order
perturbative calculation with one gluon radiation contributions.
From these plots, we can see that the above asymptotic results capture
the leading contributions at low transverse momentum where $\phi$ is
close to $\pi$ for the back-to-back azimuthal correlation region.
We would like to emphasize that the jet size dependent
terms are crucial to make these comparisons. Without them,
the agreements between the full LO calculations and our asymptotic
derivation results will not be as evident as shown in Fig.~\ref{asymptotic}.

\section{Factorization and Resummation}

Clearly, from the derivations in the last section, and the plots of
the differential cross sections shown in Fig.~\ref{asymptotic}, we see that the collinear
factorization calculations for the dijet production lead to divergent
behavior at low imbalance transverse momentum $q_\perp$, which
corresponds to the back-to-back azimuthal angular correlation between
the two final state jets. In this region, to have a reliable prediction, we
need to perform the resummation. Since we are interested in the
dependence of the differential cross sections
on the transverse momentum $q_\perp$, the TMD
resummation is the appropriate framework for carrying out
our theory calculations.

The TMD resummation was originally derived for Drell-Yan type of
hard processes, where the final state only contains color neutral particle
with large invariant mass. In order to apply this resummation
formalism to the current case of dijet production in hadronic
collisions, some modifications are needed. In particular, the two
final state jets carry color so that additional soft gluon radiation will introduce
large logarithms associated with the final state jets. Although, these
additional interactions will not modify the leading double logarithmic
contribution, they will enter at the next-to-leading logarithmic level.
This is evident from the explicit calculations discussed in the previous
section.

To carry out the TMD factorization, we extend the original CSS
formalism, and take into account the final state radiation by
assuming a soft factor in the factorization formalism. The exact
same idea has been applied to resum large logarithms associated
with the threshold logarithms for dijet productions. The similarity
between the TMD and threshold resummation is not surprising, because
they both deal with soft and collinear gluon radiations.

In this section, we will argue the TMD factorization for dijet
production based on the explicit one-loop calculations
presented in the previous
sections. The factorization is verified in the transverse momentum
space and $b_\perp$-space at the one-loop order. At this
order, we have to take into account the matrix form of the
hard and soft factors. The explicit comparisons support the
factorization we argued in this paper.

With the explicit form of the TMD factorization formulas, we
derive the resummation result by solving the relevant renormalization
group equations. In particular, the TMDs follow
the examples of color neutral particle productions,
which have studied extensively in
the literature. The additional soft factor obeys the renormalization
group equation controlled by the anomalous dimension. We calculate
the soft factor at one-loop order, from which we derive the anomalous
dimension at one-loop order. Solving this renormalization group
equation, we resume the sub-leading logarithms in dijet production
processes.

Generically, the low transverse momentum $q_\perp$ originates from
collinear gluon radiation associated with the incoming partons
and the soft gluon radiation effects. In particular, the collinear
gluon radiation from the incoming partons contribute to the total
transverse momentum of the final state particles. This is the dominant
contribution in dijet production in hadronic collisions. The collinear
gluon radiation associated with the two final state jets will not contribute
to the total transverse momentum of the dijet. This is because
they are factorized into the jet contribution which do not contribute
a finite $q_\perp$ as we have shown in Sec. III. On the other hand,
soft gluon radiation among the incoming partons and final state jets
will contribute to the low imbalance transverse momentum of the dijet. To account
for this part of contribution, we introduce the soft factor in the TMD
factorization. Since the final states carry color, the soft gluon radiation
is more complicated than that for the color neutral particle production
in hadronic collisions.

We follow the method developed for threshold resummation in dijet
productions in hadronic collisions in Ref.~\cite{Kidonakis:1997gm}, where the soft
factor is formulated in the orthogonal color space and both
the incoming and outgoing partons are represented by
Eikonal gauge links. Because of light-cone
singularity in the parton distributions, we choose off-light-front gauge
links for the incoming partons. Meanwhile, the Wilson lines associated
with the final state jets are constructed in such a way
that only out-of-cone gluon radiation
contribute to a nonzero $q_\perp$. Consequently, the soft factor in our
calculations will depend on the jet cone size~\footnote{It is
possible to factorize the soft factor in our paper into a soft
factor and a jet function for the final state jet. By doing that,
the soft factor may not depend on the jet cone size. We leave
this for a future study.}.

Following the above argument, we can write down the factorization formula
as,
\begin{eqnarray}
\frac{d^4\sigma}
{dy_1 dy_2 d P_T^2
d^2q_{\perp}}&=&\sum\limits_{ab}\sigma_0
\int d^2k_{1\perp}d^2k_{2\perp}d^2\lambda_\perp \,
x_af_a(x_a,k_{1\perp})\,
x_bf_b(x_b,k_{2\perp})
\nonumber\\
&& 
\times\textmd{Tr}\left[\mathbf{H}_{ab\to cd}(Q^2)
\mathbf{S}_{ab\to cd}(\lambda_\perp)\right]
\delta^{(2)}(\vec{k}_{1\perp}+\vec{k}_{2\perp}
            +\lambda_\perp-\vec{q}_\perp) \; ,
\label{tmdqt}
\end{eqnarray}
where $f_a(x_1,k_{1\perp})$ and $f_b(x_2,k_{2\perp})$
are TMDs and will be introduced in
the following. Here, the hard factor and soft factor are
expressed in the matrix forms in the color spaces for
the incoming and outgoing partons. We can also express
$W(b_\perp)$ in the $b_\perp$-space as
\begin{eqnarray}
W_{ab\to cd}&=&x_1\,f_a(x_1,b_\perp,\zeta_1^2,\mu^2,\rho^2)
x_2\, f_b(x_2,b_\perp,\zeta_2^2,\mu^2,\rho^2)\nonumber\\
&\times& \textmd{Tr}\left[\mathbf{H}_{ab\to cd}(Q^2,\mu^2,\rho^2,y_1-y_2,R_1,R_2)
\mathbf{S}_{ab\to cd}(b_\perp,\mu^2,\rho^2,y_1-y_2,R_1,R_2)\right]\ ,\label{tmdb}
\end{eqnarray}
where we have shown all the explicit dependence of the TMDs and hard
and soft factors.

In the following, we will first introduce the TMDs, and
then formulate the soft factors for all partonic channels. With these
factors calculated in perturbation theory, we will show the above
factorization is valid at one-loop order, by comparing to the
derivations in the last few sections.

\subsection{Transverse Momentum Dependent Parton Distributions}

In the factorization formula, $x f_i(x,b_\perp,\zeta^2,\mu^2)$ is the Fourier
transformation of quark TMD parton distribution $xf_i(x,k_\perp,\zeta^2,\mu^2)$.
We follow the Ji-Ma-Yuan scheme~\cite{Ji:2004wu} to define the TMDs.
For the quark distribution, we have~\footnote{Here, we follow the original
definition in Ji-Ma-Yuan~\cite{Ji:2004wu}, where the soft factor is
subtracted from the naive gauge invariant TMDs. If other subtraction
method would be used, the associated soft factor definition would have been
changed as well.}
\begin{eqnarray}
xf_q(x,k_\perp,\zeta^2,\mu^2)&=&\int\frac{d\xi^-d^2\xi_\perp}{P^+(2\pi)^3}
    e^{-ixP^+\xi^-+i\vec{k}_\perp\cdot \vec\xi_\perp}\nonumber\\
    &&~~\times \frac{\left\langle P|\Psi(\xi^-,\xi_\perp)
{\cal L}^\dagger_{v}(\xi^-,\xi_\perp)\gamma^+ {\cal L}_{v}(0,0_\perp)
\Psi(0)|P \right\rangle}{\langle 0|{\cal L}_{\bar
vcb'}^\dagger(b_\perp;\infty) {\cal
L}_{vb'a}^\dagger(\infty;b_\perp){\cal L}_{vab}(0;\infty){\cal
L}_{\bar vbc}(\infty;0)  |0\rangle  }\ .
\end{eqnarray}
For gluon one, it can be written as:
\begin{eqnarray}
xf_g(x,k_\perp,\zeta^2,\mu^2)&=&\int\frac{d\xi^-d^2\xi_\perp}{P^+(2\pi)^3}
    e^{-ixP^+\xi^-+i\vec{k}_\perp\cdot \vec\xi_\perp}\nonumber\\
    &&~~\times \frac{\left\langle P|F^+_{a\mu}(\xi^-,\xi_\perp)
{\cal L}^\dagger_{vab}(\xi^-,\xi_\perp)\gamma^+ {\cal L}_{vbc}(0,0_\perp)
F_{c}^{\mu+}(0)|P \right\rangle}{\langle 0|{\cal L}_{\bar
vcb'}^\dagger(b_\perp;\infty) {\cal
L}_{vb'a}^\dagger(\infty;b_\perp){\cal L}_{vab}(0;\infty){\cal
L}_{\bar vbc}(\infty;0)  |0\rangle  }\ .
\end{eqnarray}
In the above equations, the relevant gauge links have to apply, in
the fundamental and adjoint representations
for the quark and gluon distributions, respectively.
The gauge link ${\cal L}_v$ is chosen along the direction $v$,
\begin{eqnarray}
{\cal L}_v(\xi^-,\xi_\perp)=Pexp\left(-ig\int_{-\infty} ^{0}d\lambda v\cdot A(\lambda v+\xi)\right) \ ,\label{GL}
\end{eqnarray}
for both cases, in the appropriate representations of $SU(3)$, adjoint for the gluon distribution
and fundamental for the quark distribution, respectively. 
Similarly, we define the TMDs from hadron $\bar P$, which depend on the gauge link
along the direction $\bar v$.
The vectors $v$ and $\bar{v}$ are off-light-front:  $v=(v^+,v^-,0_\perp )$ with $v^- \gg v^+$ and
$\bar{v}=(\bar{v}^+,\bar{v}^-,0_\perp )$ with $\bar{v}^- \ll \bar{v}^+$, and .
Since non-light-like vectors exist, the TMD parton distribution depends on
a new large scale $\zeta^2=(2v\cdot P)/v^2$ and a free parameter
$\rho^2=\frac{v^-\bar{v}^+}{\bar{v}^-v^+}$. Based on these definitions,
we can calculate them under perturbative QCD order by order, and
can be expressed in terms of the integrated parton distributions.
At one-loop order, the gluon distribution can be written as~\cite{Ji:2004wu}:
\begin{eqnarray}
f_g(x,b,\zeta,\mu,\rho)&=&\frac{\alpha_sC_A}{2\pi}\int\frac{dx'}{x'}
\left\{\ln\frac{b_0^2}{\bar{\mu}^2b^2}\left[{\cal P}_{gg}(\xi)f_g(x',\bar\mu)+{\cal P}_{gq}(\xi) f_q(x',\bar\mu)\right]
+\delta(\xi-1)f_g(x,\bar\mu)\right.\nonumber\\
&&\left.\times\left[\left(\ln\rho^2+2\beta_0-1\right)\ln\frac{b^2\mu^2}{b_0^2}e^{2\gamma_E}- \frac{1}{2}\ln^2\frac{b^2\zeta^2}{b_0^2}\right.\right.
    \left.\left.+\frac{3}{2}\ln\frac{\zeta^2}{\mu^2}-\frac{\pi^2}{2}-\frac{7}{2}\right]\right\}\ , \label{tmdg}
\end{eqnarray}
where $\xi=x/x'$. For the quark distribution,
\begin{eqnarray}
f_q(x,b,\zeta,\mu,\rho)&=&\frac{\alpha_sC_F}{2\pi}\int\frac{dx'}{x'}
\left\{\ln\frac{b_0^2}{\bar{\mu}^2b^2}\left[{\cal P}_{qg}(\xi)f_g(x',\bar\mu)+{\cal P}_{qq}(\xi) f_q(x',\bar\mu)\right]
+\delta(\xi-1)f_q(x,\bar\mu)\right.\nonumber\\
&&\left.\times\left[\left(\ln\rho^2-\frac{1}{2}\right)\ln\frac{b^2\mu^2}{b_0^2}- \frac{1}{2}\ln^2\frac{b^2\zeta^2}{b_0}\right.\right.
    \left.\left.-\frac{\pi^2}{4}-1\right]\right\}\ . \label{tmdq}
\end{eqnarray}
In the above equations, ${\cal P}_{ij}$ are splitting kernels. At one-loop order, we have
\begin{eqnarray}
{\cal P}_{gg}(x)&=&6\left(\frac{1-x}{x}+\frac{x}{(1-x)_+}+x(1-x)+\beta_0\delta(1-x)\right)\ ,\\
{\cal P}_{gq}(x)&=&\frac{4}{3}\left(\frac{1+(1-x)^2}{x}\right)\ ,\\
{\cal P}_{qq}(x)&=&\frac{4}{3}\left(\frac{1+x^2}{(1-x)_+}+\frac{3}{2}\delta(1-x)\right)\ ,\\
{\cal P}_{qg}(x)&=&\frac{1}{2}\left(x^2+(1-x)^2\right)\ .
\end{eqnarray}
Similarly, we can obtain the $k_\perp$-dependent expressions for the TMDs
from these references.

\subsection{Color Space Decompositions for the Soft and Hard Factors}

It has been shown in Refs.~\cite{Kidonakis:1997gm}, for soft gluon radiation in dijet production, it is more
convenient to construct the factorization in the color space matrix, where
the soft and hard factors can be calculated in the orthogonal color bases. In
particular, the associated anomalous dimension for the soft factors can be
formulated and the relevant resummation can be performed accordingly.

In this paper, we consider dijet production through the partonic $2\to 2$
subprocesses. Therefore, for each partonic channel, we will construct the
color-space bases depending on the color indexes of two incoming partons
and two outgoing partons. The soft gluon radiation does not modify the
fundamental scattering structure for each partonic channel, so that the
color-space bases are constructed for all order perturbative calculations.
In addition, the color-space bases are not unique. In our calculations, we follow those
used in Ref.~\cite{Kidonakis:1997gm}. For the quark-quark scattering subprocesses, we have
fundamental representation from both incoming and outgoing partons, for example,
\begin{eqnarray}
q^k_a+\bar{q}^k_b \rightarrow q^k_c +\bar{q}^k_d, \;\;\; q^k_a+\bar{q}^k_b \rightarrow q^j_c +\bar{q}^j_d , \;\;\; q^j_a+\bar{q}^k_b \rightarrow q^j_c +\bar{q}^k_d \ ,\label{channel0}
\end{eqnarray}
where $j$ and $k$ indicate the flavors of the quarks, and $a, b, c$ and $d$ are the color
indices. For this channel, we have two independent color
configurations,
\begin{eqnarray}
C^{ab}_{1cd} =\delta_{ac}\delta_{bd}\ ,   \;\;\;\; C^{ab}_{2cd} =T^{a'}_{ac}T^{a'}_{bd} \ ,
\end{eqnarray}
corresponding to the color-singlet and the color-octet couplings,
respectively.
Similarly, we have the same decomposition for identical quark-quark
scattering subprocess,
\begin{eqnarray}
q^k_a+q^k_b \rightarrow q^k_c +q^k_d,\;\;\; q^j_a+q^k_b \rightarrow q^j_c +q^k_d  \ .\label{channel1}
\end{eqnarray}
For $gg\to q\bar q$ channel,
\begin{eqnarray}
g_a+g_b \rightarrow \bar{q}^k_c +q^k_d, \label{channel2}
\end{eqnarray}
we have three independent color bases,
\begin{eqnarray}
C^{ab}_{1cd} =T^{a'}_{ac}T^{a'}_{cd}\ ,   \;\;\;\;   C^{ab}_{2cd} =d^{abc'}T^{c'}_{cd}\ ,  \;\;\;\;   C^{ab}_{3cd} =if^{abc'}T^{c'}_{cd} \ .
\end{eqnarray}
Similarly for $q\bar q\to gg$ channel:
\begin{eqnarray}
 \bar{q}^k_c+q^k_d \rightarrow g_a +g_b, \label{channel3} \ ,
\end{eqnarray}
we have
\begin{eqnarray}
C^{cd}_{1ab} =T^{a'}_{ac}T^{a'}_{cd}\ ,   \;\;\;\;   C^{cd}_{2ab} =d^{abc'}T^{c'}_{cd}\ ,  \;\;\;\;   C^{cd}_{3ab} =if^{abc'}T^{c'}_{cd} \ .
\end{eqnarray}
That will also apply to $qg\to qg$ channel,
\begin{eqnarray}
q^k_c+g_a \rightarrow  q^k_d +g_b \label{channel4}\ ,
\end{eqnarray}
where we have
\begin{eqnarray}
C^{ac}_{1bd} =T^{a'}_{ac}T^{a'}_{cd}\ ,   \;\;\;\;   C^{ac}_{2bd} =d^{abc'}T^{c'}_{cd}\ ,  \;\;\;\;   C^{ac}_{3bd} =if^{abc'}T^{c'}_{cd} \ .
\end{eqnarray}
For $gg\to gg$ channel,
\begin{eqnarray}
g_a+g_b \rightarrow g_c +g_d \label{channel5} \ ,
\end{eqnarray}
however, it is much more complicated,
\begin{eqnarray}
C^{ab}_{1cd} &=&\frac{i}{4}(f^{abc'}d^{cdc'}-d^{abc'}f^{cdc'})\ ,   \;\;\;\;  C^{ab}_{2cd} =\frac{i}{4}(f^{abc'}d^{cdc'}+d^{abc'}f^{cdc'})\ ,   \nonumber\\
C^{ab}_{3cd} &=&\frac{i}{4}(f^{acc'}d^{bdc'}+d^{acc'}f^{bdc'})\ ,   \;\;\;\;  C^{ab}_{4cd} =\frac{1}{8}\delta^{ac}\delta^{bd}\ ,     \nonumber\\
C^{ab}_{5cd} &=&\frac{3}{5}d^{acc'}d^{bdc'} \ ,  \;\;\;\;  C^{ab}_{6cd} =\frac{1}{3}f^{acc'}f^{bdc'} \ ,    \nonumber\\
C^{ab}_{7cd} &=&\frac{1}{2}(\delta^{ab}\delta^{cd}-\delta^{ad}\delta^{bc})-\frac{1}{3}f^{acc'}f^{bdc'} \ ,   \nonumber\\
C^{ab}_{8cd} &=&\frac{1}{2}(\delta^{ab}\delta^{cd}+\delta^{ad}\delta^{bc})-\frac{1}{8}\delta^{ac}\delta^{bd}-\frac{3}{5}d^{acc'}d^{bdc'}\ ,
\end{eqnarray}
where we have eight independent color bases.

With the above color bases, we can decompose the soft factors
in the matrix form. The associated soft gluon radiation is represented
by eight gauge links. This is because all the initial and final state can radiate or absorb soft gluons.
Therefore, we can decompose the soft factor, according to the following
formula,
\begin{eqnarray}
S_{IJ}(b_\perp)&=&\int_0^\pi
\frac{(\sin\phi)^{-2\epsilon}d\phi}{\frac{\sqrt{\pi}\Gamma(\frac{1}{2}-\epsilon)}
{\Gamma(1-\epsilon)}}\; C^{bb'}_{Iii'} C^{aa'}_{Jll'}\langle 0|{\cal L}_{
vcb'}^\dagger(b_\perp) {\cal
L}_{ \bar{v}bc'} (b_\perp){\cal L}_{\bar
vc'a'}^\dagger(0) {\cal
L}_{ vac}(0)\nonumber\\
&&\times  {\cal L}_{n ji}^\dagger(b_\perp) {\cal
L}_{\bar n i'k}(b_\perp) {\cal L}_{\bar nkl}^\dagger (0) {\cal
L}_{nl'j} (0)  |0\rangle \label{soft}\ ,
\end{eqnarray}
where $IJ$ represent the color indices in the color-space bases
constructed above. Therefore, $S_{IJ}$ is the matrix element in the
associated color-space bases for a particular partonic channel.
As mentioned above, we have four gauge links associated with two
incoming partons, for which we follow the TMDs to
adopt off-light-cone vectors $v$ and $\bar v$ to construct the
gauge links. The off-light-cone vectors are applied to regulate the
light-cone singularities.
For the two outgoing partons, we apply the
off-shellness to cast the out-of-cone radiation contribution
to the soft factor. This regulation depends on the jet size. That is why
we introduce the off-shellness  $n^2=R_1^2P_T^2/Q^2$ and $\bar{n}^2=R_2^2P_T^2/Q^2$
for the two final state jets with jet size $R_1$ and $R_2$, respectively.
$C^{bb'}_{Iii'}$ and $C^{aa'}_{Jll'}$ represent the corresponding
color configurations, as introduced above.

It is straightforward to calculate the soft factor at the leading
Born order, $S_{IJ}=C_{Iii'}^{aa'}C_{Ji'i}^{a'a}$.
For the channels in Eqs.~(\ref{channel1},\ref{channel0}), we find
\begin{eqnarray}
S^{(0)}=  \left[
             \begin{array}{cc}
               C_A^2 & 0 \\
               0 & \frac{C_AC_F}{2} \\
             \end{array}
           \right] \ .
\end{eqnarray}
For the channels in Eqs.~(\ref{channel2},\ref{channel3},\ref{channel4}), we have
\begin{eqnarray}
S^{(0)}= \left[
                 \begin{array}{ccc}
                   2C_A^2C_F & 0 & 0 \\
                   \\
                   0 & (C_A^2-4)C_F & 0 \\
                   \\
                   0 & 0 & C_A^2C_F \\
                 \end{array}
               \right]  \ .
\end{eqnarray}
For the channels in Eq.~(\ref{channel5}), we obtain 
\begin{eqnarray}
S^{(0)}= \left[
       \begin{array}{cccccccc}
         5 & 0 & 0 & 0 & 0 & 0 & 0 & 0 \\
         0 & 5 & 0 & 0 & 0 & 0 & 0 & 0 \\
         0 & 0 & 5 & 0 & 0 & 0 & 0 & 0 \\
         0 & 0 & 0 & 1 & 0 & 0 & 0 & 0 \\
         0 & 0 & 0 & 0 & 8 & 0 & 0 & 0 \\
         0 & 0 & 0 & 0 & 0 & 8 & 0 & 0 \\
         0 & 0 & 0 & 0 & 0 & 0 & 20 & 0 \\
         0 & 0 & 0 & 0 & 0 & 0 & 0 & 27 \\
       \end{array}
     \right] \ .
\end{eqnarray}
The hard factor ${\bf H}$ should be expanded by the same color bases.
At the tree level, our results are consistent with those in Ref.~\cite{Kidonakis:1997gm}.
For completeness, we list these results in the following.
For the partonic channel of $q^{k}+\bar{q}^{k}\rightarrow q^{k}+\bar{q}^{k}$,
\begin{eqnarray}
H^{(0)}=\left[
        \begin{array}{cc}
          H_{11} & H_{12} \\
          H_{21} & H_{22} \\
        \end{array}
      \right] \ ,
 \end{eqnarray}
where
\begin{eqnarray}
H_{11}&=&\frac{2C_F^2}{C_A^4}\frac{t^2+u^2}{s^2}\ ,\nonumber\\
H_{12}&=&H_{21}=-\frac{2C_F}{C_A^4}\frac{t^2+u^2}{s^2}+\frac{2C_F}{C_A^3}\frac{u^2}{st}\ ,\nonumber\\
H_{22}&=&\frac{2}{C_A^4}\frac{t^2+u^2}{s^2}+\frac{4}{C_A^2}\frac{s^2+u^2}{t^2}-\frac{4}{C_A^3}\frac{u^2}{st}\ .
\end{eqnarray}
Similarly, the partonic channel $q^{k}+\bar{q}^{k}\rightarrow q^{j}+\bar{q}^{j}$ is
expressed in the $2\times 2$ matrix with
\begin{eqnarray}
H_{11}&=&\frac{2C_F^2}{C_A^4}\frac{t^2+u^2}{s^2}\nonumber\\
H_{12}&=&H_{21}=-\frac{2C_F}{C_A^4}\frac{t^2+u^2}{s^2}\nonumber\\
H_{22}&=&\frac{2}{C_A^4}\frac{t^2+u^2}{s^2}\ .
\end{eqnarray}
And for channel $q^{k}+\bar{q}^{j}\rightarrow q^{k}+\bar{q}^{j}$,
\begin{eqnarray}
H_{11}&=&H_{12}=H_{21}=0\ ,\nonumber\\
H_{22}&=&\frac{2}{C_A^2}\frac{s^2+u^2}{t^2}\ .
\end{eqnarray}
For channel $q^{k}+q^{k}\rightarrow q^{k}+q^{k}$,
\begin{eqnarray}
H_{11}&=&\frac{2C_F^2}{C_A^4}\frac{t^2+s^2}{u^2}\ ,\nonumber\\
H_{12}&=&H_{21}=-\frac{2C_F}{C_A^4}\frac{t^2+s^2}{u^2}+\frac{2C_F}{C_A^3}\frac{s^2}{ut}\ ,\nonumber\\
H_{22}&=&\frac{2}{C_A^2}\frac{s^2+u^2}{t^2}+\frac{2}{C_A^4}\frac{s^2+t^2}{u^2}-\frac{2}{C_A}\frac{s^2}{ut}\ .
\end{eqnarray}
For channel $q^{k}+q^{j}\rightarrow q^{k}+q^{j}$,
\begin{eqnarray}
H_{11}&=&H_{12}=H_{21}=0\ ,\nonumber\\
H_{22}&=&\frac{2}{C_A^2}\frac{s^2+u^2}{t^2}\ .
\end{eqnarray}
For channel $g+g\rightarrow q+\bar{q}$ and $q+\bar{q}\rightarrow g+g$,
the hard factor is calculated as $3\times 3$ matrix,
\begin{eqnarray}
H^{(0)}=\left[
        \begin{array}{ccc}
          H_{11} & H_{12} & H_{13} \\
          H_{21} & H_{22} & H_{23} \\
          H_{31} & H_{32} & H_{33} \\
        \end{array}
      \right]\ ,
\end{eqnarray}
where
\begin{eqnarray}
H_{11}&=&\frac{1}{2C_A^4}\frac{u^2+t^2}{ut}\ ,\nonumber\\
H_{12}&=&H_{21}=\frac{1}{2C_A^3}\frac{u^2+t^2}{ut}\ ,\nonumber\\
H_{22}&=&\frac{1}{2C_A^2}\frac{u^2+t^2}{ut}\ ,\nonumber\\
H_{13}&=&H_{31}=\frac{1}{2C_A^3}\frac{t^2-u^2}{ut}+\frac{1}{C_A^3}\frac{t-u}{s}\ ,\nonumber\\
H_{23}&=&H_{32}=\frac{1}{2C_A^2}\frac{t^2-u^2}{ut}+\frac{1}{C_A^2}\frac{t-u}{s}\ ,\nonumber\\
H_{33}&=&\frac{1}{2C_A^2}\frac{s^2}{ut}+\frac{4}{C_A^2}\frac{tu}{s^2}-\frac{3}{C_A^2}\ .
\end{eqnarray}
For channel $q+g\rightarrow q+g$, we find a similar $3\times 3$ matrix with
\begin{eqnarray}
H_{11}&=&\frac{1}{4C_A^4C_F}\frac{t^2-2su}{su}\ ,\nonumber\\
H_{12}&=&H_{21}=\frac{1}{4C_A^3C_F}\frac{t^2-2su}{su}\ ,\nonumber\\
H_{22}&=&\frac{1}{4C_A^2C_F}\frac{t^2-2su}{su}\ ,\nonumber\\
H_{13}&=&H_{31}=-\frac{1}{2C_A^3C_F}\frac{s^2t+4s^2u+2stu-tu^2}{stu}\ ,\nonumber\\
H_{23}&=&H_{32}=-\frac{1}{2C_A^2C_F}\frac{s^2t+4s^2u+2stu-tu^2}{stu}\ ,\nonumber\\
H_{33}&=&\frac{1}{4C_A^2C_F}\frac{(2su-t^2)(t^2-4su)}{st^2u}\ .
\end{eqnarray}
For channel $g+g\rightarrow g+g$, however, we have $8\times 8$ matrix,
\begin{eqnarray}
H^{(0)}=\left[
         \begin{array}{cccccccc}
                    0 & 0 & 0 & 0 & 0 & 0 & 0 & 0 \\
                    0 & 0 & 0 & 0 & 0 & 0 & 0 & 0 \\
                    0 & 0 & 0 & 0 & 0 & 0 & 0 & 0 \\
                    0 & 0 & 0 & H_{44} & H_{45} & H_{46} & 0 & H_{48} \\
                    0 & 0 & 0 & H_{54} & H_{55} & H_{56} & 0 & H_{58} \\
                    0 & 0 & 0 & H_{64} & H_{65} & H_{66} & 0 & H_{68} \\
                    0 & 0 & 0 & 0      &0       & 0      & 0 & 0 \\
                    0 & 0 & 0 & H_{84} & H_{85} & H_{86} & 0 & H_{88} \\
         \end{array}
      \right] \ ,
\end{eqnarray}
where
\begin{eqnarray}
H_{44}&=&-\frac{9}{16}\frac{(s^2 - t u) (s t - u^2)}{s^2 u^2}\ ,\nonumber\\
H_{45}&=&H_{54}=-\frac{9}{32}\frac{(s^2 - t u) (s t - u^2)}{s^2 u^2}\ ,\nonumber\\
H_{55}&=&-\frac{9}{64}\frac{(s^2 - t u) (s t - u^2)}{s^2 u^2}\ ,\nonumber\\
H_{46}&=&H_{64}=-\frac{9}{32}\frac{(s-u) \left(s^2+s u+u^2\right) \left(s u-t^2\right)}{s^2 t u^2}\ ,\nonumber\\
H_{56}&=&H_{65}=-\frac{9}{64}\frac{(s-u) \left(s^2+s u+u^2\right) \left(s u-t^2\right)}{s^2 t u^2}\ ,\nonumber\\
H_{66}&=&\frac{9 \left(s^2 t-2 s u^2+t^2 u\right) \left(2 s^2 u-s t^2-t u^2\right)}{64 s^2 t^2
   u^2}\ ,\nonumber\\
 H_{48}&=&H_{84}=\frac{3}{16}\frac{(s^2 - t u) (s t - u^2)}{s^2 u^2}\ ,\nonumber\\
  H_{58}&=&H_{85}=\frac{3}{32}\frac{(s^2 - t u) (s t - u^2)}{s^2 u^2}\ ,\nonumber\\
  H_{68}&=&H_{86}=-\frac{3}{32}\frac{(s-u) \left(s^2+s u+u^2\right) \left(s u-t^2\right)}{s^2 t u^2}\ ,\nonumber\\
  H_{88}&=&-\frac{1}{16}\frac{(s^2 - t u) (s t - u^2)}{s^2 u^2}\ .
\end{eqnarray}
The above hard factors are normalized to reproduce the leading order
differential cross sections in Sec.~II,
\begin{eqnarray}
h_{12\to 34}^{(0)}=S_{IJ}^{12\to 34(0)}H_{JI}^{12\to 34(0)} \ ,
\end{eqnarray}
where sum over $IJ$ is understood. Because the leading order
soft factor is diagonalized, it is simple to verify the above equation.

\subsection{Soft Factor and the Anomalous Dimension at One-loop order}

\begin{figure}[h]
\begin{center}
\includegraphics[width=0.80\textwidth]{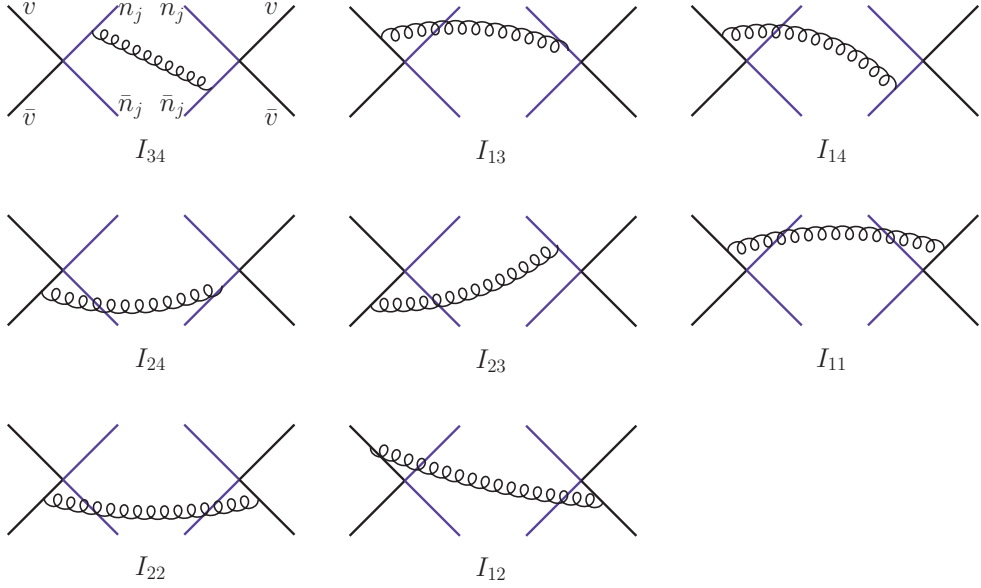}
\end{center}
    \caption{Representative diagrams contributing to the soft function at
    the NLO.    }\label{soft-hadron}
\end{figure}

At one-loop order, we will expand the soft factor definition with perturbative
corrections. The gluon radiations between all gauge links will contribute.
For convenience we separate out the common kinematic integrals from
the color factors for each soft factor calculation,
\begin{equation}
S^{(1)}=\sum\limits_{ij}W_{ij}I_{ij} \ ,
\end{equation}
where $ij$ label the associated gauge links: $12$ for the incoming two
partons and $34$ for the outgoing partons. In the above equation,
$I_{ij}$ represent the kinematic integrals for the soft gluon
radiation between $i$ and $j$ gauge links, whereas the factor $W_{ij}$
represent the associated color factor in the matrix form for particular
partonic channel.

We show in Fig.~(\ref{soft-hadron})\, a few representative
real gluon emission diagrams, labelled by the relevant kinematic
integral $I_{ij}$, which contribute to the soft function at the
next-to-leading order (NLO) in $\alpha_s$.

For the $I_{34}$ diagram, with gluon connecting
$3$ and $4$ gauge links, we find the following results:
\begin{eqnarray}
 I_{34 }(k_t)&=& \frac{\alpha_s
 }{2\pi^2} \int^\pi_0\frac{(\sin\phi)^{-2\epsilon}d\phi}{\frac{\sqrt{\pi}\Gamma(\frac{1}{2}-\epsilon)}{\Gamma(1-\epsilon)}} \int dk^+dk^-  \frac{2n_j\cdot \bar{n}_{j}}{(k\cdot n_j) (k\cdot\bar{n}_{j}  ) }\delta(k^2)\theta(k_0)\\
&=&  \frac{\alpha_s
 }{2\pi^2}  \;\frac{1}{k_t^2}\left[\ln\left(\frac{Q^4}{R_1^2R_2^2P_T^4}\right)
+
\epsilon\left(\frac{1}{2}\ln^2\frac{1}{R_1^2}+\frac{1}{2}\ln^2\frac{1}{R_2^2}
+\frac{\pi^2}{3}-4\ln\frac{s}{-t}\ln\frac{s}{-u}\right)\right]\ , \nonumber
\end{eqnarray}
in the transverse momentum space, where $k$ represents the radiated gluon momentum.
We have kept the $\epsilon$ term in the above equation, because they will
contribute to a finite term when Fourier transforming to $b_\perp$-space.
In the calculations, $n_j$ and $\bar n_j$ carry the momentum direction of
two outgoing jets, and therefore, the scalar product $n_j\cdot \bar n_j$
enters in the final result which is proportional to $Q^2/P_\perp^2$. That is
the cause of having the logarithmic dependence on this factor in
our final result. Again, the offshellness of the two outgoing gauge links
are introduced to cast the contribution that is only associated with out-of-cone
radiation. Gluon radiation within the jet cone will not induce a nonzero transverse
momentum of the dijet and its contribution needs to be subtracted out.

Similarly, for gluon connection between all other gauge links needed
in the final calculations, we find
\begin{eqnarray}
I_{13 }( k_t)
&=& 2\frac{\alpha_s
}{2\pi^2}  \int^\pi_0 \frac{(\sin\phi)^{-2\epsilon}d\phi}{\frac{\sqrt{\pi}\Gamma(\frac{1}{2}-\epsilon)}{\Gamma(1-\epsilon)}}\int dk^+dk^- \frac{\bar{v}\cdot n_j}{(k\cdot n_j ) (k\cdot \bar{v}) }\delta(k^2)\theta(k_0)\nonumber\\
&=&  \frac{\alpha_s
}{2\pi^2} \frac{1}{k_t^2}\left[\ln{\left(\frac{t^2 \rho^2}{\zeta_1^2\;P_T^2R_1^2}\right) } +\epsilon\left(\frac{1}{2}\ln^2\frac{1}{R_1^2}+\frac{\pi^2}{6}\right) \right]\ , \\
 I_{14 }( k_t)&=& 2\frac{\alpha_s
 }{2\pi^2}   \int^\pi_0\frac{(\sin\phi)^{-2\epsilon}d\phi}{\frac{\sqrt{\pi}\Gamma(\frac{1}{2}-\epsilon)}{\Gamma(1-\epsilon)}}\int dk^+dk^- \frac{\bar{v}\cdot \bar{n}_{j}}{(k\cdot \bar{n}_j) (k\cdot \bar{v}) }\delta(k^2)\theta(k_0)\nonumber\\
 &=&\frac{\alpha_s
 }{2\pi^2}\frac{1}{k_t^2} \left[\ln{\left(\frac{u^2 \rho^2}{\zeta_1^2\;P_T^2R_2^2}\right) }+\epsilon\left(\frac{1}{2}\ln^2\frac{1}{R_2^2}+\frac{\pi^2}{6}\right)\right]\ ,\\
I_{24 }( k_t)
&=& 2\frac{\alpha_s
}{2\pi^2}  \int^\pi_0\frac{(\sin\phi)^{-2\epsilon}d\phi}{\frac{\sqrt{\pi}\Gamma(\frac{1}{2}-\epsilon)}{\Gamma(1-\epsilon)}} \int dk^+dk^- \frac{v\cdot n_j}{(k\cdot n_j) (k\cdot v) }\delta(k^2)\theta(k_0)\nonumber\\
&=&  \frac{\alpha_s
}{2\pi^2} \frac{1}{k_t^2} \left[\ln{\left(\frac{t^2 \rho^2}{\zeta_2^2\;P_T^2R_2^2}\right) }+\epsilon\left(\frac{1}{2}\ln^2\frac{1}{R_2^2}+\frac{\pi^2}{6}\right)\right]\ , \\
 I_{23 }( k_t)&=& 2\frac{\alpha_s
 }{2\pi^2} \int^\pi_0\frac{(\sin\phi)^{-2\epsilon}d\phi}{\frac{\sqrt{\pi}\Gamma(\frac{1}{2}-\epsilon)}{\Gamma(1-\epsilon)}}  dk^+dk^- \frac{v\cdot \bar{n}_{j}}{(k\cdot\bar{n}_{j} ) (k\cdot v) }\delta(k^2)\theta(k_0)\nonumber\\
 &=&\frac{\alpha_s
 }{2\pi^2}\frac{1}{k_t^2} \left[\ln{\left(\frac{u^2 \rho^2}{\zeta_2^2\;P_T^2R_1^2}\right) }+\epsilon\left(\frac{1}{2}\ln^2\frac{1}{R_1^2}+\frac{\pi^2}{6}\right)\right]\ , \\
 I_{11 }( k_t)&=&I_{22 }( k_t)
= - \frac{\alpha_s
}{2\pi^2}  \int^\pi_0\frac{(\sin\phi)^{-2\epsilon}d\phi}{\frac{\sqrt{\pi}\Gamma(\frac{1}{2}-\epsilon)}{\Gamma(1-\epsilon)}} \int dk^+dk^- \frac{\bar{v}\cdot \bar{v}}{(k\cdot \bar{v})^2 }\delta(k^2)\theta(k_0)\nonumber\\
&=&  -\frac{\alpha_s
 }{2\pi^2}  \frac{1}{k_t^2 }\ ,\\
 I_{12 }( k_t)
&=& 2 \frac{\alpha_s
}{2\pi^2}  \int^\pi_0\frac{(\sin\phi)^{-2\epsilon}d\phi}{\frac{\sqrt{\pi}\Gamma(\frac{1}{2}-\epsilon)}{\Gamma(1-\epsilon)}} \int dk^+dk^- \frac{\bar{v}\cdot v}{(k \cdot \bar{v})(k\cdot v) }\delta(k^2)\theta(k_0)\nonumber\\
&=&  \frac{\alpha_s
}{2\pi^2} \frac{1}{k_t^2 }\ln \rho^2 \ .
\end{eqnarray}
Here, we did not list either $I_{33}$ or $I_{44}$,
because they will not contribute to the soft factor.

For the contribution from virtual diagrams, all the integrals are proportional to
\begin{eqnarray}
\frac{1}{\epsilon_{IR}}-\frac{1}{\epsilon_{UV}}\ .
\end{eqnarray}
After introducing the UV counter-terms which are proportional to
\begin{eqnarray}
-\frac{1}{\epsilon_{UV}} + \gamma_E +\ln\frac{1}{4\pi} \ ,
\end{eqnarray}
only IR divergence will be left and cancel the singularity in real diagrams,
after we have performed Fourier transformation from the transverse
momentum space to $b_\perp$ space.

The final expressions in the $b_\perp$-space can be written as
\begin{eqnarray}
 I^r_{34 }(b_\perp)&=&  -\frac{\alpha_s }{2\pi}  \;\left[\ln\left(\frac{Q^4}{R_1^2R_2^2P_T^4}\right)\ln\left(\frac{\mu^2\,b^2}{b_0^2}\right)
 \nonumber\right.\\
 &&\left.~~~~~~+\left(\frac{1}{2}\ln^2\frac{1}{R_1^2}+\frac{1}{2}\ln^2\frac{1}{R_1^2}
+\frac{\pi^2}{3}-4\ln\frac{s}{-t}\ln\frac{s}{-u}\right) \right] \ ,\label{i34}\\
I^r_{13 }( b)&=&-\frac{\alpha_s }{2\pi}\left[ \ln{\left(\frac{t^2 \rho^2}{\zeta_1^2\;P_T^2R_1^2}\right) }\ln\left(\frac{\mu^2\,b^2}{b_0^2}\right)+\left(\frac{1}{2}\ln^2\frac{1}{R_1^2}+\frac{\pi^2}{6}\right)\right]\ ,\\
 I^r_{14 }( b)&=&-\frac{\alpha_s }{2\pi}\left[\ln{\left(\frac{u^2 \rho^2}{\zeta_1^2\;P_T^2R_2^2}\right) }\ln\left(\frac{\mu^2\,b^2}{b_0^2}\right)+\left(\frac{1}{2}\ln^2\frac{1}{R_2^2}+\frac{\pi^2}{6}\right)\right]\ , \\
I^r_{24 }( b)
&=&- \frac{\alpha_s }{2\pi}\left[ \ln{\left(\frac{t^2 \rho^2}{\zeta_2^2\;P_T^2R_2^2}\right) }\ln\left(\frac{\mu^2\,b^2}{b_0^2}\right)+\left(\frac{1}{2}\ln^2\frac{1}{R_2^2}+\frac{\pi^2}{6}\right)\right]\ , \\
 I^r_{23 }( b)&=&-\frac{\alpha_s }{2\pi}\left[\ln{\left(\frac{u^2 \rho^2}{\zeta_2^2\;P_T^2R_1^2}\right) }\ln\left(\frac{\mu^2\,b^2}{b_0^2}\right)+\left(\frac{1}{2}\ln^2\frac{1}{R_1^2}+\frac{\pi^2}{6}\right)\right]\ ,\\
 I^r_{11 }( b)&=&I_{22 }( b)
=\frac{\alpha_s }{2\pi}\ln\left(\frac{\mu^2\,b^2}{b_0^2}\right) \ ,\\
 I^r_{12 }( b)
&=& -\frac{\alpha_s }{2\pi}\ln \rho^2\ln\left(\frac{\mu^2\,b^2}{b_0^2}\right) \ .\label{i12}
\end{eqnarray}
The color factors can be calculated diagram by diagram as well, and for the the channel of
$q+\bar{q}\rightarrow q+\bar{q}$, they are:
\begin{eqnarray}
W_{11}&=&W_{22}=C_F\,S^{(0)}\\
W_{34}&=&W_{12}= \left[
          \begin{array}{cc}
            0 & \frac{C_AC_F}{2} \\
            \\
            \frac{C_AC_F}{2} &\frac{1}{4}(C_A^2-2)C_F \\
          \end{array}
        \right]\\
W_{13}&=&W_{24}= \left[
          \begin{array}{cc}
            C_A^2C_F & 0 \\
            \\
            0 &-\frac{ C_F}{4} \\
          \end{array}
        \right]\\
W_{23}&=&W_{14}= \left[
          \begin{array}{cc}
            0 & -\frac{C_AC_F}{2} \\
            \\
           - \frac{C_AC_F}{2} &\frac{C_F}{2} \\
          \end{array}
        \right] \ .
\end{eqnarray}
For the channel of $q+q\rightarrow q+q $,
\begin{eqnarray}
W_{11}&=&W_{22}=C_F\,S^{(0)}\\
W_{34}&=&W_{12}= \left[
          \begin{array}{cc}
            0 & -\frac{C_AC_F}{2} \\
            \\
            -\frac{C_AC_F}{2} &\frac{C_F}{2} \\
          \end{array}
        \right]\\
W_{13}&=&W_{24}= \left[
          \begin{array}{cc}
            C_A^2C_F & 0 \\
            \\
            0 &-\frac{ C_F}{4} \\
          \end{array}
        \right]\\
W_{23}&=&W_{14}= \left[
          \begin{array}{cc}
            0 & \frac{C_AC_F}{2} \\
            \\
           \frac{C_AC_F}{2} &\frac{1}{4}(C_A^2-2)C_F \\
          \end{array}
        \right]\ .
\end{eqnarray}
For the channel $g+g\rightarrow \bar{q} +q$, we have
\begin{eqnarray}
W_{11}&=&W_{22}=C_A\,S^{(0)}\\
W_{12}&=& \left[
                 \begin{array}{ccc}
                   2C_A^3C_F & 0 & 0 \\
                   \\
                   0 & \frac{C_A(C_A^2-4)C_F}{2} & 0 \\
                   \\
                   0 & 0 & \frac{C_A^3C_F}{2} \\
                 \end{array}
               \right]
               \end{eqnarray}
\begin{eqnarray}
W_{34}&=& \left[
                 \begin{array}{ccc}
                   2C_A^2C_F^2 & 0 & 0 \\
                   \\
                   0 & \frac{(C_A^2-4)C_F(2C_F-C_A)}{2} & 0 \\
                   \\
                   0 & 0 & -\frac{C_AC_F}{2} \\
                 \end{array}
               \right]\\
W_{13}&=&W_{24}= \left[
                 \begin{array}{ccc}
                   0 & 0 & -C_A^2C_F \\
                   \\
                   0 & \frac{C_A(C_A^2-4)C_F}{4} & -\frac{C_A(C_A^2-4)C_F}{4} \\
                   \\
                   -C_A^2C_F & -\frac{C_A(C_A^2-4)C_F}{4} & \frac{C_A^3C_F}{4} \\
                 \end{array}
               \right]\\
W_{23}&=&W_{14}= \left[
                 \begin{array}{ccc}
                   0 & 0 & C_A^2C_F \\
                   \\
                   0 & \frac{C_A(C_A^2-4)C_F}{4} & \frac{C_A(C_A^2-4)C_F}{4} \\
                   \\
                   C_A^2C_F & \frac{C_A(C_A^2-4)C_F}{4} & \frac{C_A^3C_F}{4} \\
                 \end{array}
               \right] \ .
\end{eqnarray}
For the channel $q+\bar{q}\rightarrow g +g$,
\begin{eqnarray}
W_{11}&=&W_{22}=C_F\,S^{(0)}\\
W_{34}&=& \left[
                 \begin{array}{ccc}
                   2C_A^3C_F & 0 & 0 \\
                   \\
                   0 & \frac{C_A(C_A^2-4)C_F}{2} & 0 \\
                   \\
                   0 & 0 & \frac{C_A^3C_F}{2} \\
                 \end{array}
               \right]
               \end{eqnarray}
\begin{eqnarray}
W_{12}&=& \left[
                 \begin{array}{ccc}
                   2C_A^2C_F^2 & 0 & 0 \\
                   \\
                   0 & \frac{(C_A^2-4)C_F(2C_F-C_A)}{2} & 0 \\
                   \\
                   0 & 0 & -\frac{C_AC_F}{2} \\
                 \end{array}
               \right]\\
W_{13}&=&W_{24}= \left[
                 \begin{array}{ccc}
                   0 & 0 & -C_A^2C_F \\
                   \\
                   0 & \frac{C_A(C_A^2-4)C_F}{4} & -\frac{C_A(C_A^2-4)C_F}{4} \\
                   \\
                   -C_A^2C_F & -\frac{C_A(C_A^2-4)C_F}{4} & \frac{C_A^3C_F}{4} \\
                 \end{array}
               \right]\\
W_{23}&=&W_{14}= \left[
                 \begin{array}{ccc}
                   0 & 0 & C_A^2C_F \\
                   \\
                   0 & \frac{C_A(C_A^2-4)C_F}{4} & \frac{C_A(C_A^2-4)C_F}{4} \\
                   \\
                   C_A^2C_F & \frac{C_A(C_A^2-4)C_F}{4} & \frac{C_A^3C_F}{4} \\
                 \end{array}
               \right]\ .
\end{eqnarray}
For the channel $q+ g \rightarrow q +g$, the color matrixes are,
\begin{eqnarray}
W_{11}&=&C_F\,S^{(0)}\\
W_{22}&=&C_A\,S^{(0)}\\
W_{34}&=&W_{12}=  \left[
                 \begin{array}{ccc}
                   0 & 0 & -C_A^2C_F \\
                   \\
                   0 & \frac{C_A(C_A^2-4)C_F}{4} & -\frac{C_A(C_A^2-4)C_F}{4} \\
                   \\
                   -C_A^2C_F & -\frac{C_A(C_A^2-4)C_F}{4} & \frac{C_A^3C_F}{4} \\
                 \end{array}
               \right]
               \end{eqnarray}
\begin{eqnarray}
W_{13}&=&\left[
                 \begin{array}{ccc}
                   2C_A^2C_F^2 & 0 & 0 \\
                   \\
                   0 & \frac{(C_A^2-4)C_F(2C_F-C_A)}{2} & 0 \\
                   \\
                   0 & 0 & -\frac{C_AC_F}{2} \\
                 \end{array}
               \right]\\
W_{24}&=& \left[
                 \begin{array}{ccc}
                   2C_A^3C_F & 0 & 0 \\
                   \\
                   0 & \frac{C_A(C_A^2-4)C_F}{2} & 0 \\
                   \\
                   0 & 0 & \frac{C_A^3C_F}{2} \\
                 \end{array}
               \right]\\
W_{23}&=&W_{14}= \left[
                 \begin{array}{ccc}
                   0 & 0 & C_A^2C_F \\
                   \\
                   0 & \frac{C_A(C_A^2-4)C_F}{4} & \frac{C_A(C_A^2-4)C_F}{4} \\
                   \\
                   C_A^2C_F & \frac{C_A(C_A^2-4)C_F}{4} & \frac{C_A^3C_F}{4} \\
                 \end{array}
               \right]\ .
\end{eqnarray}
For the channel $g+g\rightarrow g+g$, they are
\begin{eqnarray}
W_{34}&=&W_{12}=  \left[
                  \begin{array}{cccccccc}
                    \frac{15}{2} & 0 & 0 & 0 & 0 & 0 & 0 & 0 \\
                    0 & \frac{15}{2} & 0 & 0 & 0 & 0 & 0 & 0 \\
                    0 & 0 & 0 & 0 & 0 & 0 & 0 & 0 \\
                    0 & 0 & 0 & 0 & 0 & 3 & 0 & 0 \\
                    0 & 0 & 0 & 0 & 6 & 6 & 12 & 0 \\
                    0 & 0 & 0 & 3 & 6 & 6 & 0 & 9 \\
                    0 & 0 & 0 & 0 & 12 & 0 & 30 & 18 \\
                    0 & 0 & 0 & 0 & 0 & 9 & 18 & 54 \\
                  \end{array}
                \right]
\end{eqnarray}
\begin{eqnarray}
W_{13}&=&W_{24}=  \left[
                  \begin{array}{cccccccc}
                    \frac{15}{2} & 0 & 0 & 0 & 0 & 0 & 0 & 0 \\
                    0 & 0 & 0 & 0 & 0 & 0 & 0 & 0 \\
                    0 & 0 & \frac{15}{2} & 0 & 0 & 0 & 0 & 0 \\
                    0 & 0 & 0 & 3 & 0 & 0 & 0 & 0 \\
                    0 & 0 & 0 & 0 & 12 & 0 & 0 & 0 \\
                    0 & 0 & 0 & 0 & 0 & 12 & 0 & 0 \\
                    0 & 0 & 0 & 0 &0 & 0 & 0 & 0 \\
                    0 & 0 & 0 & 0 & 0 & 0 & 0 & -27 \\
                  \end{array}
                \right]\\
W_{23}&=&W_{14}=  \left[
                  \begin{array}{cccccccc}
                    0 & 0 & 0 & 0 & 0 & 0 & 0 & 0 \\
                    0 & \frac{15}{2} & 0 & 0 & 0 & 0 & 0 & 0 \\
                    0 & 0 & \frac{15}{2} & 0 & 0 & 0 & 0 & 0 \\
                    0 & 0 & 0 & 0 & 0 & -3 & 0 & 0 \\
                    0 & 0 & 0 & 0 & 6 & -6 & -12 & 0 \\
                    0 & 0 & 0 & -3 & -6 & 6 & 0 & -9 \\
                    0 & 0 & 0 & 0 &-12 & 0 & 30 & -18 \\
                    0 & 0 & 0 & 0 & 0 & -9 & -18 & 54 \\
                  \end{array}
                \right]\ .
\end{eqnarray}
With the above results of $I_{nm}$ and $W_{nm}$, we will be able to calculate the
final results for the soft factor,
\begin{eqnarray}
S^{(1)}(k_t)=\sum_{nm}I_{nm}W_{nm}\ ,
\end{eqnarray}
for  all partonic channels, where $nm$ run from 1 to 4.
It is important to note that after summing the contribution in every diagram,
the jet cone size $R_{1,2}$ and $\rho$ dependence will be canceled by each other
in all the non-diagonal elements in $S^{(1)}$ and their color factors are probational to $S^{(0)}$.
For example, for the contribution from $q+\bar{q}\rightarrow q+\bar{q}$ channel,
we find the soft factor can be written as
\begin{eqnarray}
S^{(1)}(k_t)&=& \frac{\alpha_s(4\pi^2 \mu^2)^\epsilon }{2\pi}\frac{1}{k_t^2} \left\{S^{(0)}\;\left[C_F(\ln\rho^2-2)
+C_F\ln\left(\frac{s}{R_1^2P_T^2}\right)
+C_F\ln\left(\frac{s}{R_2^2P_T^2}\right)\right.\right.\nonumber\\
&+&\left.\epsilon\left(\frac{C_F}{2}\ln^2\left(\frac{1}{R_1^2}\right)
+\frac{C_F}{2}\ln^2\left(\frac{1}{R_2^2}\right)+C_F\frac{\pi^2}{3}\right)\right] -4\epsilon W_{34}\,U\,T\nonumber\\
&+&\left.2\;\Xi_{q+\bar{q}\rightarrow q+\bar{q}}\right\} \ ,
\end{eqnarray}
in the transverse momentum space,
where, for convenience, we have defined
\begin{eqnarray}
T=\ln\left(\frac{-t}{s}\right) \;\;\;\; U=\ln\left(\frac{-u}{s}\right) \ .
\end{eqnarray}
In the above equation, we have also introduced a short notation $\Xi_{q\bar q\to q\bar q}$ for
an additional term, which will be defined later, together with those for all other channels.

For $q+q\rightarrow q+q$, we have
\begin{eqnarray}
S^{(1)}(k_t)&=& \frac{\alpha_s(4\pi^2 \mu^2)^\epsilon }{2\pi}\frac{1}{k_t^2} \left\{S^{(0)}\;\left[C_F(\ln\rho^2-2)
+C_F\ln\left(\frac{s}{R_1^2P_T^2}\right)
+C_F\ln\left(\frac{s}{R_2^2P_T^2}\right)\right.\right.\nonumber\\
&+&\left.\epsilon\left(\frac{C_F}{2}\ln^2\left(\frac{1}{R_1^2}\right)
+\frac{C_F}{2}\ln^2\left(\frac{1}{R_2^2}\right)+C_F\frac{\pi^2}{3}\right)\right] -4\epsilon W_{34}\,U\,T\nonumber\\
&+&\left.2\;\Xi_{q+q\rightarrow q+q}\right\} \ .
\end{eqnarray}
For $g+g\rightarrow \bar{q}+q$, we have
\begin{eqnarray}
S^{(1)}(k_t)&=& \frac{\alpha_s(4\pi^2 \mu^2)^\epsilon }{2\pi}\frac{1}{k_t^2} \left\{S^{(0)}\;\left[C_A(\ln\rho^2-2)
+C_F\ln\left(\frac{s}{R_1^2P_T^2}\right)
+C_F\ln\left(\frac{s}{R_2^2P_T^2}\right)\right.\right.\nonumber\\
&+&\left.\epsilon\left(\frac{C_F}{2}\ln^2\left(\frac{1}{R_1^2}\right)
+\frac{C_F}{2}\ln^2\left(\frac{1}{R_2^2}\right)+C_F\frac{\pi^2}{3}\right)\right] -4\epsilon W_{34}\,U\,T\nonumber\\
&+&\left.2\;\Xi_{g+g\rightleftharpoons \bar{q}+q}\right\} \ .
\end{eqnarray}
For $q+\bar{q}\rightarrow g+g$, we have
\begin{eqnarray}
S^{(1)}(k_t)&=& \frac{\alpha_s(4\pi^2 \mu^2)^\epsilon }{2\pi}\frac{1}{k_t^2} \left\{S^{(0)}\;\left[C_F(\ln\rho^2-2)
+C_A\ln\left(\frac{s}{R_1^2P_T^2}\right)
+C_A\ln\left(\frac{s}{R_2^2P_T^2}\right)\right.\right.\nonumber\\
&+&\left.\epsilon\left(\frac{C_A}{2}\ln^2\left(\frac{1}{R_1^2}\right)
+\frac{C_A}{2}\ln^2\left(\frac{1}{R_2^2}\right)+C_A\frac{\pi^2}{3}\right)\right] -4\epsilon W_{34}\,U\,T\nonumber\\
&+&\left.2\;\Xi_{g+g\rightleftharpoons \bar{q}+q}\right\}\ .
\end{eqnarray}
For $q+g\rightarrow q+g$,  we have
\begin{eqnarray}
S^{(1)}(k_t)&=& \frac{\alpha_s(4\pi^2 \mu^2)^\epsilon }{2\pi}\frac{1}{k_t^2} \left\{S^{(0)}\;\left[\frac{C_F+C_A}{2}(\ln\rho^2-2)
+C_F\ln\left(\frac{s}{R_1^2P_T^2}\right)
+C_A\ln\left(\frac{s}{R_2^2P_T^2}\right)\right.\right.\nonumber\\
&+&\left.\epsilon\left(\frac{C_A}{2}\ln^2\left(\frac{1}{R_1^2}\right)
+\frac{C_F}{2}\ln^2\left(\frac{1}{R_2^2}\right)+\frac{C_F+C_A}{2}\frac{\pi^2}{3}\right)\right] -4\epsilon W_{34}\,U\,T\nonumber\\
&+&\left.2\;\Xi_{q+g\rightarrow q+g}\right\}\ .\nonumber\\
\end{eqnarray}
And finally, for $g+g\rightarrow g+g$, we obtain
\begin{eqnarray}
S^{(1)}(k_t)&=& \frac{\alpha_s(4\pi^2 \mu^2)^\epsilon }{2\pi}\frac{1}{k_t^2} \left\{S^0\;\left[C_A(\ln\rho^2-2)
+C_A\ln\left(\frac{s}{R_1^2P_T^2}\right)
+C_A\ln\left(\frac{s}{R_2^2P_T^2}\right)\right.\right.\nonumber\\
&+&\left.\epsilon\left(\frac{C_A}{2}\ln^2\left(\frac{1}{R_1^2}\right)
+\frac{C_A}{2}\ln^2\left(\frac{1}{R_2^2}\right)+C_A\frac{\pi^2}{3}\right)\right] -4\epsilon W_{34}\,U\,T\nonumber\\
&+&\left.2\;\Xi_{g+g\rightarrow g+g}\right\}\ .\nonumber\\
\end{eqnarray}
In the above equations, the $\Xi$ matrixes for all different channels are
defined as
\begin{eqnarray}
\Xi_{q+\bar{q}\rightarrow q+\bar{q}}\normalsize=\left[
     \begin{array}{cc}
       2C_FC_A^2\,T &\;\; -C_FC_A\,U \\\\
       -C_FC_A\,U &\;\; C_F\,U-\frac{C_F}{2}\,T \\
     \end{array}
   \right]\ ,
\end{eqnarray}
\begin{eqnarray}
\Xi_{q+q\rightarrow q+q}=\left[\begin{array}{cc}
      2C_FC_A^2\,T &\;\; C_FC_A\,U \\\\
       C_FC_A\,U &\;\; \frac{1}{2}(C_A^2-2)C_F\,U-\frac{C_F}{2}\,T \\
     \end{array} \right ]\ ,
\end{eqnarray}
\begin{eqnarray}
\Xi_{g+g \rightleftharpoons \bar{q}+q}=\left[
     \begin{array}{ccc}
        0 &\;\; 0 &\;\; 2C_A^2C_F\,(U-T) \\\\
                           0 &\;\; \frac{C_A(C_A^2-4)C_F}{2}\,(U+T) &\;\; \frac{C_A(C_A^2-4)C_F}{2}\,(U-T) \\\\
                           2C_A^2C_F\,(U-T) &\;\; \frac{C_A(C_A^2-4)C_F}{2}\,(U-T) &\;\; \frac{C_A^3C_F}{2}\,(U+T) \\
     \end{array}
   \right]\ ,
\end{eqnarray}
\begin{eqnarray}
\Xi_{q+g\rightarrow q+g}=\left[
     \begin{array}{ccc}
         2C_A^2C_F(C_A+C_F)\,T &\;\; 0 &\;\; 2C_A^2C_F\,U \\\\
                           0 &\;\; (C_A^2-4)C_F(\frac{C_A}{2}\,U+C_F\,T) &\;\; \frac{C_A(C_A^2-4)C_F}{2}\,U \\\\
                           2C_A^2C_F\,U &\;\; \frac{C_A(C_A^2-4)C_F}{2}\,U &\;\; (C_A^2-1)(\frac{C_A}{2}\,U+C_F\,T)  \\
     \end{array}
   \right]\ ,\nonumber\\
\end{eqnarray}
\begin{eqnarray}
\Xi_{g+g\rightarrow g+g}=\left[
      \begin{array}{cccccccc}
                    15T & 0 & 0 & 0 & 0 & 0 & 0 & 0 \\\\
                    0 & 15U  & 0 & 0 & 0 & 0 & 0 & 0 \\\\
                    0 & 0 & 15(U+T) & 0 & 0 & 0 & 0 & 0 \\\\
                    0 & 0 & 0 & 6T & 0 & -6U & 0 & 0 \\\\
                    0 & 0 & 0 & 0 & 12(2T+U) & -12U & -24U & 0 \\\\
                    0 & 0 & 0 & -6U & -12U & 12(2T+U) & 0 & -18U \\\\
                    0 & 0 & 0 & 0 &-24U & 0 & 60U & -36U \\\\
                    0 & 0 & 0 & 0 & 0 & -18U & -36U & 54(2U-T) \\
     \end{array}
   \right]\ .\nonumber\\
\end{eqnarray}
Similarly, we obtain the soft factors in the $b_\perp$-space,
\begin{eqnarray}
S^{(1)}(b)=\sum_{nm}I^r_{nm}(b)W_{nm}  \ ,
\end{eqnarray}
where $I_{nm}(b)$ have been calculated in Eqs.~(\ref{i34}-\ref{i12}).
For channel $q+\bar{q}\rightarrow q+\bar{q}$, we find
\begin{eqnarray}
S^{(1)}&=&-S^{(0)}\; \frac{\alpha_s}{2\pi} \left\{\left[C_F(\ln\rho^2-2)
+C_F\ln\left(\frac{s}{R_1^2P_T^2}\right)
+C_F\ln\left(\frac{s}{R_2^2P_T^2}\right)\right]\ln\left(\frac{\mu^2\,b^2}{b_0^2}\right)\right.\nonumber\\
&-&\left.\left[\frac{C_F}{2}\ln^2\left(\frac{1}{R_1^2}\right)
+\frac{C_F}{2}\ln^2\left(\frac{1}{R_2^2}\right) +C_F\frac{\pi^2}{3}\right] \right\}+\frac{2\alpha_s}{\pi}W_{34}\,U\,T\nonumber\\
&-&\frac{\alpha_s}{\pi}\;\Xi_{q+\bar{q}\rightarrow q+\bar{q}}\ln\left(\frac{\mu^2\,b^2}{b_0^2}\right) \ .
\end{eqnarray}
For $q+q\rightarrow q+q$, we find
\begin{eqnarray}
S^{(1)}&=&-S^{(0)}\; \frac{\alpha_s}{2\pi} \left\{\left[C_F(\ln\rho^2-2)
+C_F\ln\left(\frac{s}{R_1^2P_T^2}\right)
+C_F\ln\left(\frac{s}{R_2^2P_T^2}\right)\right]\ln\left(\frac{\mu^2\,b^2}{b_0^2}\right)\right.\nonumber\\
&-&\left.\left[\frac{C_F}{2}\ln^2\left(\frac{1}{R_1^2}\right)
+\frac{C_F}{2}\ln^2\left(\frac{1}{R_2^2}\right)+C_F\frac{\pi^2}{3}\right] \right\}+\frac{2\alpha_s}{\pi}W_{34}\,U\,T\nonumber\\
&-&\frac{\alpha_s}{\pi}\;\Xi_{q+q\rightarrow q+q}\ln\left(\frac{\mu^2\,b^2}{b_0^2}\right) \ .
\end{eqnarray}
For $g+g\rightarrow \bar{q}+q$, we find
\begin{eqnarray}
S^{(1)}&=&-S^{(0)}\; \frac{\alpha_s}{2\pi} \left\{\left[C_A(\ln\rho^2-2)
+C_F\ln\left(\frac{s}{R_1^2P_T^2}\right)
+C_F\ln\left(\frac{s}{R_2^2P_T^2}\right)\right]\ln\left(\frac{\mu^2\,b^2}{b_0^2}\right)\right.\nonumber\\
&-&\left.\left[\frac{C_F}{2}\ln^2\left(\frac{1}{R_1^2}\right)
+\frac{C_F}{2}\ln^2\left(\frac{1}{R_2^2}\right)+C_F\frac{\pi^2}{3}\right] \right\}+\frac{2\alpha_s}{\pi}W_{34}\,U\,T\nonumber\\
&-&\frac{\alpha_s}{\pi}\;\Xi_{g+g\rightleftharpoons \bar{q}+q}
\ln\left(\frac{\mu^2\,b^2}{b_0^2}\right)\ .\nonumber\\
\end{eqnarray}
For $q+\bar{q}\rightarrow g+g$, we find
\begin{eqnarray}
S^{(1)}&=&-S^{(0)}\; \frac{\alpha_s}{2\pi} \left\{\left[C_F(\ln\rho^2-2)
+C_A\ln\left(\frac{s}{R_1^2P_T^2}\right)
+C_A\ln\left(\frac{s}{R_2^2P_T^2}\right)\right]\ln\left(\frac{\mu^2\,b^2}{b_0^2}\right)\right.\nonumber\\
&-&\left.\left[\frac{C_A}{2}\ln^2\left(\frac{1}{R_1^2}\right)
+\frac{C_A}{2}\ln^2\left(\frac{1}{R_2^2}\right)+C_A\frac{\pi^2}{3}\right] \right\}+\frac{2\alpha_s}{\pi}W_{34}\,U\,T\nonumber\\
&-&\frac{\alpha_s}{\pi}\;\Xi_{g+g\rightleftharpoons \bar{q}+q}
\ln\left(\frac{\mu^2\,b^2}{b_0^2}\right)\ .\nonumber\\
\end{eqnarray}
For $q+g\rightarrow q+g$, we find
\begin{eqnarray}
S^{(1)}&=&-S^{(0)}\; \frac{\alpha_s}{2\pi} \left\{\left[\frac{C_F+C_A}{2}(\ln\rho^2-2)
+C_F\ln\left(\frac{s}{R_1^2P_T^2}\right)
+C_A\ln\left(\frac{s}{R_2^2P_T^2}\right)\right]\ln\left(\frac{\mu^2\,b^2}{b_0^2}\right)\right.\nonumber\\
&-&\left.\left[\frac{C_F}{2}\ln^2\left(\frac{1}{R_1^2}\right)
+\frac{C_A}{2}\ln^2\left(\frac{1}{R_2^2}\right)+\frac{C_F+C_A}{2}\frac{\pi^2}{3}\right] \right\}+\frac{2\alpha_s}{\pi}W_{34}\,U\,T\nonumber\\
&-&\frac{\alpha_s}{\pi}\;\Xi_{q+g\rightarrow q+g}
\ln\left(\frac{\mu^2\,b^2}{b_0^2}\right)\ .\nonumber\\
\end{eqnarray}
For $g+g\rightarrow g+g$, we find
\begin{eqnarray}
S^{(1)}&=&-S^{(0)}\; \frac{\alpha_s}{2\pi} \left\{\left[C_A(\ln\rho^2-2)
+C_A\ln\left(\frac{s}{R_1^2P_T^2}\right)
+C_A\ln\left(\frac{s}{R_2^2P_T^2}\right)\right]\ln\left(\frac{\mu^2\,b^2}{b_0^2}\right)\right.\nonumber\\
&-&\left.\left[\frac{C_A}{2}\ln^2\left(\frac{1}{R_1^2}\right)
+\frac{C_A}{2}\ln^2\left(\frac{1}{R_2^2}\right)+\frac{C_F+C_A}{2}\frac{\pi^2}{3}\right] \right\}+\frac{2\alpha_s}{\pi}W_{34}\,U\,T\nonumber\\
&-&\frac{\alpha_s}{\pi}\;\Xi_{g+g\rightarrow g+g}
\ln\left(\frac{\mu^2\,b^2}{b_0^2}\right)\ .\nonumber\\
\end{eqnarray}
We would like to emphasize again that the nontrivial results of the soft
factor calculations in the above. In particular, the $\rho$ and $R_{1,2}$-dependence
only appear in terms proportional to $S^{(0)}$. This is consistent with
the factorization we have argued in the beginning of this section.
We will show explicitly how the factorization works in the following
sub-section.

\subsection{Factorization at One-loop Order}

In this subsection, we will apply the one-loop results for the soft factor and TMDs
calculated in the last subsection to verify the factorization formalism we have proposed.
We will show the factorization in both transverse momentum space and Fourier conjugate
$b_\perp$-space. At his order, the TMDs can be calculated,
and have been listed in Sec. VIB.
The soft factors have also been calculated above in Sec. VIE.

\subsubsection{Transverse Momentum Space Factorization}

We find from Eq.~(\ref{tmdqt}) that the dijet differential cross section
at a nonzero transverse momentum $q_\perp$ receives contributions
from the two incoming parton distributions and the soft factor.
The hard factor, for a given $2 \to 2$ process, does not depend
on $q_\perp$, as expected. Therefore, we can write down the
expansion of the finite $q_\perp$
contribution from the TMD factorization formula at one-loop order:
\begin{eqnarray}
\frac{d^4\sigma}
{dy_1 dy_2 d P_T^2
d^2q_{\perp}}&=&\sum\limits_{ab}\sigma_0\left\{
\left(x_1f_1^{(1)}(x_1,q_{\perp})\,
x_2f_b(x_2)+x_1f_1(x_1)\,
x_2f_b^{(1)}(x_2,q_\perp)\right)\textmd{Tr}\left[\mathbf{H}_{ab\to cd}^{(0)}
\mathbf{S}_{ab\to cd}^{(0)}\right]\right.\nonumber\\
\nonumber\\
&& 
\left.+x_1f_1(x_1)\,
x_2f_b(x_2)\textmd{Tr}\left[\mathbf{H}_{ab\to cd}^{(0)}
\mathbf{S}_{ab\to cd}^{(1)}(q_\perp)\right]\right\} \ ,
\label{tmdqt1}
\end{eqnarray}
where we have expanded the TMDs and the
soft factor at one-loop order. We have calculated all these factors
in previous subsections. Substituting these results into the above
equation, we can derive the finite $q_\perp$ expression for dijet
production via each channel, and reproduce the
results shown in Sec.~IV. Especially, the $\rho$-dependence
from both the TMDs and the soft factors is cancelled out
by each other, which is a nontrivial illustration of the factorization formula.

\subsubsection{Factorization in $b_\perp$-space}
Similarly, the factorization can be demonstrated in the $b_\perp$-space, for
which we carry out the calculations for $W(b_\perp)$ in Eq.~(\ref{tmdb}).
\begin{eqnarray}
W_{ab\to cd}(b_\perp)&=&x_1\,f_a^{(1)}(x_1,b_\perp,\zeta_1^2,\mu^2,\rho^2)
x_2\, f_b(x_2)\textmd{Tr}\left[\mathbf{H}_{ab\to cd}^{(0)}
\mathbf{S}_{ab\to cd}^{(0)}\right]\nonumber\\
&&+x_1\,f_a(x_1)x_2\, f_b^{(1)}(x_2,b_\perp,\zeta_2^2,\mu^2,\rho^2)\textmd{Tr}\left[\mathbf{H}_{ab\to cd}^{(0)}
\mathbf{S}_{ab\to cd}^{(0)}\right]\nonumber\\
&&+x_1\,f_a(x_1)x_2\, f_b(x_2)\textmd{Tr}\left[\mathbf{H}_{ab\to cd}^{(0)}(Q^2)\mathbf{S}_{ab\to cd}^{(1)}
(b_\perp)\right]\ .\label{tmdb1}
\end{eqnarray}
With the expressions calculated in previous sub-sections, we will be
able to reproduce the logarithmic terms in Sec.~IV. In principle, we should be
able to extract the hard factors at one-loop order as well. However, to do that,
we need the expressions of the virtual diagrams contributions in the color-space
constructed in Sec.~VIC. We hope these calculations can be performed in
the future, from which we shall obtain the hard factors at one-loop order.
However, to show the factorization, we do not need the hard factors at
the one-loop order.

\subsection{Resummation}
According to the definition of the $W$ function, cf. Eq. (\ref{tmdb}),
when we choose factorization
scale $\mu=Q^2$, there will exist
two classes of scales in TMDs  and soft factor, one is
$b^2_\perp$, which comes from the Fourier
transformation of transverse momentum $q_\perp$, hence it is a small scale, the
other class contains scales of $Q^2$ and $\zeta^2$, where $\zeta_1^2\zeta_2^2=Q^4\rho^2$,
they are large scales. Our one-loop results for the TMDs
and soft factor show that there are large double and single
logarithms of $\ln Q^2b^2$ in them.
(Here, we use $b_\perp$ and $b$ exchangeably.)
These large logarithms appear at
every order of perturbative expansion as
the form of $\alpha_s^n\ln^{2n} Q^2b^2$ or $\alpha_s^n\ln^n Q^2b^2$,
when $\ln Q^2b^2 > 1/\alpha_s$, the convergence of the conventional
perturbative expansion is
impaired and no longer gives a correct prediction. In order to make a
reliable calculation for the $W$ function, these large logs have to be resummed.
Based on the factorization theorem, we could find all the relevant scales present in
each factorized factor, which can be evolved individually and satisfies
a renormalisation group equation(RGE).
For example, for the TMD $f(x,b,\zeta,\mu,\rho)$,
the relevant Collins-Soper evolution equation reads as:
\begin{equation}
\frac{\partial}{\partial\ln\zeta}f(x,b,\zeta,\mu,\rho)= (K(b,\mu)+G(\zeta,\mu) ) \;f(x,b,\zeta,\mu,\rho)
\ ,\
\end{equation}
According to the one-loop result, we can get
\begin{equation}
K(b,\mu)+G(\zeta,\mu)=-\frac{\alpha_sC_I}{\pi}\ln \frac{\zeta^2
b^2}{4}e^{2\gamma_E-\frac{3}{2}} \ ,
\end{equation}
where $C_I=C_A$ for gluon distribution and $C_I=C_F$ for quark distribution. $K(b,\mu)$ satisfies:
\begin{equation}
K(b,\mu)=\frac{\partial\ln {S(b,\mu,\rho)} }{\partial \ln {\rho}}=-\frac{\alpha_sC_I}{\pi}\ln \frac{\mu^2
b^2}{4}e^{2\gamma_E} \ .
\end{equation}
The renormalization group equation for $K(b,\mu)$ is governed by the
associated anomalous dimension $\gamma_K$,
\begin{equation}
\gamma_K=-\frac{\partial K(b,\mu)}{\partial \ln\mu}=\frac{\partial G(\zeta,\mu)}{\partial \ln\mu}=\frac{2\alpha_sC_I}{\pi}\ .
\end{equation}
Then we can solve the above renormalization group equation,
\begin{equation}
K(b,\mu)+G(\zeta,\mu)= K(b,Q_0)-  \int^\mu_{Q_0}\frac{d\mu'}{\mu'}\gamma_K(\mu')-\frac{\alpha_sC_I}{\pi}\ln \frac{\zeta^2
}{ \mu^2}e^{-\frac{3}{2}} \ ,
\end{equation}
where in the end we will choose $Q_0=b_0/b$ to resum the large logarithms.
In addition, we can evolve the factorization scale in the TMD distribution as well,
\begin{equation}
\frac{\partial}{\partial\ln\mu}f(x,b,\zeta,\mu,\rho)=\gamma_F \;f(x,b,\zeta,\mu,\rho) \ ,
\end{equation}
with the anomalous dimension
\begin{equation}
\gamma_F= 2\frac{\alpha_sC_I}{\pi}\left(\ln\rho+B_I-\frac{1}{2}-\frac{3}{4}\right) \ .
\end{equation}
Here, for gluon distribution we have $B_I=\beta_0$, and for quark distribution we have $B_I=\frac{3}{4}$.
By solving the above evolution equations, we resum the large logarithms associated with the
TMDs,
\begin{eqnarray}
&&f(x,b,\zeta^2=\rho Q^2,\mu=Q,\rho)\nonumber\\
&&~=f(x,b,\frac{b_0^2}{\rho b^2},\frac{b_0}{b},\rho)\textmd{exp}\left[\int_{b_0/b}^Q\frac{d\mu}{\mu}\left( K(b,\mu)+G(\zeta,\mu) +\gamma_F  \right) \right]\nonumber\\
&&~=f(x,b,\frac{b_0^2}{\rho b^2},\frac{b_0}{b},\rho)
\textmd{exp}\left[\int_{b_0/b}^Q\frac{d\mu}{\mu}\left(-\ln{\frac{Q}{\mu}}\gamma_K(\mu) + \frac{2\alpha_sC_I}{\pi}B_I-\frac{\alpha_sC_I}{\pi}\left(1-\frac{\ln\rho^2}{2}\right) \right)  \right]\ , \label{resump}
\end{eqnarray}
where we have chosen $\zeta^2=\rho Q^2$. All the large logarithms
have been resummed into the Sudakov form factors.
In addition, applying the one-loop results in Eqs.~(\ref{tmdq},\ref{tmdq}),
we find that the large logarithms ($\ln\frac{b_0^2}{b_\perp^2\bar\mu^2}$)
associated with the integrated parton distributions can be resummed by choosing the
relevant scale as $\bar \mu=b_0/b$. By doing that, the first factor
in the above equation can be replaced with
\begin{equation}
f(x,b,\frac{b_0^2}{\rho b^2},\frac{b_0}{b},\rho)\rightarrow f(x,\bar\mu=b_0/b) \left(1+\alpha_s\cdots\right)\ ,
\end{equation}
where the right hand side is the integrated parton
distributions at the scale of $\bar\mu=b_0/b$, and we have also
neglect all the constant terms, such as $\rho$-dependent terms.
These terms are beyond the accuracy of NLL we are considering
in this paper.

For the soft factor, it satisfies the evolution equation of
\begin{equation}
\frac{\partial}{\partial\ln\mu}S_{IJ}(b,\mu, \rho)=-\sum_L S_{IL}\;\Gamma^S_{LJ}-\sum_L \Gamma^{S\dagger}_{IL}\;S_{LJ}\ ,
\end{equation}
where
\begin{eqnarray}
\Gamma_S&=&  \frac{\alpha_s}{2\pi} \left[A\;(\ln\rho^2-2)
+D_{I1}\ln\left(\frac{s}{R_1^2P_T^2}\right)
+D_{I2}\ln\left(\frac{s}{R_2^2P_T^2}\right)\right]\Gamma^E +\gamma^s \ . \label{gammas}
\end{eqnarray}
Here $\Gamma^E$ is an identity matrix, for di-gluon initial states $A=C_A$, for di-quark initial states $A=C_F$ and
for quark-gluon initial states $A=(C_F+C_A)/2$. For quark jet $D_I=C_F$, while for gluon jet $D_I=C_A$. According to
the one-loop correction of $S^{(1)}$, the $\gamma^s$ factor for each channel can be directly read out.
For example, for the production channel $q + \bar{q} \rightarrow q + \bar{q}$, we find
\begin{eqnarray}
\gamma^s=\frac{\alpha_s}{\pi}\left[
           \begin{array}{cc}
             2C_F\,T & -\frac{C_F}{C_A}\,U \\\\
             -2U     & -\frac{1}{C_A}(T-2U) \\
           \end{array}
         \right] \ .
\end{eqnarray}
For the channel $q + q \rightarrow q + q$,
\begin{eqnarray}
\gamma^s=\frac{\alpha_s}{\pi}\left[
           \begin{array}{cc}
             2C_F\,T & \frac{C_F}{C_A}\,U \\\\
             2U     & (C_A-2/C_A)\,U-\frac{1}{C_A}\,T \\
           \end{array}
         \right]\ .
\end{eqnarray}
For the channels $g + g \rightarrow \bar{q} + q$ and $q + \bar{q} \rightarrow g + g$,
\begin{eqnarray}
\gamma^s=\frac{\alpha_s}{\pi}\left[
           \begin{array}{ccc}
             0 & 0 & U-T \\\\
             0 & \frac{C_A}{2}(T+U) & \frac{C_A}{2}(U-T) \\\\
             2(U-T) & \frac{(C_A^2-4)}{2C_A}(U-T) & \frac{C_A}{2}(T+U) \\
           \end{array}
         \right] \ .
\end{eqnarray}
For the channel $q + g \rightarrow q + g$,
\begin{eqnarray}
\gamma^s=\frac{\alpha_s}{\pi}\left[
           \begin{array}{ccc}
             (C_A+C_F)T & 0 & U \\\\
             0 & C_F\,T+\frac{C_A}{2}U & \frac{C_A}{2}U \\\\
             2U & \frac{(C_A^2-4)}{2C_A}(U) & C_F\,T+\frac{C_A}{2}U \\
           \end{array}
         \right] \ .
\end{eqnarray}
For the channel $g + g \rightarrow g + g$,
\begin{eqnarray}
\gamma^s=\frac{\alpha_s}{\pi}\left[
                         \begin{array}{cccccccc}
                    3T & 0 & 0 & 0 & 0 & 0 & 0 & 0 \\\\
                    0 & 3U  & 0 & 0 & 0 & 0 & 0 & 0 \\\\
                    0 & 0 & 3(U+T) & 0 & 0 & 0 & 0 & 0 \\\\
                    0 & 0 & 0 & 6T & 0 & -6U & 0 & 0 \\\\
                    0 & 0 & 0 & 0 & 3T+\frac{3}{2}U & -\frac{3}{2}U & -3U & 0 \\\\
                    0 & 0 & 0 & -\frac{3}{4}U & -\frac{3}{2}U & 3T+\frac{3}{2}U & 0 & -\frac{9}{4}U \\\\
                    0 & 0 & 0 & 0 &-\frac{6}{5}U & 0 & 3U & -\frac{9}{5}U \\\\
                    0 & 0 & 0 & 0 & 0 & -\frac{2}{3}U & -\frac{4}{3}U & 2(2U-T) \\
                         \end{array}
                       \right] \ .
\end{eqnarray}
By solving the evolution equation, we obtain
\begin{eqnarray}
S(b,\mu=Q)&=& \textmd{exp}\left\{ -\int^Q_{b_0/b}\frac{d\mu }{\mu}\frac{\alpha_s}{\pi} \left[A\;(\ln\rho^2-2)
+D_{I1}\ln\left(\frac{s}{R_1^2P_T^2}\right)
+D_{I2}\ln\left(\frac{s}{R_2^2P_T^2}\right)\right] \right\}  \nonumber\\
& \times &\textmd{exp}\left[ -\int^Q_{b_0/b}\frac{d\mu }{\mu}\gamma^{s\dagger}(\alpha_s(\mu))\right]S(b,\mu=b_0/b)\textmd{exp}\left[ -\int^Q_{1/b}\frac{d\mu }{\mu}\gamma^{s}(\alpha_s(\mu))\right] \ ,\label{resums}
\end{eqnarray}
where the first factor comes from the contribution of the $\Gamma_E$ term in Eq.~(\ref{gammas}).
We note that since there is only one scale present in $S(b,\mu=b_0/b)$,
it does not contain any large logarithm.

Substituting the above solutions, Eqs.~(\ref{resump}) and (\ref{resums}), into the
factorization formula, Eq.~(\ref{tmdb}), we obtain the final resummation
results for $W$. In particular, since we have set the factorization scale
$\mu=Q$, there will be no large logarithms in the hard factor ${\bf H}$.
Therefore, all the large logarithms have been resummed into the
Sudakov form factors, and we have
\begin{eqnarray}
W\left(x_1,x_2,b\right)&=&x_1\,f_a(x_1,b_0/b_\perp)
x_2\, f_b(x_2,b_0/b_\perp) e^{-S_{\rm Sud}(Q^2,b_\perp)} \nonumber\\
&\times& \textmd{Tr}\left[\mathbf{H}(Q)
\mathrm{exp}[-\int_{b_0/b_\perp}^{Q}\frac{d
\mu}{\mu}\mathbf{\gamma}_{}^{s\dag}]\mathbf{S}(b_0/b)
\mathrm{exp}[-\int_{b_0/b_\perp}^{Q}\frac{d
\mu}{\mu}\mathbf{\gamma}_{}^{s}]\right]\ ,\label{resumf}
\end{eqnarray}
with
\begin{eqnarray}
S_{\rm Sud}(Q^2,b_\perp)=\int^{Q^2}_{b_0^2/b_\perp^2}\frac{d\mu^2}{\mu^2}
\left[\ln\left(\frac{Q^2}{\mu^2}\right)A+B+D_1\ln\frac{Q^2}{P_T^2R_1^2}+
D_2\ln\frac{Q^2}{P_T^2R_2^2}\right]\ , \label{suf}
\end{eqnarray}
which are exact the results we showed in the Introduction section.

\subsection{Contributions from the Non-global Logarithms}

In Refs.~\cite{Banfi:2008qs,Banfi:2003jj}, the so-called non-global logarithms were discussed
for the dijet correlation in hadronic collisions. They further take an
example of dijet production in DIS processes, and estimate their
contributions~\cite{Banfi:2003jj}. This non-global logarithm comes from the kinematics
of two gluon radiations, where one is within the jet and another
soft gluon outside the jet. Since it happens at ${\cal O}(\alpha_s^2)$
order, the one-loop calculations in this paper do not encounter
this non-global logarithm~\footnote{It seems possible that this
contribution might belong to the soft factor in our factorization formula of
Eqs.~(\ref{tmdqt},\ref{tmdb}) at two-loop order. At this order, the soft factor contribution
will have similar kinematics as described in~\cite{Banfi:2003jj} for the non-global
logarithm, where one gluon is within the jet (the
Wilson line in the soft factor definition) and one gluon is soft
and outside the jet. Needless to say that this has to be verified
by an explicit calculations with two gluon radiations. We plan to
come back to this issue in the future.  }. Numerically, the non-global logarithms
are negligible. This is because it starts at ${\cal O}(\alpha_s^2)$,
and in the kinematics of low imbalance transverse momentum,
the resummation is overwhelmingly dominated by leading double
logarithms contributions. In the following calculations, we will
not consider their contributions when we compare to the
experimental data.

\section{Phenomenology of Dijet Correlations at Tevatron and LHC}

In this section, we will apply our resummation formula to the dijet
production from collider experiments, including Tevatron and the LHC.
In these experiments, the leading jet energy is large, and we
expect that the resummation is dominated by the perturbative
form factors. That means that our predictions are not sensitive
to the non-perturbative form factors. 

\subsection{Non-perturbative Form Factors in the Resummation}

To apply the resummation formula for phenomenological applications,
we follow the $b_*$-prescription to introduce the non-perturbative
form factors~\cite{Collins:1984kg}, {\it i.e.},
\begin{equation}
b_\perp\to b_*=b/\sqrt{1+b^2/b_{max}^2}\ ,
\end{equation}
in the $b$-space cross section contribution $W(b_\perp)$
with $b_{max}$ a parameter which will be set as $b_{max}=0.5\,{\rm GeV}^{-1}$.
By doing that, it is guaranteed that $b_*$ is always in the perturbative
region. Therefore, $W(b_\perp)$ is replaced by
\begin{equation}
W(x_1,x_2,b)\to W(x_1,x_2,b_*)e^{-S_{\rm NP}(Q,b)} \ .
\end{equation}
The non-perturbative form factors follow the parameterizations in
Ref.~\cite{Sun:2012vc}. Since we have quark and gluon from
the initial state, we decompose the non-perturbative form factor
into the quark and gluon contributions,
\begin{equation}
S_{\rm NP}^{ab\to cd}(Q,b)=S_{NP}^{(a)}(Q,b)+S_{NP}^{(b)}(Q,b)\ ,
\end{equation}
for a partonic channel $ab\to cd$. In the right hand side
of the above equation, $S_{\rm NP}$ depends on the
flavor of the incoming partons,
\begin{eqnarray}
S_{\rm NP}^{(q)}&=&\frac{g_1^{(q)}}{2} b^2+\frac{g_2^{(q)}}{2}\ln\left(\frac{Q}{2Q_0}\right) b^2+ g_3^{(q)}\ln(10x_1) b^2 \ ,\nonumber\\
S_{\rm NP}^{(g)}&=&\frac{g_1^{(g)}}{2} b^2+\frac{g_2^{(g)}}{2}\ln\left(\frac{Q}{2Q_0}\right) b^2+ g_3^{(g)}\ln(10x_1) b^2 \ ,
\end{eqnarray}
with the following parameters:
\begin{eqnarray}
&g_1^{(q)}=0.21, ~~~g_2^{(q)}=0.68, ~~~g_3^{(q)}=-0.29, \nonumber\\
&g_1^{(g)}=0.03, ~~~g_2^{(g)}=0.87, ~~~g_3^{(g)}=-0.17 \ .
\end{eqnarray}
These parameters are fitted to the hard scattering processes in the
relevant $q\bar q$ and $gg$ processes in Ref.~\cite{Sun:2012vc}.
In our calculations, we assume that these non-perturbative
form factors apply to the dijet production processes as well. This
is an approximation. However, we would like
to emphasize that because the jet energy is so large that our final
results are not sensitive to the non-perturbative form factors at all.
We have also checked several recent proposals for the non-perturbative
form factors~\cite{Konychev:2005iy,Su:2014wpa}, and found that
all of them predict almost the same distribution for dijet
production in the following phenomenological studies.

\subsection{Dijet Correlations at the Tevatron}

With all the ingredients calculated above, we now compare our resummation
results to the experimental data from the Tevatron. In experiments, the normalized
differential cross sections are measured,
\begin{eqnarray}
\frac{1}{\sigma_{\rm dijet}}\frac{d\sigma_{\rm dijet}}{d\phi_{\rm dijet}}\ ,
\end{eqnarray}
where $\phi_{\rm dijet}$ is the azimuthal angle between the leading two jets. The
leading jets are in the separate transverse momentum bins, with the second
leading jet transverse momentum $P_T>40$ GeV. The events are selected
in the mid-rapidity, $|y_{jet}|<0.5$.

\begin{figure}[tbp]
\centering
\includegraphics[width=6cm]{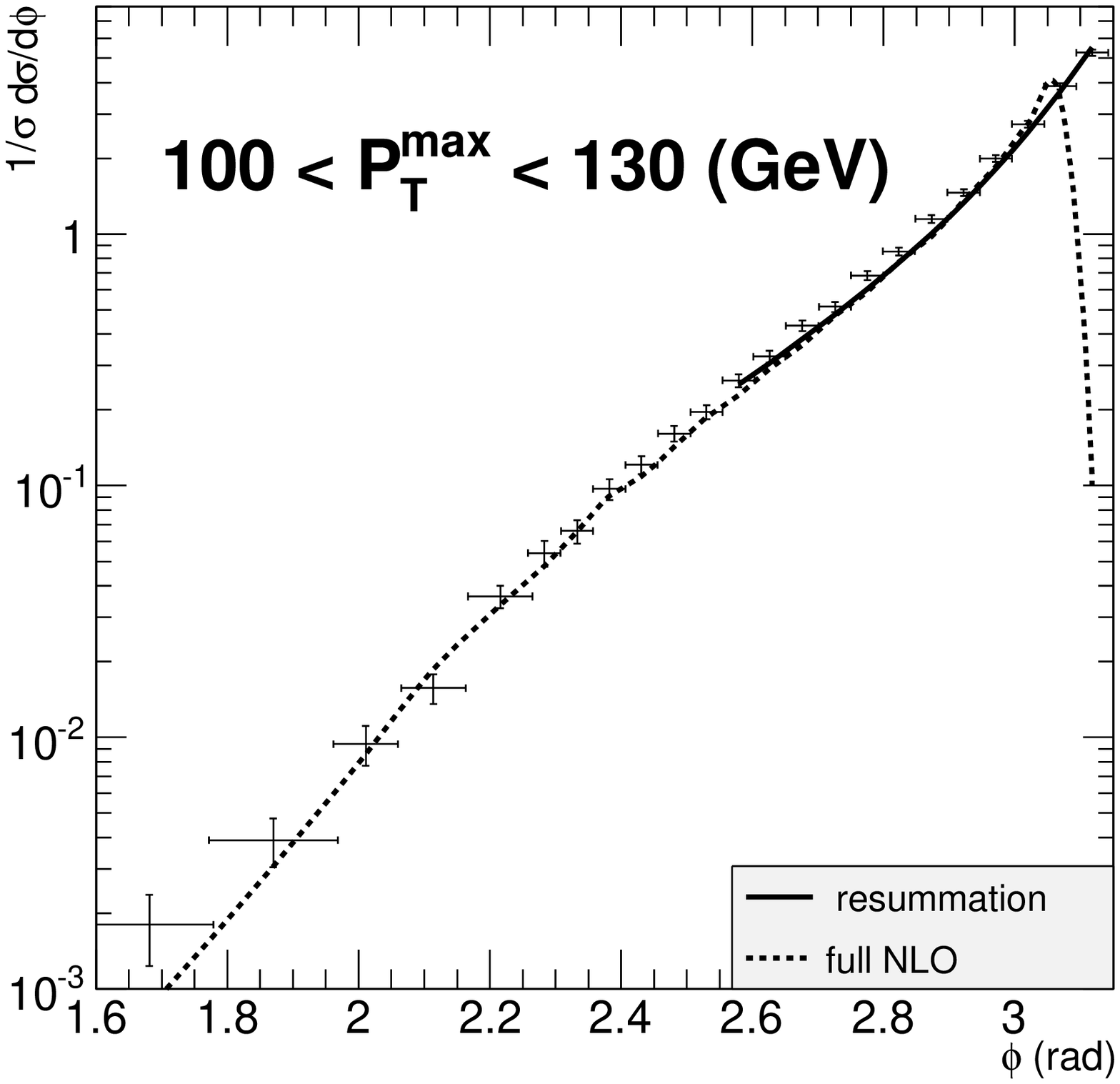}
\includegraphics[width=6cm]{Dijet_100_130_res_da.eps}
\includegraphics[width=6cm]{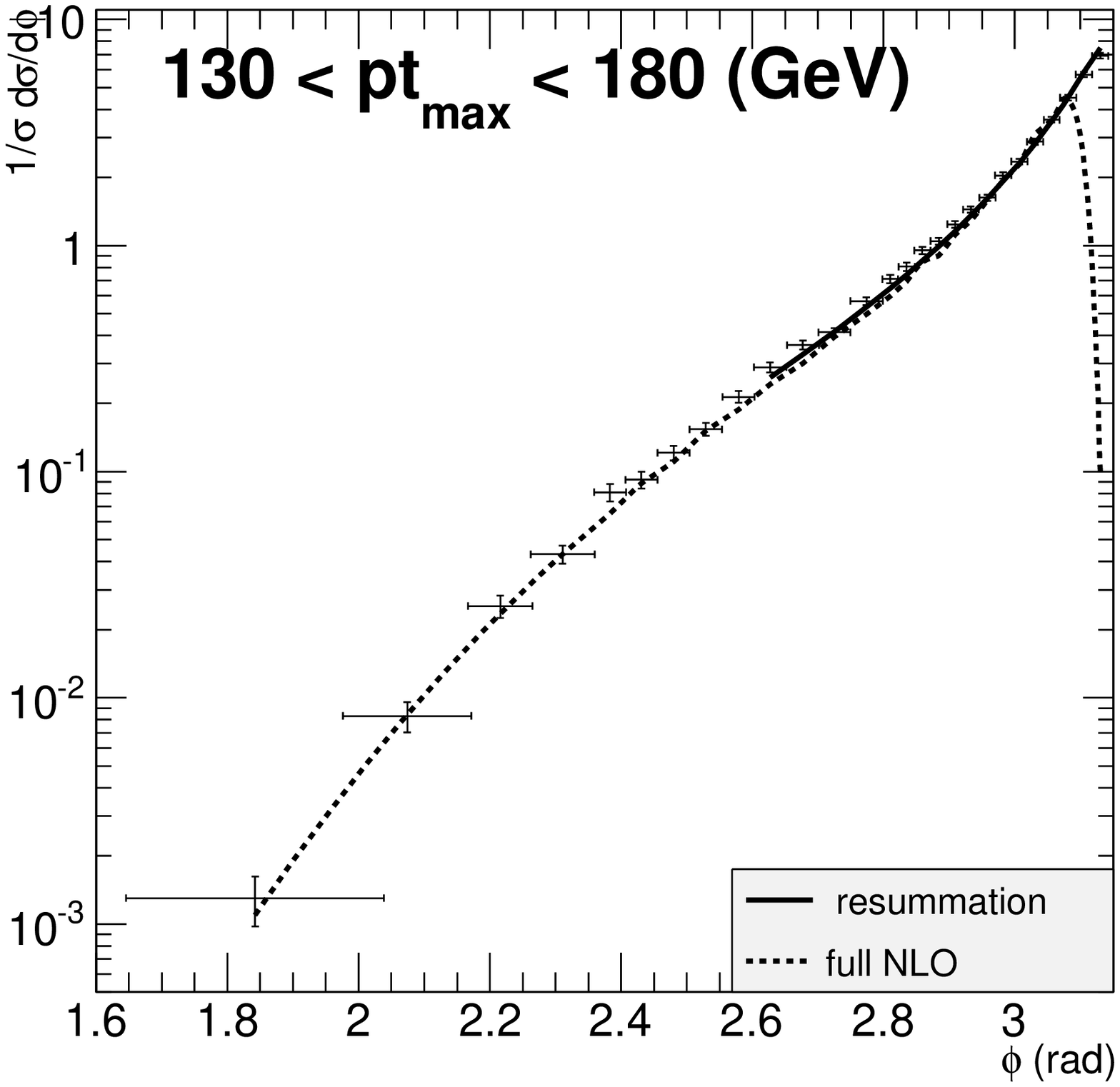}
\includegraphics[width=6cm]{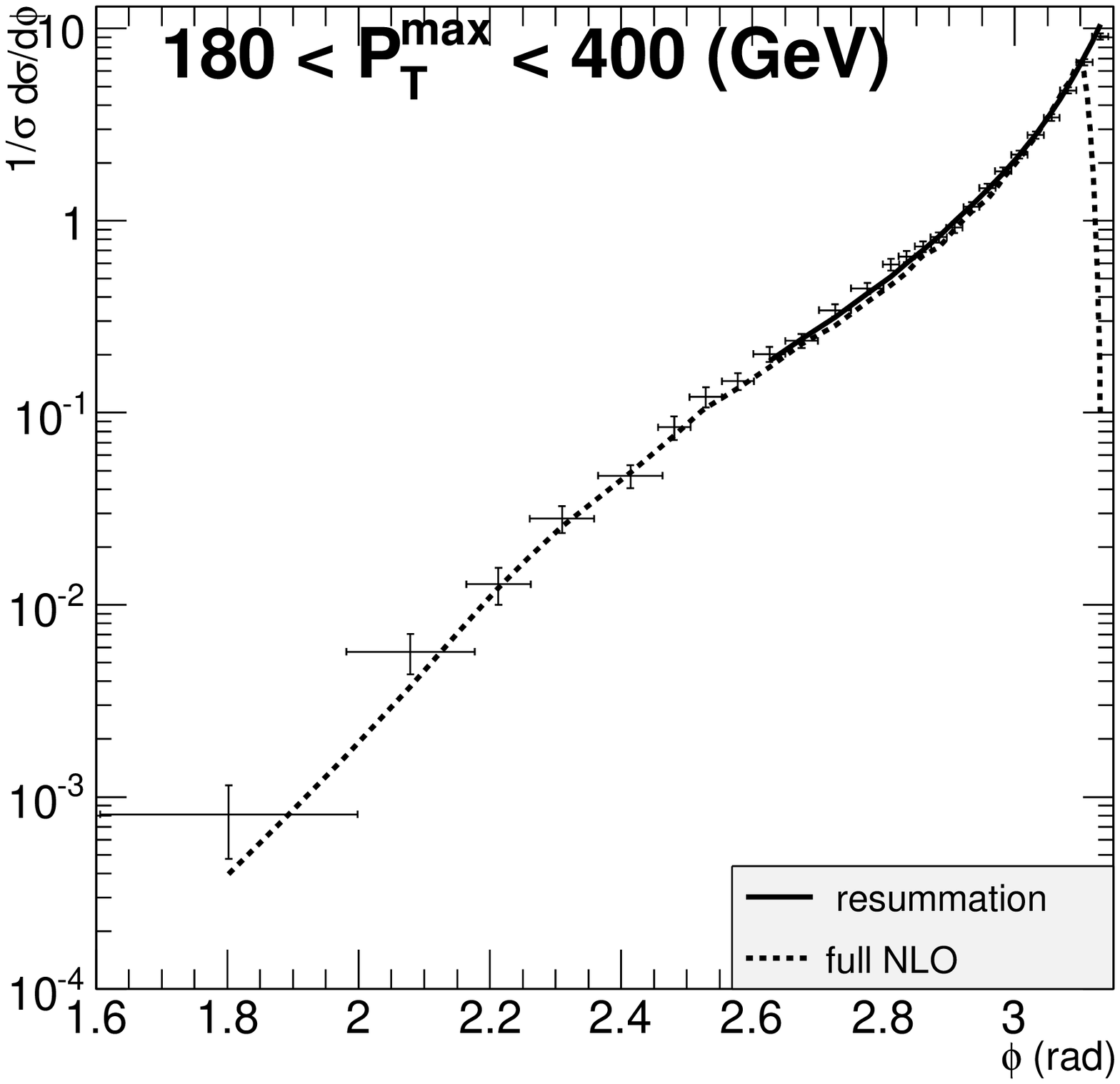}
\caption{The comparisons between the theory calculations and the
experimental data from the D0 collaboration at the Tevatron.}
\label{exp-tevatron}
\end{figure}

In Fig.~\ref{exp-tevatron}, we plot the comparisons between our
resummation result and the experimental data from D0 collaboration
at Tevatron. For completeness, we also show the prediction of
a fixed order calculation at the NLO~\cite{Nagy:2001fj},
which includes both one-loop $2 \to 3$ and
tree level $2\to 4$ contributions.
To compare with the normalized differential cross section, we
have normalized the result of our resummation calculation in
$\phi_{\rm dijet}$ distribution by the LO dijet cross section, with both
jets in the specified $p_T$ and $y$ bins of the data point.
Namely, the normalization factor $\sigma_{\rm dijet}$ is taken to be
the NLO dijet cross section in the NLO prediction, and the LO dijet
cross section in the resummation prediction. This is because in our
resummation calculation, cf. Eq.~(\ref{resumf}), the
hard factor ${\bf H}$ is only kept at the LO in this calculation.
From these plots, we can clearly see that the resummation
results agree well with the experimental data around the back-to-back
correlation region of $\phi\sim \pi$.
For smaller value of $\phi$
(away from the back-to-back configuration), the resummation calculations match to
the fixed order results at NLO~\cite{Nagy:2001fj}, which has
also been separately shown in Fig.~\ref{exp-tevatron}.
We note that a full NLO calculation cannot describe
experimental data for $\phi\sim \pi$~\cite{Abazov:2004hm},
where the fixed order calculation becomes divergent.
Our resummation calculation, after being matched with
the NLO result
clearly improves the theory prediction and can describe the experimental data
in a wider kinematic region.
This demonstrates
the importance of all order resummation in perturbative calculations
for these type of hard QCD processes.

\subsection{Dijet Correlations at the LHC}

Dijet production processes are among the first few measurements
of $pp$ collisions at the LHC. Both CMS and ATLAS have reported
the experimental results on the azimuthal angular correlations
of dijet productions, as done by the
D0 Collaboration at the Tevatron.

\begin{figure}[tbp]
\centering
\includegraphics[width=10cm]{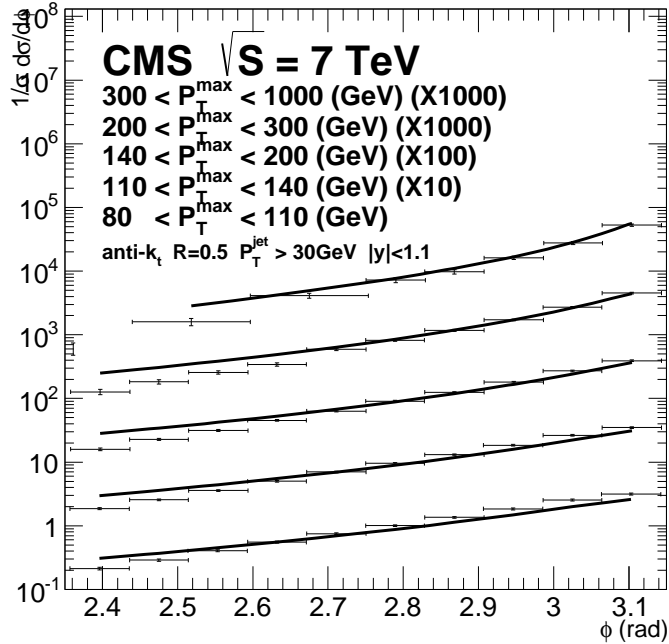}
\caption{The comparisons between the resummation results
and the experimental data from the CMS collaboration at the LHC.
The kinematics of each bins are specified according to the experiment~\cite{Khachatryan:2011zj}.}
\label{exp-cms}
\end{figure}

In Fig.~\ref{exp-cms}, we compare our resummation results
with the experimental data from the CMS collaboration at the
LHC. Similar to the D0 measurements, the dijet measurements
are presented in several kinematic bins, with the leading jet transverse
momentum labelled by $P_T^{max}$ in the figure. The second
jet transverse momentum is chosen to be larger than $30$ GeV.
Both jets are in the mid-rapidity region, $|y_{jet}|<1.1$.
Anti-$k_t$ jet algorithm with jet size $R=0.5$ was used in the data analysis. We have
also applied this algorithm in our calculations. In this figure,
we limit the comparisons in the back-to-back correlation
region, where we find perfect agreements between the
resummation calculations and the experimental data over
all transverse momentum bins.
Similar to that in Fig.~\ref{exp-tevatron}, away from
the back-to-back region, the resummation calculations will
match to the fixed order results.

\begin{figure}[tbp]
\centering
\includegraphics[width=10cm]{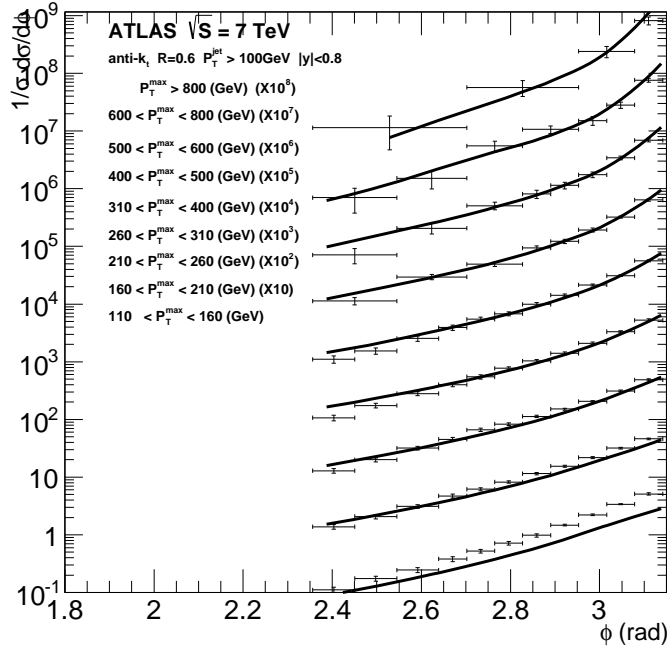}
\caption{The comparisons between the theory calculations and the
experimental data from the ATLAS Collaboration at the 7 TeV LHC.}
\label{exp-atlas}
\end{figure}

In Fig.~\ref{exp-atlas}, we compare to the measurements form the
ATLAS collaboration~\cite{daCosta:2011ni}. In this experiment,
the same anti-$k_t$ algorithm has been used, however, with
jet size $R=0.6$. The two jets are selected from the mid-rapidity region
($|y_{jet}|<0.8$) with minimum transverse momentum of 100 GeV. The data sets
are chosen according to different leading jet transverse momentum
region as indicated in the figure. From this figure, we can see that
the agreements between the resummation results and the experimental
data are very well around the back-to-back correlation regions, except
in the lowest $P_T^{max}$ bin.
The apparent poor agreement between the resummation prediction and the
ATLAS data in the lowest $P_T^{max}$ bin (between 110 GeV and 160 GeV)
is caused by the stronger
kinematic cut made on the second jet $P_T$,
which is required to be above 100 GeV at ATLAS and 30 GeV at CMS.
With a much tighter cut on this second jet $P_T$, the phase space for multiple
soft gluon emission is limited so that our resummation calculation
(which allows all possible soft gluon radiation) becomes less reliable
in this case. We note that in the lowest $P_T^{max}$ bin, the
cross section is dominated by $P_T^{max}$ around 110 GeV which is close to
the 100 GeV cut on the second jet made by the ATLAS.

\section{Summary and Discussions}

In summary, in this paper, we have investigated all order soft
gluon resummation in dijet production processes in hadronic collisions.
The procedure and methodology follow the original CSS resummation for
massive neutral particle production. Because the final state jets
carry color, the resummation formulas have to be modified to include
the effect of soft gluon radiation associated with the final state jets.
In the derivation,
we calculated the complete one-loop contributions from soft
and collinear gluon radiations in all partonic channels in dijet productions.
The soft divergences are shown to be cancelled out completely
between real and virtual graphs, which provides an important check
to our one-loop calculations.
In order to derive an analytic
expression with the jet cone size dependence to demonstrate the
cancellations in the final results and to derive the resummation formula, we
apply the narrow jet approximation in our calculations.
We have also implement the (anti-$k_t$)
jet algorithm to separate out the out-of-jet cone radiation
from the gluon radiation inside the cone jet.
Hence, the final results of our calculation
depend on the jet algorithm and the jet size, which
can then be directly compared to experimental data.
As an important cross check, we have compared our
derivation of soft and collinear gluon radiation contribution to
the fixed order calculations at this order, and the numeric
comparisons show that they agree very well
in the kinematics of back-to-back correlation regions.
In this region, because the predictions of fixed order calculations
are divergent, we have to take into account all order
resummation effects.

We have compared our resummation results to the experimental
data from the measurements at the Tevatron and LHC. All these
comparisons demonstrate that the resummation results are
crucial to improve the theory descriptions of the experimental data
around the azimuthal back-to-back kinematic region. The combination
of the NLO perfurbative calculations (including both one-loop $2 \to 3$ and
tree level $2\to 4$ contributions) and our resummation
results provide the most adequate theory descriptions to these
experimental data.

Our calculations are the first systematic derivations of the
TMD resummation for dijet production in hadronic collisions
at the NLL order. The results have been cross checked through
various perspectives, and they are consistent within the theoretical
framework our calculations are built on. These cross checks
are nontrivial supports for the factorization arguments used
in our derivations. A number of extensions can be performed
along this direction. For example, we shall be able to calculate the soft
gluon resummation effects in the vector boson (or Higgs boson)
plus a high $P_T$ jet production at the colliders where
the total transverse momentum of the boson and the jet is much
smaller than the invariant mass of the final state particles~\cite{Sun:2014lna}.
These processes are important channels to study the Standard
Model physics at the LHC.

Finally, we would like to comment on the applications of our
results to the dijet production with large rapidity separation.
This particular kinematics is very interesting to study the QCD resummation
physics. It has long been realized that the so-called BFKL resummation
\cite{Balitsky:1978ic} will be important
in this kinematics, which is referred as the Mueller-Navelet dijet
production~\cite{Mueller:1986ey}. In particular, recently, the CMS collaboration at the LHC
has measured the dijet azimuthal correlation with large rapidity separation
between the jets, which has been interpreted as the BFKL resummation
effects~\cite{Ducloue:2013bva}. However, in this calculation, only
BFKL-type resummation has been taken into account. We would like to
argue that there should be Sudakov resummation as well. Theoretically,
how to resum large logarithms from both types of physics effects is an important
question. We will not address it in this paper. Instead, we will discuss
below the physics of our resummation formula when considered in this kinematics.

When the two jets are produced with large rapidity separation, we are
in a special kinematic region, where the physics is dominated by $t$-channel
diagrams. Therefore, we can apply the following kinematic approximations,
$ s\sim -u\gg -t $,  which also implies that $P_T^2=tu/s\approx -t$.
More importantly, all the partonic channels with $t$-channel gluon exchange
will be the most important contributions. This is because they all have terms
which are proportional to $s^2/t^2$. This includes the following 
channels: $qq'\to qq'$, $qg\to qg$, and $gg\to gg$.
In addition, by applying the above approximation, we find out that the
anomalous dimensions for the associated soft factors derived in the last section
become diagonalized. The direct consequence is that we can simplify
the final resummation formula, by absorbing the soft factor anomalous
dimension of Eq.~(\ref{resumf}) into the overall Sudakov perturbative form factor
of Eq.~(\ref{suf}).
This much simplified result, as compared to that presented in the last section,
may indicate a consistent resummation formula for the Mueller-Navelet
dijet production.
The remaining task is to develop a consistent theoretical
framework to include both physics effects induced by the
BFKL and Sudakov resummation dynamics~\cite{xiao}.

\begin{acknowledgments}
This material is based upon work supported by the U.S. Department of Energy,
Office of Science, Office of Nuclear Physics, under contract number
DE-AC02-05CH11231, and by the U.S. National Science Foundation under
Grant No. PHY-1417326.

\end{acknowledgments}

\end{document}